\documentclass[10pt,journal,compsoc]{IEEEtran}
\usepackage{graphics}
\usepackage{graphicx}
\usepackage[center]{caption}
\usepackage{romannum}
\makeatletter
\g@addto@macro\normalsize{%
  \setlength\abovedisplayskip{40pt}
  \setlength\belowdisplayskip{40pt}
  \setlength\abovedisplayshortskip{40pt}
  \setlength\belowdisplayshortskip{40pt}
}
\makeatother
\usepackage[T1]{fontenc}
\usepackage[ruled,vlined,algo2e]{algorithm2e}
\usepackage{algorithm}
\usepackage{algpseudocode}
\usepackage{float}
\usepackage[left=1.2cm, right=1.2cm, top=1.5cm,bottom=1.5cm]{geometry}
\usepackage{amsmath}
\usepackage{url}
\usepackage{hyperref}
\usepackage[caption=false]{subfig}
\usepackage{enumitem}
\usepackage{array}
\usepackage{multirow}
\usepackage[table,xcdraw]{xcolor}
\usepackage[symbol]{footmisc}
\usepackage{ragged2e}
\usepackage{verbatim}
\usepackage{enumitem}
\usepackage{amssymb}
\usepackage{booktabs}

\newcommand{\eqnspacebefore}{\vspace{-1cm}}
\newcommand{\eqnspaceafter}{\vspace{-1.25cm}}
\usepackage{xcolor}
\usepackage{soul}
\usepackage{mathtools}
\usepackage{pifont}
\usepackage{fix-cm} 
\usepackage[
    n,
    operators,
    advantage,
    sets,
    adversary,
    landau,
    probability,
    notions,    
    logic,
    ff,
    mm,
    primitives,
    events,
    complexity,
    asymptotics,
    keys]{cryptocode}

\title{SecureVAX: A Blockchain-Enabled Secure Vaccine Passport System}
\author{Debendranath Das, Indian Statistical Institute, debendra\_das@hotmail.com, India\\Sushmita Ruj,  UNSW Sydney, sushmita.ruj@unsw.edu.au, Australia\\Subhamoy Maitra, Indian Statistical Institute, subho@isical.ac.in, India}

\IEEEtitleabstractindextext{%
    \begin{abstract}
    \justify
          A vaccine passport serves as documentary proof, providing passport holders with greater freedom while roaming around during pandemics. It confirms vaccination against certain infectious diseases like COVID-19, Ebola, and flu. The key challenges faced by the digital vaccine passport system include passport forgery, unauthorized data access, and inaccurate information input by vaccination centers. Privacy concerns also need to be addressed to ensure that the user's personal identification information (PII) is not compromised. Additionally, it is necessary to track vaccine vials or doses to verify their authenticity, prevent misuse and illegal sales, as well as to restrict the illicit distribution of vaccines. To address these challenges, we propose a \emph{Blockchain-Enabled Secure Vaccine Passport System}, leveraging the power of smart contracts. Our solution integrates off-chain and on-chain cryptographic computations, facilitating secure communication among various entities. We have utilized the InterPlanetary File System (IPFS) to store encrypted vaccine passports of citizens securely. Our prototype is built on the Ethereum platform, with smart contracts deployed on the Sepolia Test network, allowing for performance evaluation and validation of the system's effectiveness. By combining IPFS as a distributed data storage platform and Ethereum as a blockchain platform, our solution paves the way for secure, efficient, and globally interoperable vaccine passport management, supporting comprehensive vaccination initiatives worldwide.

    \end{abstract}
    
}

\begin{document}

\maketitle

\section{Introduction}
A vaccine passport, also known as an immunity passport, serves as documentary proof, implying a person has been vaccinated against certain infectious diseases. It can be digital, like a phone app, or physical, like a small paper card. People can carry it with them and show it whenever required, like before entering the office, boarding an airplane, or visiting a restaurant, movie theatre, or gym.
In the recent past, we have witnessed the global impact of the COVID-19 pandemic \cite{ali2020covid}, underscoring the critical importance of vaccine passports in managing and mitigating the spread of infectious diseases. However, the relevance of vaccine passports extends beyond the COVID-19 pandemic to encompass a broader spectrum of contagious illnesses that pose significant public health risks.
Historical outbreaks such as the Spanish flu in 1918 \cite{park2021fighting}, the H1N1 influenza pandemic in 2009 \cite{patel2010pandemic}, and the Ebola outbreak in West Africa in 2014 \cite{gatherer20142014} highlight the devastating consequences of infectious diseases on global populations. These events serve as stark reminders of the urgent need for robust public health measures, including vaccination and disease containment strategies.
As the COVID-19 pandemic continues to persist in recent memory, the implementation of a vaccine passport system emerges as a crucial tool in safeguarding public health and promoting safe mobility. The prospect of fully reopening businesses, facilitating international travel, and reviving economies hinges on the adoption of vaccine passports as a means to verify vaccination status and mitigate the transmission of contagious diseases.
As international passenger traffic gradually rebounds from the impacts of the COVID-19 pandemic, the role of vaccine passports becomes increasingly prominent in facilitating safe movement across borders. By providing individuals with proof of vaccination, vaccine passports not only enable access to various venues and activities but also contribute to broader public health objectives by arresting the spread of infectious diseases.

\subsection{Challenges of the Vaccine Passport System}

Vaccine passports play a crucial role in managing infectious diseases, yet they also present several challenges. 

\begin{enumerate}[leftmargin=*]
    \item \textbf{Authentication Concerns:} The issuance of vaccine certificates by healthcare clinics or third parties introduces the risk of counterfeit or fraudulent passports. Instances of counterfeit vaccine certificates have been reported in various disease outbreaks, such as the Ebola outbreak in West Africa. For example, during the COVID-19 pandemic, there were reports of individuals purchasing fake vaccine certificates on the black market to bypass travel restrictions or gain access to venues. Such incidents underscore the importance of robust authentication mechanisms to prevent forgery and ensure the integrity of vaccine passports.
    
    \item \textbf{Regulatory Ambiguity:} The proliferation of various vaccines introduces complexity in determining their efficacy, dosage requirements, and international recognition. Lack of standardization across different diseases complicates travel restrictions and entry requirements based on vaccine type. For instance, certain vaccines may not be universally accepted for international travel during outbreaks of diseases such as yellow fever or Zika virus. In the recent past, during the COVID-19 outbreak, we have seen that the European Union does not allow Indian citizens vaccinated with Covishield to enter their countries \cite{covirestriction}; likewise, the US has not approved Sputnik-V and Co-vaxin. Such inconsistencies highlight the need for harmonization of vaccine regulations to facilitate global mobility and ensure equitable access to travel opportunities.
    
    \item \textbf{Equity Issues:} Vaccine distribution disparities worsen existing inequalities worldwide. Limited vaccine availability, along with unequal distribution, means some privileged people receive doses while others do not. Unequal access to healthcare, coupled with restrictions on the import and export of medical supplies, exacerbates global health inequalities. For example, during the H1N1 influenza pandemic, wealthier nations stockpiled vaccines, leaving developing countries with limited access. Additionally, concerns about fake vaccine certificates and unethical practices at vaccination centers undermine fairness and trust in the vaccination process.
    
    \item \textbf{Privacy Considerations:} Privacy concerns emerge as individuals may be reluctant to disclose personal information beyond vaccination status. A vaccine passport should solely serve the purpose of verifying vaccination without compromising individuals' privacy by revealing unnecessary personal details. Furthermore, individuals with medical exemptions may hesitate to disclose their health conditions due to fears of social stigma or discrimination.
\end{enumerate}

By acknowledging and addressing these challenges, the development and implementation of vaccine passport systems can better serve their intended purpose while safeguarding public health and individual rights across various infectious diseases.
\subsection{Limitations of Existing Legacy Technology}

\begin{enumerate}[leftmargin=*]

\item \textbf{Data Tampering Vulnerabilities:} Traditional centralized databases are vulnerable to unauthorized data changes, which can occur due to insider manipulation or external hacking attempts. These systems often lack robust auditing mechanisms, making it difficult to detect and prevent fraudulent alterations. As a result, data integrity is compromised, leading to potential inaccuracies in critical information such as vaccination records.

\item \textbf{Lack of Transparency and Auditability:} Centralized systems operate as opaque entities managed by a single trusted authority. This lack of transparency means that users must place blind trust in the administrators maintaining these systems. Without open mechanisms for independent verification and auditing of data integrity, concerns arise regarding the accuracy and reliability of the information stored within these databases.

\item \textbf{Centralized Trust Model:} Current solutions rely on a centralized trust model, wherein a single entity controls the entire system. While this may simplify management, it introduces significant vulnerabilities, such as single points of failure and susceptibility to insider threats or external attacks. Moreover, the lack of decentralized data integrity undermines the overall reliability of these systems.

\item \textbf{Data Silos and Lack of Interoperability:} Data fragmentation across various institutions, organizations, or nations creates silos that hinder global information sharing. Incompatible data formats, standards, and policies further exacerbate this issue, impeding efforts to establish seamless interoperability. Without standardized and universally accessible platforms for managing immunization records, the efficient exchange of critical data remains a challenge.

\item \textbf{Privacy Concerns:} Centralized databases containing sensitive healthcare information raise significant privacy concerns. The potential for data leakage or misuse poses risks to individuals' privacy rights and confidentiality. Inadequate privacy-preserving controls and access policies further compound these concerns, highlighting the need for enhanced safeguards to protect users' personal information while still enabling efficient verification of vaccination status.

\end{enumerate}

\subsection{How does Blockchain help to solve these problems?}
In the Vaccine Passport System, we want the following criteria to be satisfied - 
\begin{itemize}
    \item various countries are involved that allow their foreign guest to enter only if a valid vaccine certificate is produced by the traveler. Here, the basic assumption is that the verifier (i.e. the entity that verifies the vaccine passport) does not trust the user or the traveler.
    \item At the same time, the involved parties in the system would like to take a common decision on whether the travelers should produce a valid vaccine certificate, and they can not forge the certificate in anyways.
    \item The users or travelers do not want to reveal their personal information due to privacy concerns - they intend to prove that they are vaccinated and nothing more.
    \item Also, from the global perspective, we want an equitable distribution of vaccine doses across all countries to guarantee fairness.
\end{itemize}
Blockchain integrated with smart contracts can be a perfect fit to satisfy the above-mentioned criteria.
\begin{enumerate}
    \item \textbf{Immutability :} Storing the records of vaccine passports in a Blockchain helps to protect from any unauthorized modifications of data. The immutable property of Blockchain guarantees that once the data is stored inside a Blockchain, it persists permanently  - no one can tamper with the data.
    \item \textbf{Decentralized Distributed System :} Since Blockchain is a distributed system, that does not solely depend on a single centralized authority for verification of transactions or records - Here, multiple nodes of different countries located across the globe would take part to validate a transaction. So, the probability to forge a vaccine passport by a single entity is almost negligible. Any certain specific country also can not be able to function maliciously in such a distributed network spanning across the entire world.
    \item \textbf{Privacy :} In a public blockchain system, as data exists across the various nodes,  so one might be concerned about privacy issues. However, we can cleverly store the record in a blockchain so that we can easily verify that a person is vaccinated without revealing any details about his/her private information.
    \item \textbf{Fairness :} Smart Contracts can be written in such a way that we can obtain statistics about how many people of the total population of a country have been vaccinated. This information essentially helps to identify which countries are lagging in the vaccination process, and in turn, it helps the Governments of so many countries take necessary actions immediately. Also, a smart contract ensures that no one can forge a vaccine certificate colluding with the vaccination center and a malicious vaccination center cannot put wrong records regarding the vaccination status inside the blockchain, intending to sell vaccine doses to a third party for money.
    \end{enumerate}

    \subsection{Major Contribution}
    In this work, we propose a vaccination passport system using a blockchain framework to ensure the following - 
    \begin{itemize}
        \item \emph{Fairness:} 
        \begin{enumerate}
            \item An individual can not create a fake vaccine certificate itself or colluding with the vaccination center.
            \item A vaccination center can not record any false information into the blockchain that a person has got vaccinated while is not the case. Also, it can not sell vaccine doses to third parties for extra money.
            \item Global perspective - Obtaining statistics regarding the proportion of the citizen who got vaccinated out of the total population for a specific country and also to get the figure about the number of available vaccine doses in different centers. It in turn allows distributing the vaccine doses where the supply is insufficient - ensuring every part across the globe should get a uniform/fair share of the vaccine doses. 
        \end{enumerate}
        \item \emph{Immutability:} Record stored in the blockchain regarding the vaccination status of a person, can not be altered by any malicious or unauthorized access.
        \item \emph{Privacy:} We will store information in the blockchain in such a way that it does not compromise the private information of a person, but still a person can verify to others that he/she has got vaccinated. Also, we incorporate an access control policy; it is up to the person to decide who can verify the vaccination record of the person. The person can grant (or revoke) access permission to (or from) other parties occasionally.
    \end{itemize}
    Although most of the crucial problems are solved in our proposed work, however, the problem caused due to lack of regulation still persists. We believe this can not be solved unless and until WHO specifies and standardizes the set of valid vaccines.
    
\subsection{Organization}
    The rest of the paper is structured as follows - \\
	Section~\ref{Related Work} briefly discusses the current state of the art. In Section~\ref{Preliminaries}, we discuss the basic building blocks. Section~\ref{System Model} describes our system model. Here, we have discussed our system components, security goal, adversarial model, assumptions, major procedures, and other technical details. We addressed our security claims in Section~\ref{Security Analysis}. The outcomes of our proposed system are shown in Section~\ref{Results and Discussions}. Finally, Section~\ref{Conclusion and Future Scope} concludes the paper and showcases some future directions.
	
\section{Related Work}
\label{Related Work}
\begin{table}[!ht]
\centering
\caption{Drawbacks of Various Vaccine Passport Projects Initiated by Governments and Private Organizations}
\label{tab:drawbacks}
\scalebox{0.80}{
\begin{tabular}{m{3cm} m{5cm}}
\toprule
\rowcolor{gray!30}
\textbf{System} & \textbf{Drawbacks} \\
\toprule
Vaccine Credential Initiative (VCI) &
\begin{itemize}[leftmargin=*]
    \item Requires healthcare providers to integrate with the VCI system, which may be a barrier to adoption
\end{itemize}
\\ \midrule
CommonPass &
\begin{itemize}[leftmargin=*]
    \item Limited adoption, currently being trialed by a small number of airlines and governments
\end{itemize}
\\ \midrule
Health Pass by IBM & 
\begin{itemize}[leftmargin=*]
    \item Limited adoption, currently being used by a small number of organizations
\end{itemize}
\\ \midrule
Digital Green Certificate &
\begin{itemize}[leftmargin=*]
    \item Only applicable within the European Union
    \item Limited adoption, still under development
\end{itemize}
\\ \midrule
Denmark's Corona Pass &
\begin{itemize}[leftmargin=*]
    \item Requires a centralized authority to manage and verify the information
    \item Limited adoption, only applicable within Denmark
\end{itemize}
\\ \midrule
Israel's Green Pass & 
\begin{itemize}[leftmargin=*]
    \item Requires a centralized authority to manage and verify the information 
    \item Limited adoption, only applicable within Israel 
\end{itemize}          
\\ \bottomrule
\end{tabular}
}
\end{table}

\begin{table*}[!ht]
	\centering
		\caption{A Comparative Analysis with Current State-of-the-Art}
		\label{tab:Related}
		\scalebox{0.62}{
		\begin{tabular}{c|cccccccccc}
		\toprule
            \rowcolor{gray!30}
            \textbf{Article} & \textbf{Algorithm} & \textbf{Implementation} & \textbf{Fairness} & \textbf{Privacy} & \begin{tabular}{c} \textbf{Access Control} \\ \textbf{Policy} \end{tabular} & \multicolumn{5}{c}{\textbf{Additional Security Features}} \\ \cline{7-11} 
            \rowcolor{gray!30}
            & & & & & & \begin{tabular}{c} \textbf{Prevention of} \\ \textbf{Passport Forgery} \end{tabular} & \begin{tabular}{c} \textbf{Prohibition of} \\ \textbf{Black Marketing} \\ \textbf{of Vaccine Vials} \end{tabular} & \begin{tabular}{c} \textbf{Validation of Vaccine} \\ \textbf{vial Authenticity} \end{tabular} & \begin{tabular}{c} \textbf{Use of Distributed} \\ \textbf{Storage System - IPFS} \end{tabular} & \begin{tabular}{c} \textbf{Reward/Penalty based} \\ \textbf{System for honest/} \\ \textbf{malicious entities} \end{tabular} \\ \midrule 
			\cite{hasan2020blockchain} & \ding{52} & \ding{52} & \ding{56} & \ding{52} & \ding{52} & \ding{52} & \ding{56} & \ding{56} & \ding{52} & \ding{56}\\ 
			\cite{barati2021privacy} & \ding{56} & \ding{52} & \ding{56} & \ding{52} & \ding{52} & \ding{52} & \ding{56} & \ding{56} & \ding{52} & \ding{56} \\ 
            \cite{haque2021towards} & \ding{56} & \ding{56} & \ding{56} & \ding{52} & \ding{52} & \ding{52} & \ding{56} & \ding{56} & \ding{56} & \ding{56} \\ 
            \cite{shaikh2022block} & \ding{56} & \ding{52} & \ding{56} & \ding{52} & \ding{56} & \ding{56} & \ding{56} & \ding{56} & \ding{52} & \ding{56} \\ 
            \cite{agbedanu2022blocovid} & \ding{56} & \ding{52} & \ding{56} & \ding{52} & \ding{56} & \ding{52} & \ding{56} & \ding{56} & \ding{56} & \ding{56} \\ 
            \cite{bradish2023covichain} & \ding{52} & \ding{52} & \ding{56} & \ding{52} & \ding{56} & \ding{56} & \ding{56} & \ding{56} & \ding{56} & \ding{56} \\ 
            \cite{wang2022blockchain} & \ding{52} & \ding{52} & \ding{56} & \ding{52} & \ding{56} & \ding{56} & \ding{56} & \ding{56} & \ding{56} & \ding{56} \\ 
            \cite{nabil2022blockchain} & \ding{52} & \ding{52} & \ding{52} & \ding{56} & \ding{56} & \ding{56} & \ding{56} & \ding{56} & \ding{56} & \ding{56} \\ 
            \cite{shih2022international} & \ding{52} & \ding{56} & \ding{56} & \ding{52} & \ding{56} & \ding{56} & \ding{56} & \ding{56} & \ding{56} & \ding{56} \\
            \cite{rashid2022block} & \ding{56} & \ding{52} & \ding{56} & \ding{52} & \ding{52} & \ding{56} & \ding{56} & \ding{56} & \ding{52} & \ding{56} \\
            \cite{pericas2022highly} & \ding{52} & \ding{52} & \ding{56} & \ding{52} & \ding{56} & \ding{52} & \ding{56} & \ding{56} & \ding{52} & \ding{56} \\
            \cite{cao2022blockchain} & \ding{56} & \ding{56} & \ding{52} & \ding{52} & \ding{56} & \ding{52} & \ding{56} & \ding{56} & \ding{56} & \ding{56} \\
            \cite{fugkeaw2023efficient} & \ding{52} & \ding{52} & \ding{56} & \ding{52} & \ding{52} & \ding{56} & \ding{56} & \ding{56} & \ding{52} & \ding{56}\\
            \cite{koyama2023decentralized} & \ding{56} & \ding{52} & \ding{56} & \ding{56} & \ding{52} & \ding{52} & \ding{52} & \ding{56} & \ding{56} & \ding{56} \\ 
            \cite{masood2024developing} & \ding{52} & \ding{52} & \ding{56} & \ding{52} & \ding{52} & \ding{56} & \ding{56} & \ding{56} & \ding{56} & \ding{56} \\
			Proposed Model & \ding{52} & \ding{52} & \ding{52} & \ding{52} & \ding{52} & \ding{52} & \ding{52} & \ding{52} & \ding{52} & \ding{52} \\ \bottomrule
		\end{tabular}
	}
\end{table*}

After the World Health Organization (WHO) declared COVID-19 a global pandemic in March 2020, governments, NGOs, and corporations worldwide made concerted efforts to combat the spread of the disease. One potential solution that gained significant attention was the use of immunity or vaccine passports to help manage the spread of infectious diseases. This approach contrasts sharply with contact tracing apps, which focus on tracking and isolating infected individuals during outbreaks, specifically in the pre-vaccination era when no proven treatments or vaccines were available, making isolation one of the most effective strategies to control outbreaks \cite{chakraborty2020contact}. The shift from isolating infected individuals to enabling safe interactions among vaccinated individuals in the post-vaccination context underscores a fundamental change in strategy. Although the concept of vaccine passports has existed for some time, the COVID-19 pandemic brought it to the forefront as a means to facilitate a return to normalcy. This renewed interest has highlighted the broader applicability of vaccine passports beyond the COVID-19 context for managing public health and enabling safe travel and access to services.

One of the earliest proposals for a blockchain-enabled vaccine passport system was the Vaccine Credential Initiative (VCI) \cite{vci} established in 2020 by the Argonaut Project. The VCI aims to provide individuals with secure, privacy-preserving, and interoperable access to their immunization information. The VCI uses a combination of blockchain and secure enclaves to create a tamper-proof record of an individual's vaccine status.

Numerous non-governmental organizations, professional associations, and private companies have been developing health and identity documents to address the COVID-19 pandemic. For example, the World Economic Forum and the Commons Project worked on an online platform called "CommonPass" \cite{commonpass} that documents an individual's COVID-19 status, including vaccinations, PCR tests, and health declarations. IBM has also created a "Digital Health Pass" \cite{ibmhealthpass} for health verification of employees, customers, or visitors. Meanwhile, the International Air Transport Association launched the "IATA Travel Pass Initiative" \cite{iata} smartphone app to inform airline staff and passengers about testing and vaccination requirements. 

In addition, the European Commission has proposed the creation of a "Digital Green Certificate" \cite{dgc} to facilitate safe travel within the European Union. The "Digital Green Certificate" would use blockchain technology to provide a secure and tamper-proof record of an individual's COVID-19 vaccine status and test results. In a similar direction, various other countries have developed their own vaccine passport systems, including Denmark’s "Corona Pass" \cite{coronapass} and Israel’s "Green Pass" \cite{greenpass}. 

Governments and private organizations rushed to finish these projects to combat the pandemic, resulting in shoddy applications with significant issues. Furthermore, these applications were only tested in a limited environment. For example, "CommonPass" was trialled by a handful of airlines and governments, and "Health Pass" by a few organizations. In addition, the "Digital Green Certificate" only applied within the European Union, while Denmark's "Corona Pass" and Israel's "Green Pass" were only valid within their respective countries. Furthermore, these "Corona Pass" and "Green Pass" systems rely on a centralized authority to manage and verify information, creating the potential for further problems.

The downsides of these project initiatives are listed in Table~\ref{tab:drawbacks}.
However, these proposals demonstrate the increasing interest in using secure vaccine passport systems across multiple geographical regions. So, gradually, researchers have come up with their proposals. 

The paper \cite{hasan2020blockchain} proposes a blockchain-based solution for COVID-19 management using digital immunity certificates. It highlights the benefits of data security, privacy, and cost-efficiency. However, the paper does not thoroughly address challenges like implementation difficulties, scalability, and real-world effectiveness. Further research is needed before this solution can be practically implemented.

The authors of the paper \cite{barati2021privacy} proposed a privacy-preserving distributed platform for COVID-19 vaccine passports using blockchain and smart contracts. This platform securely creates, stores, and verifies digital vaccine certificates. 

The author \cite{haque2021towards} proposed a blockchain-based COVID-19 vaccine passport system called VacciFi. It offers benefits such as integrity and verifiability. However, it faces significant challenges, including privacy concerns, potential inequality and exclusion, technical implementation issues, and hurdles in achieving widespread adoption. 

The paper \cite{shaikh2022block} introduces a blockchain-based system for secure vaccination records using smart contracts and IPFS storage. This system enhances data integrity and prevents forgery with unique hash values for each certificate. 

The paper \cite{agbedanu2022blocovid} discusses the challenges faced by Africa in procuring COVID-19 vaccines and authenticating certification. The authors suggest a blockchain-based system called BLOCOVID to secure and verify vaccination certificates using distributed ledger technology. Vaccine serial numbers and certificates are stored on the blockchain as hash values to ensure that they cannot be altered and are authentic. 

The paper \cite{bradish2023covichain} addresses privacy issues related to the personal data of users by proposing a two-factor authentication system. One part of the system relies on information that the user possesses, such as biometrics like retina scans and fingerprints. The other part relies on information that the user knows, such as personal details like date of birth, gender, and country.

In the paper \cite{wang2022blockchain}, the authors propose modifying the process of vaccine user verification by validating the user's physical license with the information provided by the user's QR code.

Another interesting work \cite{nabil2022blockchain} demonstrates the use of priority-based vaccine distribution in areas with higher positive test results.

The authors \cite{shih2022international} suggest a Hyperledger Fabric-based consortium blockchain solution to create digital vaccine passports (DVPs) for combating counterfeit paper passports during the COVID-19 pandemic. They recommend federated identity management for secure verification across different trust realms. 

The paper \cite{rashid2022block} introduces a blockchain-based solution called Block-HPCT that incorporates smart contracts for digital health passports and contact tracing using proof of location. It utilizes trusted oracles, IPFS, and Hyperledger Fabric for secure data storage, aiming for transparency in COVID-19 information management. 

In the paper \cite{pericas2022highly}, the authors introduce a protocol for managing digital COVID-19 certificates. It allows users to control data sharing in a hierarchical system using proxy re-encryption and blockchain. 

The study \cite{cao2022blockchain}  presents a blockchain-based vaccine passport system using a dual-chain framework: a public blockchain with IoT for transparent cold-chain logistics and a consortium blockchain for privacy and auditing. It employs distributed threshold signatures to prevent collusion in vaccine qualification and cryptographic tools to protect user privacy during customs checks. 

The paper \cite{fugkeaw2023efficient} introduces a system named UniVAC for verifying universal vaccine passports. It uses ciphertext policy attribute-based encryption and blockchain to provide secure access control to COVID-19 vaccine data. This setup ensures that transactions are transparently recorded and data indexing is reliable. 

The paper  \cite{koyama2023decentralized} introduces Vacchain, a blockchain-based system to improve the security and traceability of vaccine distribution. It presents a SYS-MAN mechanism for role verification, mutual agreement protocols for ownership transfer, and blockchain-based vaccine passports. The aim is to enhance data reliability and prevent counterfeiting. 

The study \cite{masood2024developing} builds a blockchain application using Solidity smart contracts to enhance vaccine traceability and certificate reliability. Tested on various networks, it shows high performance and successful deployment. 

The existing vaccine passport systems have notable limitations and research gaps. Most systems rely on a central authority, which requires trust in that authority. Not much research has been done on using blockchain technology to remove this centralization. The papers on blockchain-based solutions for COVID-19 management highlight significant challenges that impede practical implementation. Several studies, including those by \cite{hasan2020blockchain, barati2021privacy, shaikh2022block, agbedanu2022blocovid, shih2022international, pericas2022highly}, emphasize issues related to scalability, integration with diverse healthcare systems, and regulatory compliance \cite{haque2021towards, bradish2023covichain, rashid2022block, cao2022blockchain, fugkeaw2023efficient, koyama2023decentralized}. Privacy concerns and the need for robust security measures are recurrent themes in the research by \cite{haque2021towards, bradish2023covichain, rashid2022block, cao2022blockchain, fugkeaw2023efficient, koyama2023decentralized}. Additionally, several studies, such as those by \cite{wang2022blockchain, nabil2022blockchain}, highlight the challenges of ensuring data integrity and preventing forgery. The need for widespread adoption, comprehensive data management, and user consent is also noted in the works by \cite{masood2024developing, fugkeaw2023efficient}. There is significant untapped potential in using smart contracts integrated with blockchain. Critical issues such as fairness among involved parties, verification of vaccine authenticity, and prevention of black-market trading remain inadequately addressed. Our comparison with state-of-the-art systems, as found in Table~\ref{tab:Related}, highlights these gaps and emphasizes the need for further research in this area. 

In this paper, we propose a smart contract-powered blockchain system that addresses various security aspects. Instead of storing users' vaccination information on the blockchain or in a centralized database, we store it in decentralized IPFS storage. This approach provides greater security and privacy for users and reduces costs associated with storing large amounts of data on the blockchain. Our system uses smart contracts to ensure fairness among parties and verifies the authenticity of vaccine doses, thus addressing key security concerns.

    \section{Preliminaries}
    \label{Preliminaries}
        In this section, we discuss the building blocks and other proposed methodologies that we have used in our system. The set of all binary strings of length $n$ is denoted as $\{0,1\}^n$, and the set of all finite binary strings as $\{0,1\}^*$. The output \(x\) of an algorithm \( \mathcal{A} \) is denoted by \(x\) $\leftarrow$ \( \mathcal{A} \). \cite{das2024bisection}

	\subsection{Basic Cryptographic Primitives}
	\begin{itemize}[leftmargin=*]
	
		\item \textbf{Encryption Scheme:} Encryption schemes are used to ensure privacy and confidentiality of messages exchanged between a sender and recipient. Two types of encryption are commonly used: 
        \begin{itemize}
        	\item Private Key Encryption (also known as Symmetric Key Encryption or SKE): An encryption scheme where the same key is used for both encryption and decryption. It means that both the sender and the recipient must share and keep the same secret key secure. This method is efficient and fast, but it requires a secure method to exchange the key and keep it confidential to ensure the security of the communication.
        	\item Public Key Encryption (also known as Asymmetric Key Encryption or ASKE): An encryption scheme where different keys are used for encryption and decryption, typically a public key for encryption and a corresponding private key for decryption. The public key and private key are mathematically related but distinct pairs. The relationship ensures that data encrypted with a public key can only be decrypted with the corresponding private key, providing secure communication. The private key must be kept secret, while the public key can be freely distributed.
        \end{itemize}
		
		\item \textbf{Digital Signature Scheme:} Digital Signature (DS) is used for user authentication in the system, allowing recipients to verify the sender of a message. \cite{merkle2001certified} \cite{kerry2013digital}
		
		\item \textbf{Hash Function:} A hash function maintains data integrity within the system. It is defined as \textbf{$H:\{0,1\}^* \rightarrow \{0,1\}^k$}, mapping messages of arbitrary length to a fixed-size message digest of length k. An ideal hash function must possess the following three properties \cite{Preneel2011}:
        \begin{itemize}
	   \item \emph{One Way or Pre-image Resistant}
	   \item \emph{ Second Pre-image Resistant}
	   \item \emph{Collision Resistant} 
        \end{itemize}
	
	    \item \textbf{Merkle Tree:} Merkle Trees are binary trees used to verify the membership of data in a set. Each leaf node contains data from the set, and its output is the hash of the data. For non-leaf nodes, the output is the hash of the concatenated outputs of the child nodes. The output of the top node is called the \textit{Merkle Root}, denoted by the acronym \emph{MR}, which stands for Merkle Tree Root.\cite{Liu2021MT, merkle1988digital, jing2021merkle}
	

        \item \textbf{Proxy Re-Encryption(PRE): }Proxy Re-Encryption (PRE) is a cryptographic technique that enables a proxy to transform (i.e. re-encrypt) a ciphertext encrypted under one key into another key-encrypted version of the same message so that the receiver can decrypt the message using their private key \cite{nabeels2011proxy}. PRE can be helpful when the original encryptor wishes to delegate the decryption of the ciphertext to a third party without revealing the original decryption key.
	
	\end{itemize}

	\subsection{Other Building Blocks}
	
    \begin{itemize}[leftmargin=*]
    
        \item \textbf{Blockchain:} Blockchain is a decentralized, distributed, verifiable, immutable digital ledger that offers high levels of security. Its structure is based on a chain of blocks, where each block comprises a series of transactions. The blocks are interlinked through hashing (i.e. every block contains the hash of its previous block in the chain), meaning that any changes to a block's data or transactions will alter its block hash, causing mismatches in the subsequent block hashes \cite{narayanan2016bitcoin}. The security of blockchain lies in the fact that altering the hash of a block would require changing the hash of every subsequent block as well, making it virtually impossible to tamper with the data in the ledger. The most common consensus mechanism used in the blockchain is Proof-of-Work (PoW), which both Bitcoin and Ethereum use. Other consensus algorithms are also being explored, such as Proof-of-Stake, Proof-of-Burn, and Proof-of-Authority. In our work, we have opted for a PoW-based consensus mechanism to ensure the integrity and security of our blockchain platform.
        
        \item \textbf{Smart Contract:} Smart Contracts allow for the creation of self-executing contracts between parties without the need for intermediaries or mediators. These contracts are written in code, using programming languages like solidity, and can enforce the terms of the agreement automatically \cite{das2022blockchain}. The use of Smart Contracts on a blockchain network has numerous benefits, including increased security and transparency. In addition, by eliminating the need for intermediaries, the smart contract can help to build trust in the system and reduce the potential for fraud or manipulation. Integrating Smart Contracts with blockchain technology has opened up new opportunities for developing decentralized applications that can solve real-world problems in a secure and trustworthy manner. This has the potential to disrupt traditional industries and revolutionize the way that transactions are conducted.

        \item \textbf{InterPlanetary File System:} InterPlanetary File System or IPFS is a decentralized protocol for storing and sharing files on a distributed network. Instead of relying on a centralized server, IPFS uses a peer-to-peer network of nodes to store and retrieve files, making it more resilient to censorship and failure. In IPFS, files are addressed by their content rather than their location. When a file is added to the IPFS network, it is split into multiple smaller pieces; each piece is assigned a unique cryptographic hash. These hashes are then used to create a unique identifier for the entire file, called a Content Identifier (CID) \cite{benet2014ipfs}.
        When a user requests a file from the IPFS network, they use the CID to locate the file. The IPFS network uses a distributed hash table (DHT) to store information about which nodes in the network have a copy of the file. The DHT is a decentralized system that allows nodes to communicate with each other to find the file's location. The file is then retrieved from the node that has a copy of it.
    \end{itemize}    
    
\section{System Model}
\label{System Model}
\begin{figure}[!ht]
    \begin{center}  
    \includegraphics[width=\linewidth]{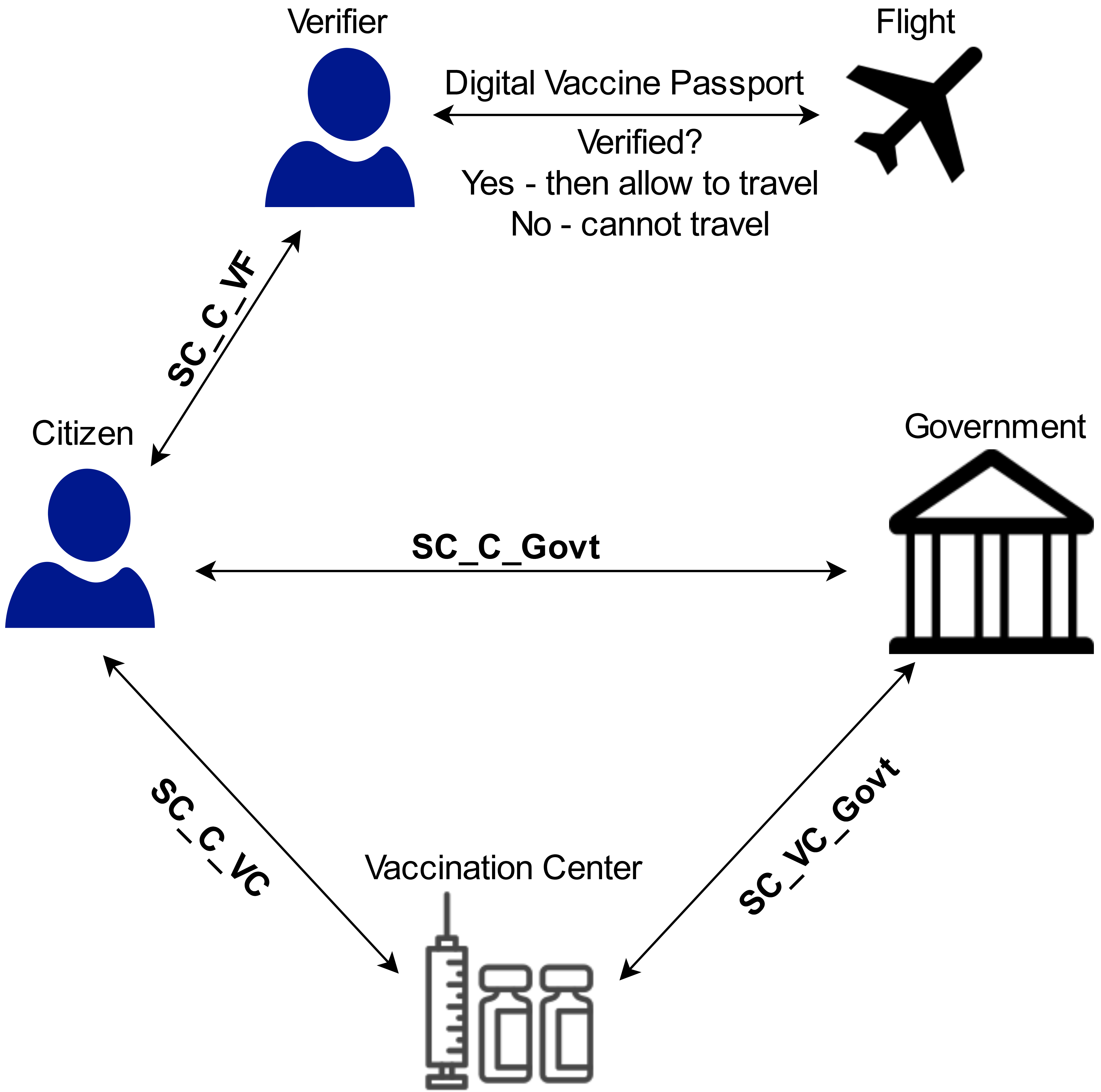}
    \caption{\small \sl System Model - Vaccination Passport System }		
    \label{Fig1}
    \end{center}  
\end{figure}

\subsection{Components}
In our system, we have the following entities/parties (Figure~\ref{Fig1}):
\begin{enumerate}
    \item \textbf{Government ($Govt$):} It is one of the major entities in our system and has several significant functionalities.
    \begin{itemize}
        \item Govt is responsible for registering VCs in the system after verifying their credentials.
        \item Govt validates citizens' identity before generating $tokenID$ for taking vaccine dose.
        \item Govt locks money in the $SC$s that gets transferred to the VCs for their service charges as per the policy.
        \item Govt also ensures the adequate supply of vaccine doses to various VCs so that the entire process can run smoothly.
    \end{itemize}
    \item \textbf{Vaccination Center ($VC$):} It actively involved in the vaccination program. Once registered into the system, VC can start their job and obtain fees for the service offered.
    \item \textbf{Citizen/User/Traveller ($C$):} It is the user (sometimes also referred to as $Citizen$ or $Traveller$) who takes the vaccine dose from the VC and then obtains the vaccine passport.
    \item \textbf{Verifier ($VF$):} It is another entity that verifies if the citizens' vaccine passports are valid.
    \item \textbf{Blockchain ($BC$):} Our proposed model employs a public blockchain e.g. Ethereum, which is a permissionless blockchain that allows anybody to join the network. $BC$ is used as a tamper-proof log of records distributed across multiple nodes.
    \item \textbf{Smart Contract ($SC$):} These are globally accessible executable pieces of code that regulate the key operations within the blockchain.
    \item \textbf{InterPlanetary File System ($IPFS$):} We have used IPFS to store the citizens' vaccine passports in a distributed manner.
\end{enumerate}

\subsection{Security Goal}
\label{Security Goal}
We state the security properties which must be realized by our proposed protocol:
\begin{itemize}[label=$\bullet$]
    \item \emph{Fairness}: To ensure that
    \begin{itemize}
        \item A person/citizen can not forge a vaccine certificate.
        \item A person should be able to validate the authenticity of the vaccine vial.
        \item Vaccination centers can not misuse the vaccine doses for their own profit. Black marketing is prevented.
        \item Global statistics of the vaccination process can be obtained.
    \end{itemize}
    \item \emph{Privacy}: To ensure that a person's private information, like name, address, etc, would not be compromised or leaked to the outside world.
    \item \emph{Data Security}: To ensure that no one can tamper with the vaccine data or records.
\end{itemize}

\subsection{Adversarial Model}
\begin{itemize}
    \item \textbf{User/Party/Player:} A user/party/player is said to be \textbf{honest} if they adhere to the system protocol. Otherwise, It is said to be \textbf{malicious} (i.e. when a user performs sporadically or deviates arbitrarily).
    \item \textbf{Adversary:} In the context of security, an adversary is a polynomial time algorithm that can compromise any user at any given point. This algorithm has an upper bound. The adversary is dynamic, meaning it can coordinate attacks through message exchanges on behalf of malicious users. On the other hand, the adversary cannot interfere with honest users' message exchanges, nor can it break cryptographic primitives like encryption schemes, digital signatures or hash functions except with negligible probability.  Additionally, the adversary is limited in computational power and storage capabilities. Finally, we assume that all participants in our system are \textbf{rational}. By \textbf{rational}, we mean that when an adversary wants to exploit certain flaws in the underlying system, its primary and sole objective is to gain sufficient benefits - either monetary or useful information. Other sorts of malevolent functionality are not considered adversarial conduct here (e.g. wasting system resources - CPU, memory, time and so on). 
\end{itemize}

\subsection{Assumption}
    \begin{enumerate}
        \item Every entity in our system has a unique $<SK, PK>$ pair where $PK$ and $SK$ represent the entity's address and authentication factor, respectively.
        \item The Government of every country is responsible for distributing the vaccine doses to its country's vaccination centers. How the Govt gets the vaccine doses from the manufacturer (i.e. supply-chain system) is out of the scope of this work.
        \item A Vaccination center (VC) must satisfy certain prerequisite conditions (e.g. it should be a hospital or healthcare unit) and appeal to the respective Government expressing their interest. The requirements or criteria may vary for different Governments of various countries. If they satisfy all the necessary measures, the Government introduces VC into the system by providing a unique ID.
        \item To ensure that every citizen has access to the vaccine, we are not charging any fees for taking vaccine doses. However, Governments pay vaccination centers for their services on a per-vaccine basis.
        \item Our system primarily uses the blockchain to store hashes of records, timestamps of significant events, issuing authority signatures, and other context-specific auxiliary information to ensure accountability, integrity, authenticity, and fairness. Citizen vaccination information is stored in IPFS, a decentralized peer-to-peer storage system.
        \item Information about registered vaccination centers, such as their location and contact details, is displayed on a public dashboard. Any updates to this information are available to the public in real-time.
        \item To simplify the problem caused by regulatory issues, we do not specify any particular vaccines by name. Instead, we assume that a person must take a dose of the vaccine. In reality, citizens must receive multiple vaccine doses periodically. 

    \end{enumerate}
    \subsection{Protocol Design}
    \label{Subsection: Protocol Design}
     \begin{table}[!ht]
		\centering
		\caption{Terminology \& Notation used in our Scheme}
		\label{tab:Table1}
		\scalebox{0.8}{
			\begin{tabular}{r|l}
                    \toprule
				\rowcolor{gray!30}
				\textbf{Abbreviation} & \textbf{Interpretation} \\
                    \toprule
				$BC$ & Blockchain \\ \hline
				$SC$ & Smart Contract \\ \hline
				V & Vaccine Dose/Vial \\ \hline
				VP & Vaccine Passport \\ \hline
				Govt & Government \\ \hline
				VC & Vaccination Center \\ \hline
				C & Citizen \\ \hline
				VF & Vaccine Passport Verifier \\ \hline	
                $<SK_{Entity}, PK_{Entity}>$ & Key pair used by an $Entity$, $Entity \in \{Govt, VC, C, VF\}$ \\ \hline
				$vID$ & Vaccine Vial ID \\ \hline
				$vcID$ & Vaccination Center ID \\ \hline
				$tokenID$ & Token ID \\ \hline
				$applID$ & Application ID\\ \hline
                $cID$ & Content Identifier in IPFS System \\ \hline
                $MD_x$ & Message Digest of x \\ \hline
                $commit_x$ & Commitment of x \\ \hline
				$T_{Event}$ & Timestamp, when the Event occurs \\ \bottomrule
			\end{tabular}
		}
	\end{table}
 
    Our proposed scheme is divided into 6 major modules, which are discussed below.\\
    
    \noindent\textbf{Module 1: Registration of Vaccination Centers (VCs)}\\
    Before participating in the vaccination program, VCs must acquire a licence. And they must meet certain pre-requisite conditions imposed by the individual country's government to obtain a licence. After validating the required credentials, the government registers an entity as VC in the system, and the VC is assigned a unique vaccination center ID ($vcID$) generated by the smart contract.
    \begin{figure}[!ht]
        \begin{center}  
        \includegraphics[width=\linewidth]{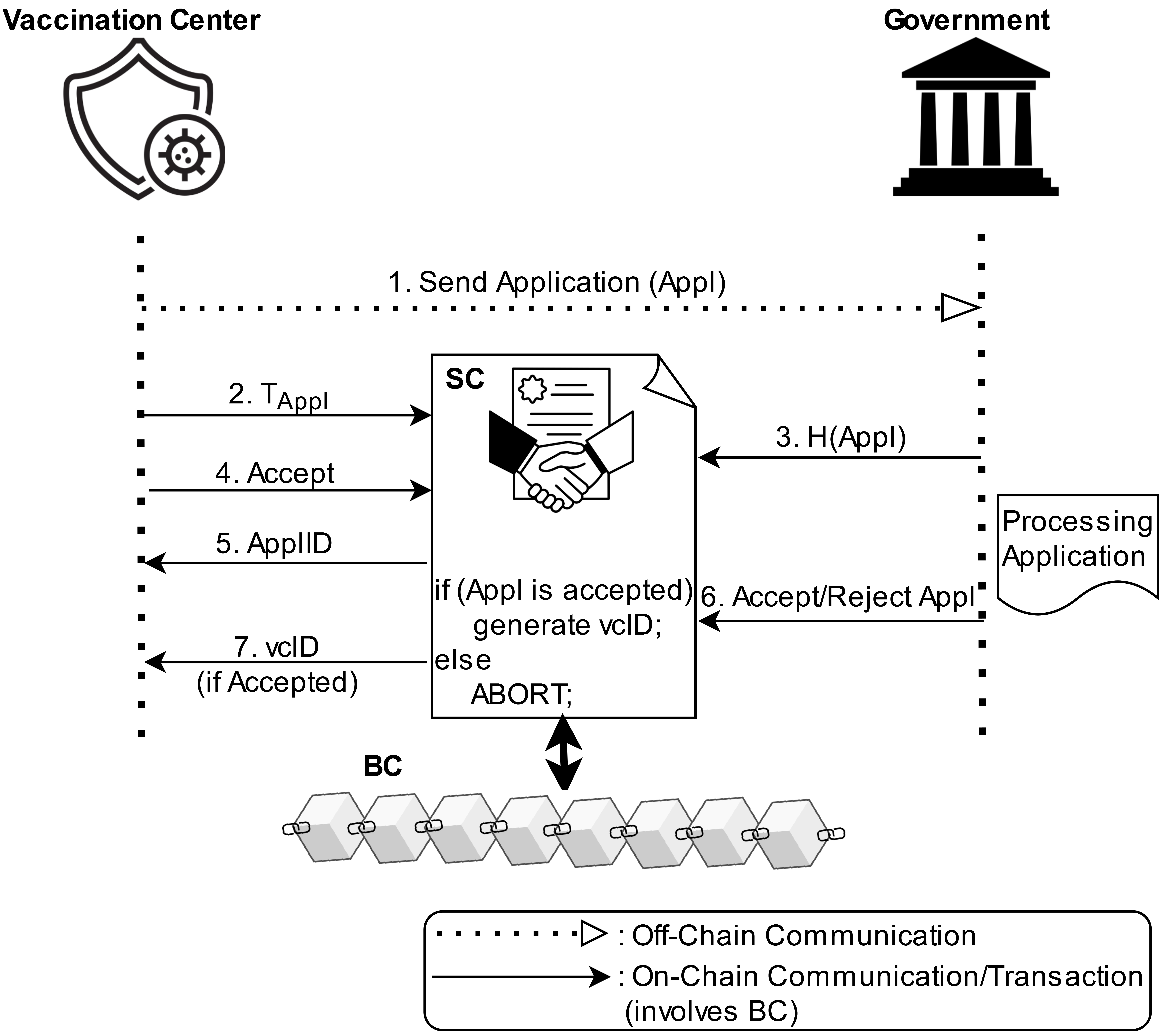}
        \caption{\small \sl Registration of Vaccination center}		
         \label{Fig2}
        \end{center}  
    \end{figure}
    Figure~\ref{Fig2} depicts the VC registration process.
    \begin{enumerate}
        \item[I.] At first, the VC sends an application ($Appl$) to the Govt, furnishing all the required details in off-chain communication and then register the timestamp of the application ($T_{Appl}$) on-chain through $SC$.
        \item[II.] Govt, in turn, creates a digest of the received application and puts it on the $BC$.
        \item[III.] Once VC agrees to the hash value, $SC$ generates an Application ID ($ApplID$) for future reference. On the contrary, if the VC does not consent to the hash value, the protocol terminates, and the VC needs to send a new application again.
        \item[IV.] Next, the Govt processes the application within a fixed period and verifies if the application satisfies the necessary requirements. Accordingly, Govt accepts or rejects the application on-chain. If the application gets accepted, the $SC$ will generate and assign a unique $vcID$ against the applicant.\\ 
    \end{enumerate}

    \noindent\textbf{Module 2: Refill of Vaccine Vials/Doses}\\
    Govt distributes the vaccine vials to the registered vaccination centers. Let us say that $V$ is the set of vaccine vial IDs, which will be transferred to the VC with ID $vcID$ upon VC applying for refilling its vaccine stock. The cardinality of the set $V$ is $|V| = n$.
    \eqnspacebefore
    \[V=\{v_{i}\}, i \in \mathbb{N} \wedge i\in[1,n]\]
    \eqnspaceafter
    \begin{enumerate}[leftmargin=*]
        \item[I.] First, the VC applies to the Govt to refill its vaccine stock. The timestamp of the application ($T_{RefillAppl}$) is recorded on $BC$.
        \item[II.] Before transferring the vaccine vials (corresponding to the set $V$) to the VC, the Govt will compute the $MR$ (i.e. Merkle Tree Root Hash) for the set $V$, organizing the member elements in ascending order. Figure~\ref{Fig3} demonstrate the computation of $MR$ for the set $V$, where $|V|$ = 8.

        \begin{figure}[!ht]
    	\begin{center}              	
            \includegraphics[width=\linewidth]{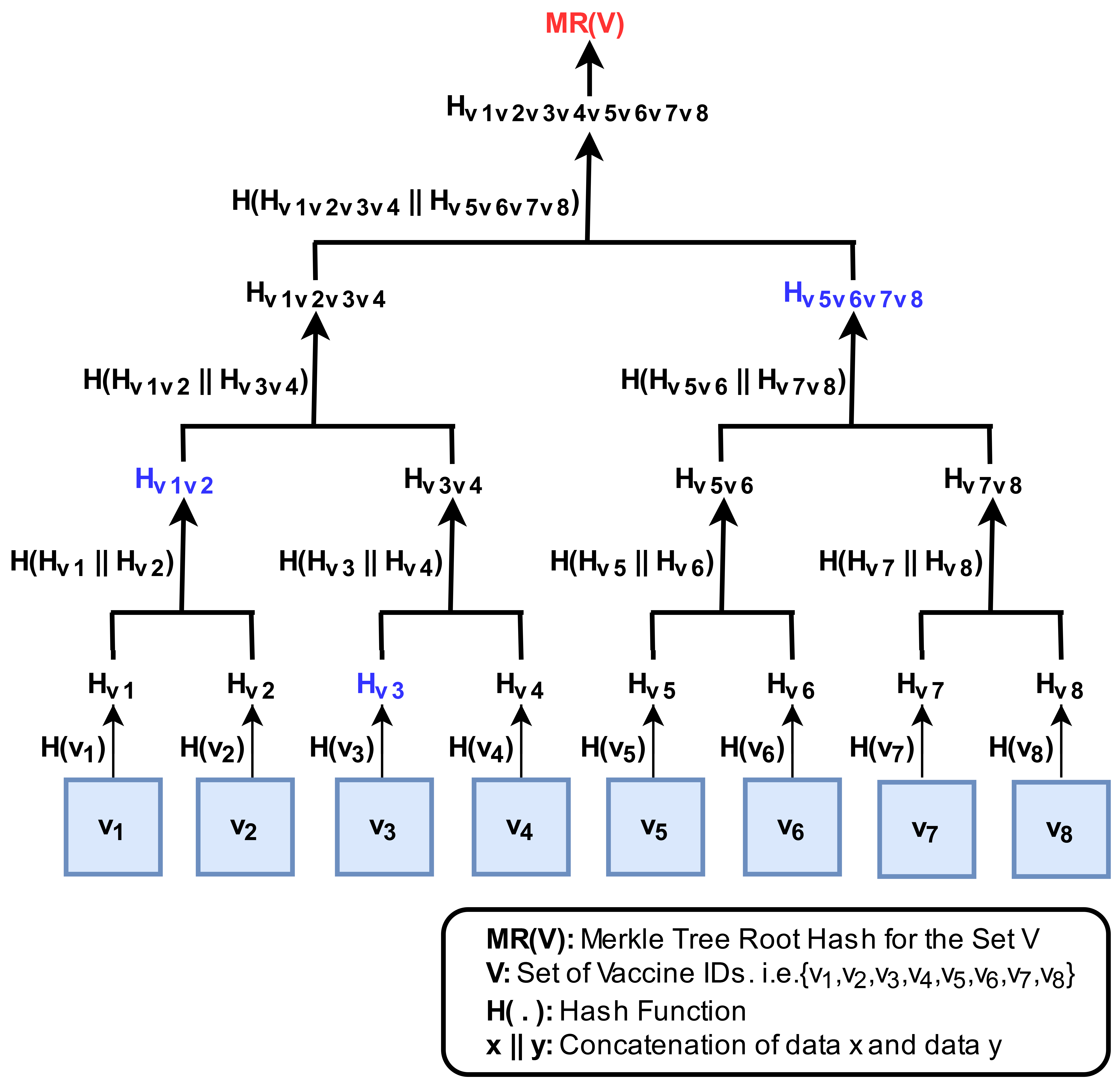}
            \caption{\small \sl Computation of $MR$ for set $V$}		
            \label{Fig3}
    	\end{center}  
        \end{figure}

        Computing $MR$, Govt puts the value on the $BC$ as the commitment of set $V$ (i.e. $BC \leftarrow MR(V)$) and then sends the corresponding vaccine vials to the VC. Govt also locks the service charge of the VC in the $SC$ apriori.

        \item[III.] Receiving the vaccine vials, VC also computes the $MR$ based on the vial IDs it got and then checks whether the calculated value matches the one that the Govt stored on $BC$. If the values match, VC accepts the vials; otherwise, it refuses the delivery and returns to the Govt again.

        \begin{figure}[!ht]
    	\begin{center}              	
            \includegraphics[width=\linewidth]{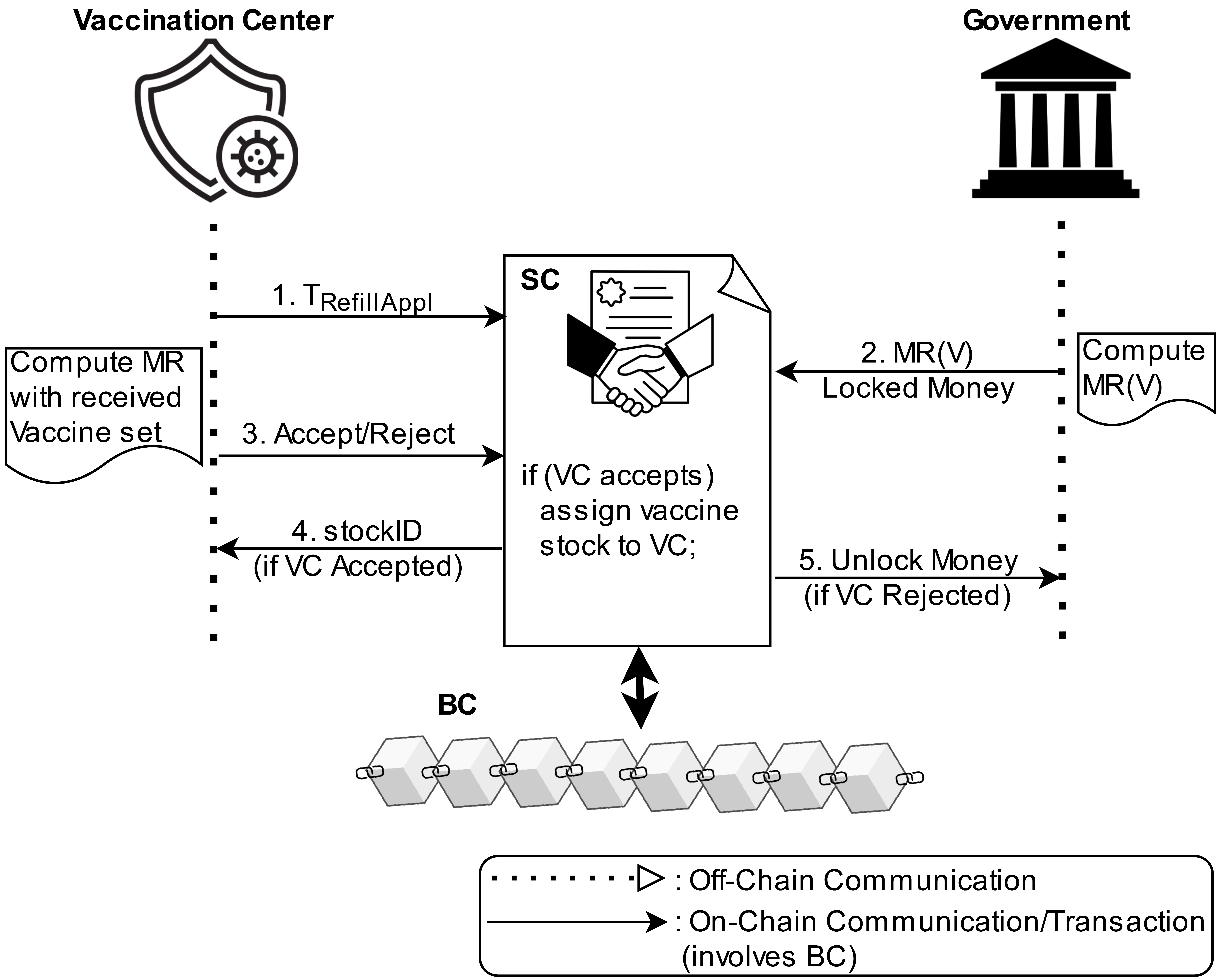}
            \caption{\small \sl Refilling Vaccine Stock}
            \label{Fig4}
    	\end{center}  
        \end{figure}

        \item[IV.] If VC accepts the vaccine stock, $SC$ generates a unique $stockID$ and assigns the stock to the VC. Otherwise, $SC$ unlocks the locked money and transfers it to the Govt.
    \end{enumerate}
    Figure~\ref{Fig4} illustrates the process of refilling vaccine stock.\newline
        
\begin{figure}[!ht]
        \begin{center}              	
        \includegraphics[width=\linewidth]{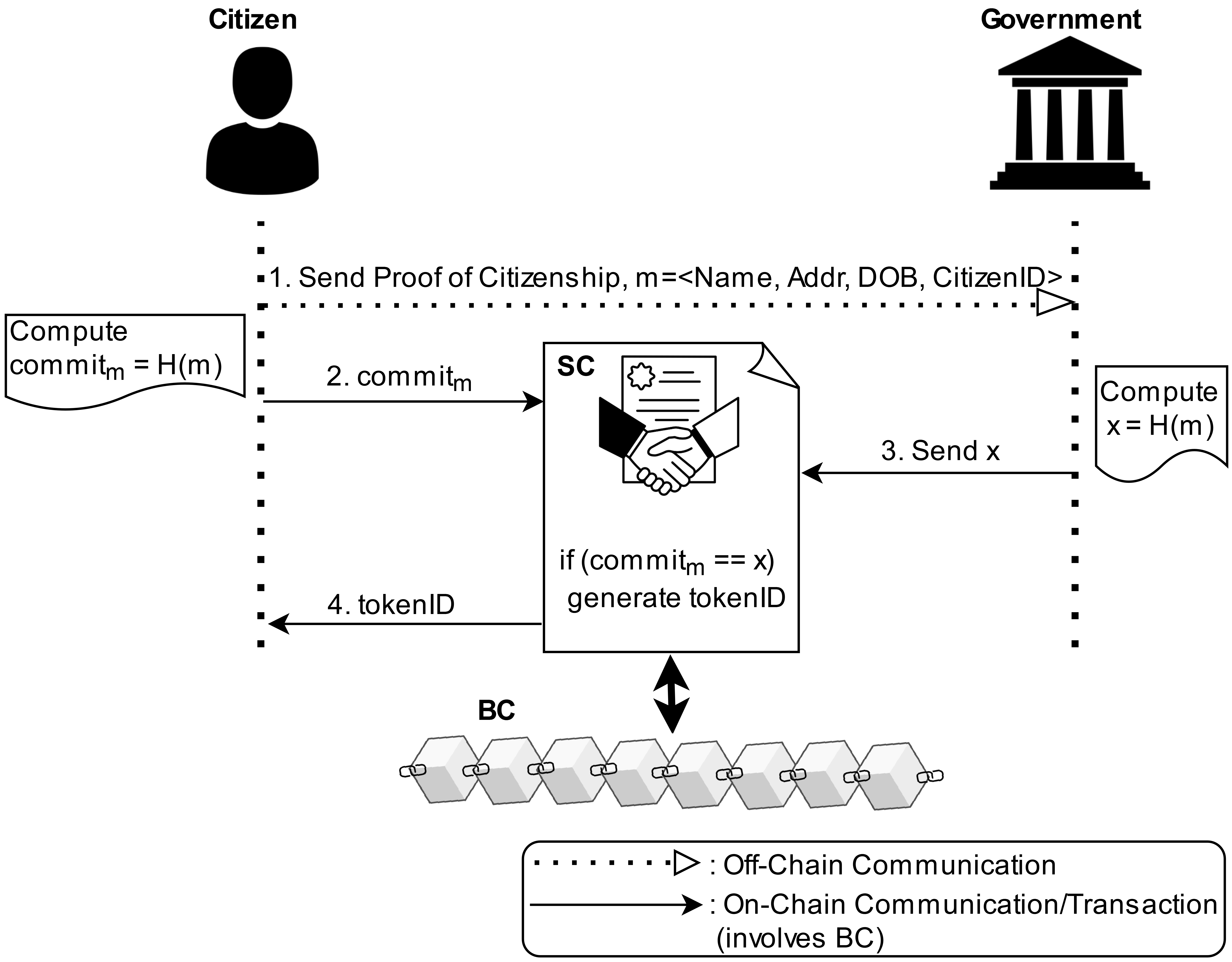}
        \caption{\small \sl Obtaining $TokenID$}		
        \label{Fig5}
        \end{center}  
    \end{figure}

    \noindent \textbf{Module 3: Obtaining \textit{TokenID}}\\ 
    If citizens want to receive the vaccine, they must obtain a unique token ID from the Govt. The process typically involves three steps.
    \begin{enumerate}[leftmargin=*]
        \item[I.]  The citizen must contact the appropriate Govt authority and provide valid proof of citizenship to express their interest in being vaccinated. This communication usually occurs offline.
        \item[II.] Simultaneously, the citizen should also generate a message digest of their private information, which includes their name, address, date of birth, and citizen ID, and record it on the $BC$. Since the hash is stored on $BC$, it preserves data privacy.
        \eqnspacebefore
        \begin{align*}
            m &\leftarrow (Name || Addr || DOB || CitizenID) \\
            commit_m &\leftarrow H(m) \\
            BC &\leftarrow commit_m
        \end{align*}
        \eqnspaceafter
        \item[III.] Once the Govt verifies that the citizen has properly recorded the message digest on the $BC$ and that it matches the information provided offline, the Govt will issue the citizen a unique $tokenID$ through the $SC$.
    \end{enumerate}
    Figure~\ref{Fig5} illustrates the process of $tokenID$ generation.\newline
    
    \noindent \textbf{Module 4: Injecting Vaccine to Citizen by Vaccination Center}\\
    \noindent Once a citizen has received a unique $tokenID$, they are eligible to receive a vaccine at a convenient VC. Information about the VCs, such as their address, contact details, and vaccine availability, is publicly available. When a citizen goes to the VC to receive the vaccine, a protocol runs between the citizen and the VC to ensure that the vaccine is administered correctly.
    \begin{figure*}[t]
        \begin{center}
        \includegraphics[keepaspectratio, width=0.796\textwidth]{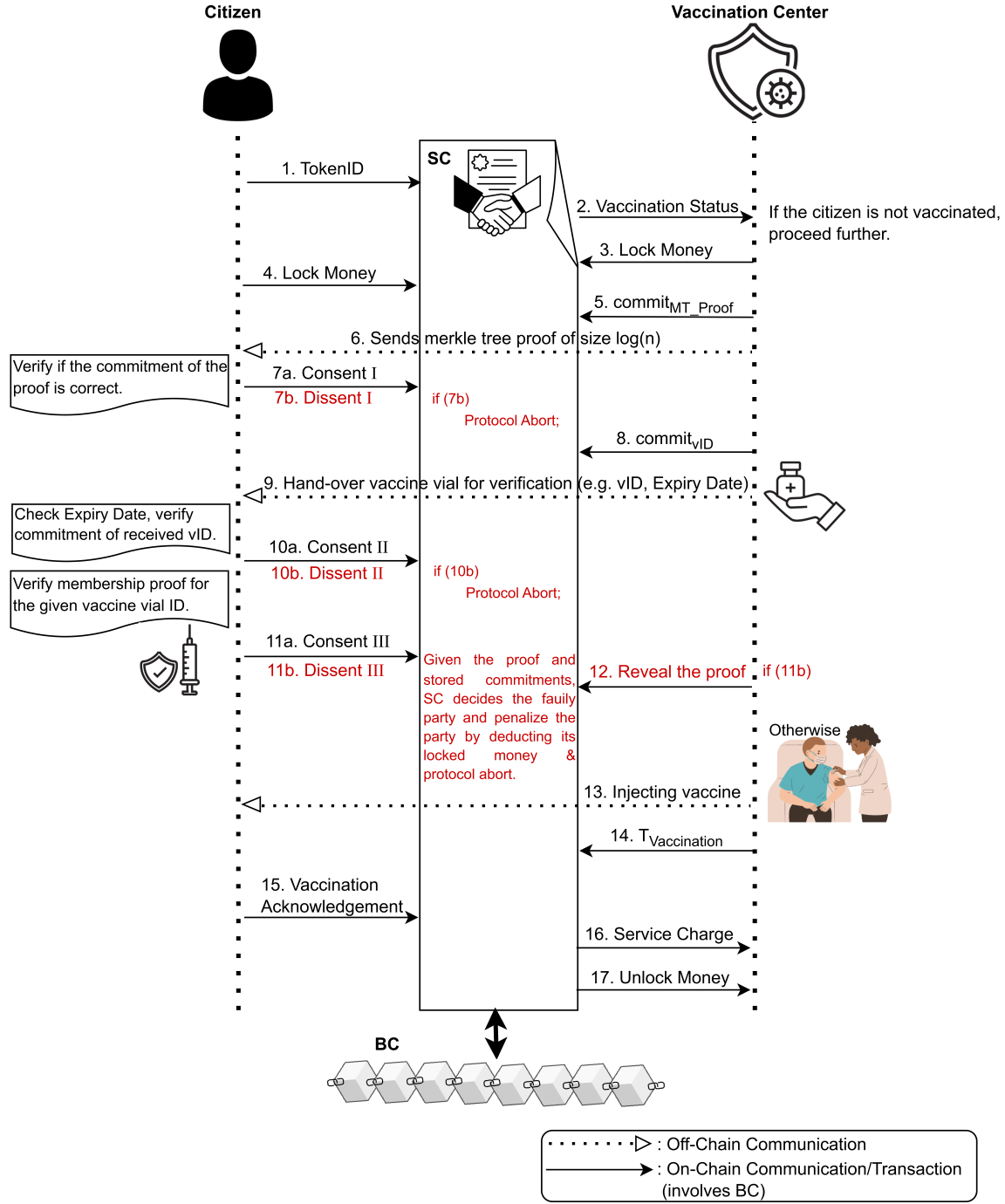}
        \caption{\small \sl Injecting Vaccine to Citizen by Vaccination Center}		
        \label{Fig6}
	\end{center}  
    \end{figure*}
    \begin{figure*}[t]
        \begin{center}
        \includegraphics[keepaspectratio, width=0.796\textwidth]{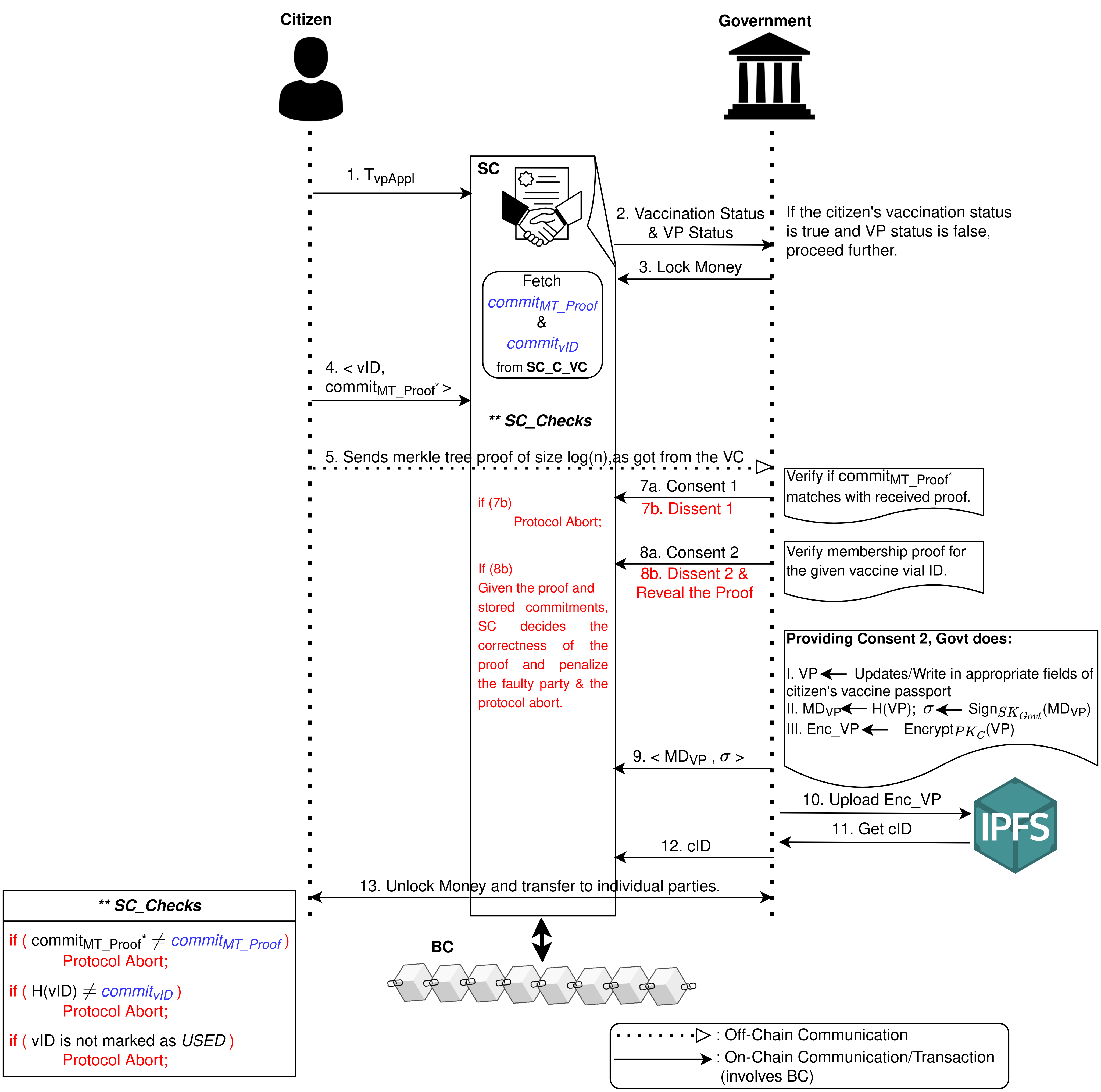}
	\caption{\small \sl Generating and Storing Citizen's Vaccine Passport on IPFS}		
        \label{Fig7}
	\end{center}  
    \end{figure*}
    \begin{figure*}[t]
        \begin{center}
        \includegraphics[keepaspectratio, width=0.7\textwidth]{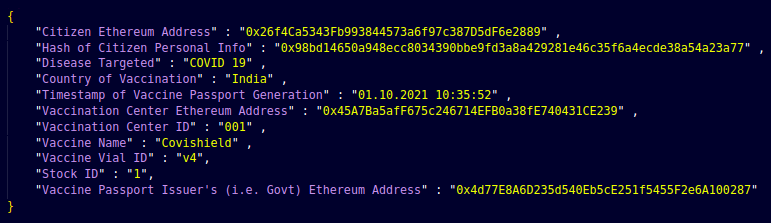}
	\caption{\small \sl Vaccine Passport as a JSON File}		
        \label{Fig8}
	\end{center}  
    \end{figure*}
    \begin{figure*}[!ht]
        \begin{center}              	
        \includegraphics[keepaspectratio, width=0.75\textwidth]{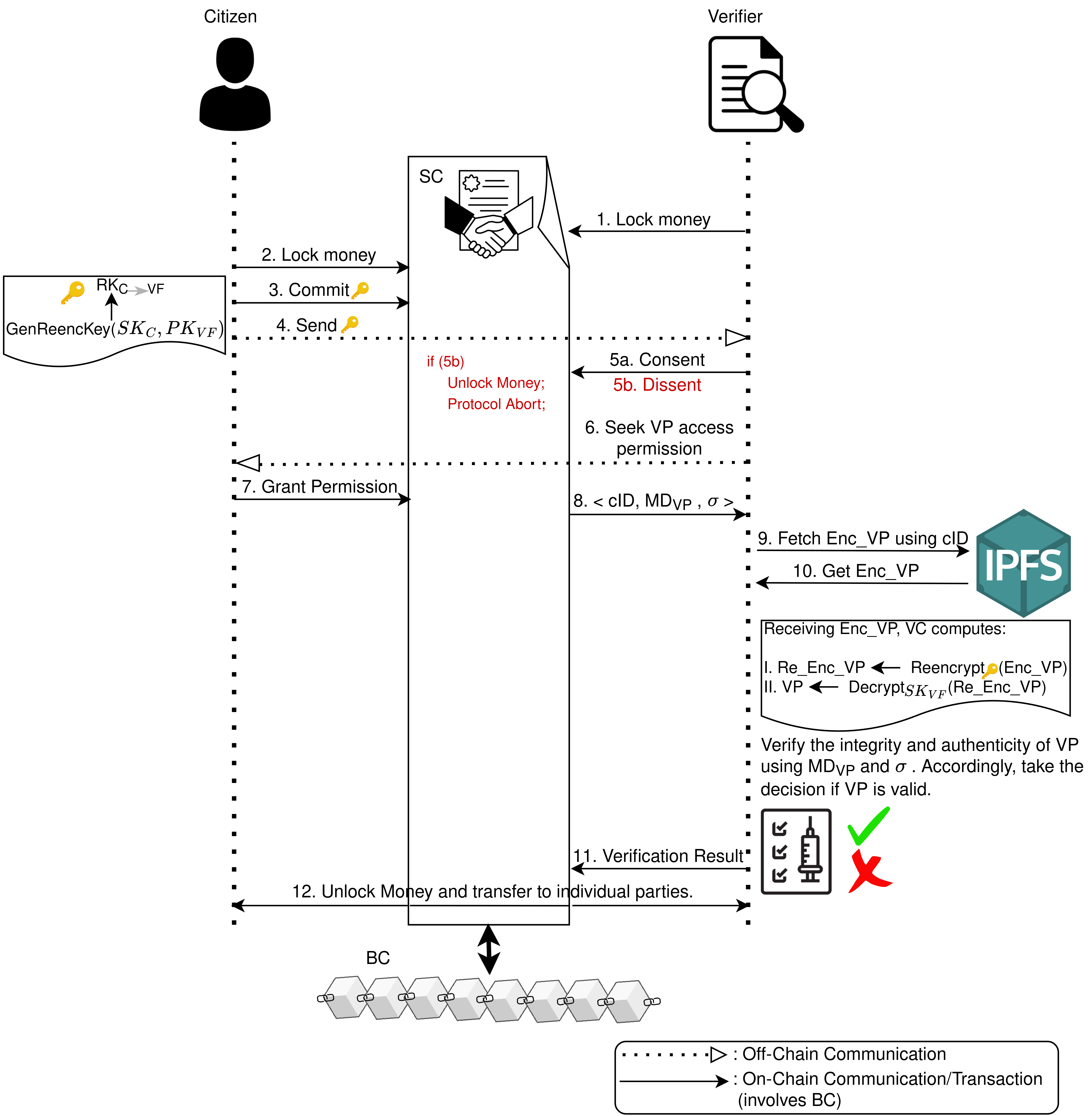}
        \caption{\small \sl Verification Process of Citizen's $VP$}		
        \label{Fig9}
        \end{center}  
    \end{figure*} 
    \begin{enumerate}[leftmargin=*]
        \item[I.] First, the citizen reveals their $tokenID$ to VC. Then VC obtains information regarding the vaccination status of the citizen from the $BC$. If the citizen is not vaccinated, the protocol proceeds further; otherwise, abort.
        \item[II.] In the next stage, both the parties, i.e. VC and citizen, lock a certain amount into the $SC$. Locking money ensures fairness. If the parties behave maliciously later in the protocol, they will be penalized by deducting their locked money. Honest parties will eventually recover their locked money at later stages.
        \item[III.] Once the money gets locked, VC proceeds further to inject the vaccine dose. VC picks the vaccine vial. Next, VC proves towards the citizen that the chosen vial is authentic and received from the Govt. Essentially, VC convinces this fact by providing a set membership proof corresponding to the selected vaccine vial ID. VC generates a Merkle Tree Proof of $\log(n)$ in size. VC sends the $\log(n)$ size proof to the citizen in offline mode and puts the hash of the proof on $BC$ as a commitment of the proof. 
        Without loss of generality, let's say VC picks the vaccine vial with $vID$ - $v4$ to be injected (Figure~\ref{Fig3}). In this case, the proof consists of $<H_{v3}, H_{v1v2}, H_{v5v6v7v8}>$ (blue coloured internal nodes in Figure~\ref{Fig3}). 
        \eqnspacebefore
        \begin{align*}
            MT\_Proof &\leftarrow\  <H_{v3}, H_{v1v2}, H_{v5v6v7v8}> \\
            commit_{MT\_Proof} &\leftarrow H(H_{v3} || H_{v1v2} || H_{v5v6v7v8}) \\
            BC &\leftarrow commit_{MT\_Proof}
        \end{align*}
        \eqnspaceafter
        
        VC sends the $MT\_Proof$ for vial ID - $v_4$ to the citizen and keeps the commitment of the proof $commit_{MT\_Proof}$ on the $BC$.
        \item[IV.] Receiving $MT\_Proof$, the citizen verifies if it matches the $commit_{MT\_Proof}$ stored on $BC$. If so, citizen provides their consent (\textbf{Consent I} as mentioned in Figure~\ref{Fig6}); else, protocol aborts.
        \item[V.] Next, VC hands over the vaccine vial to the citizen, keeping the commitment of vaccine vial ID on $BC$. Subsequently, $SC$ conducts a thorough check to confirm the freshness of the vaccine vial associated with the given commitment, ensuring it has not been used previously. In our example, VC does the following:
        \eqnspacebefore
        \begin{align*}
            commit_{vID} &\leftarrow H(v4) \\
            BC &\leftarrow commit_{vID}
        \end{align*}
        \eqnspaceafter
        \item[VI.] Then citizen verifies if the given vaccine vial satisfies the $commit_{vID}$. Also, citizen checks the expiry date and other important information printed on the vial. If the citizen finds everything is right, s/he provides their consent (\textbf{Consent II} as mentioned in Figure~\ref{Fig6}); else, protocol aborts.
        \item[VII.] Next, the citizen again provides their consent for the third time (\textbf{Consent III} as mentioned in Figure~\ref{Fig6}). This time citizen verifies whether the given $MT\_Proof$ for the specified vial with $vID$ matches the Merkle Tree Root hash, i.e. $MR$. Notably, Govt stored this $MR$ on the $BC$ while delivering the vaccine vials to the VC. 
        
        On the contrary, if the citizen finds that the $MR$ does not match the given proof for the vial, then the citizen complains to the $SC$. If citizen dissents, VC must reveal the proof to the $SC$ within a specific time window. $SC$ verifies the correctness of the complaint and judges the faulty party. Consequently, the malicious party gets penalized and it will lose its locked money.
        \item[VIII.] Upon receiving \textbf{Consent III} from the citizen, the VC administers the vaccine dose and records the vaccination timestamp on the $BC$.
        \item[IX.] Following vaccination, the citizen is required to acknowledge receipt within a specified time window. In the event of a negative acknowledgement, where the citizen denies receiving the vaccine despite the VC registering a timestamp of vaccination, legal intervention may be necessary to address the discrepancy. Although real-time image capture during vaccination could potentially resolve this issue using cameras/IoT devices, it falls outside the scope of this work.
        \item[X.] Upon receiving a positive acknowledgement from the citizen, the corresponding \textit{vID} is marked as \textit{USED}. Simultaneously, the VC is granted its service charge, which had been previously locked in the $SC$ by the Govt during the initial dispatch of the vaccine stock. Additionally, the SC unlocks the VC's security deposit. However, the locked amount from the citizen is not immediately released for security reasons. It will be released after the receipt of the Vaccine Passport ($VP$) as part of the subsequent protocol between the citizen and the Govt.
        \newline         
    \end{enumerate}
   
    \noindent\textbf{Module 5: Generating and Storing of Citizen's Vaccine Passport}\\
    Following the successful vaccine administration, the citizen's vaccination status is promptly updated on the $BC$. Subsequently, citizens are required to apply for a $VP$ from the government, encompassing crucial details such as the vaccination date, time, vaccine vial ID, and VC information. For a visual representation of a typical VP's contents, please refer to Figure~\ref{Fig8}.

    \noindent In our system, $VP$ is securely stored off-chain through the IPFS system. A citizen must initiate the $VP$ application process to unlock the funds held during the upfront vaccination protocol. Moreover, from a security perspective, this protocol assumes significance as it empowers the $SC$ to verify the accuracy of information provided by the citizen. The process of $VP$ application, generation, and storage generally involves the following steps:

    \begin{enumerate}[leftmargin=*]
    \item[I.] A citizen applies for a $VP$ from the government by locking a certain amount on the $SC$ and recording the application timestamp on the $BC$.
    \item[II.] If the $SC$ verifies the applicant's vaccination status as true and confirms the absence of the $VP$, the government proceeds to the next stage by locking a certain amount on the $SC$.
    \item[III.] Subsequently, the citizen needs to substantiate the truth of their vaccination by disclosing the vaccine vial ID (\textit{vID}) and committing the Merkle Tree Proof ($commit_{MT\_Proof^*}$) on the $SC$ for vID membership verification. Simultaneously, the citizen sends the proof to the government offline.
    \item[IV.] The $SC$ retrieves the stored values for vial ID commitment and Merkle tree proof commitment, previously shared by the VC, upon which the citizen provided its Consent I and Consent II (Figure~\ref{Fig6}). If the commitment values match those shared by the citizen, the protocol proceeds; otherwise, it terminates. The $SC$ also checks if the vial with \textit{vID} is marked as \textit{USED}.
    \item[V.] Upon receiving the Merkle tree proof offline, the government verifies its match with the commitment. Consequently, the government provides its initial response - \textbf{Consent 1} or \textbf{Dissent 1}.
    \item[VI.] Once Consent 1 is given, the government verifies if the Merkle tree proof validates the vial's membership with \textit{vID}. Subsequently, the government provides its \textbf{Consent 2} or \textbf{Dissent 2}. If Dissent 2 is given, the government must submit the proof to the $SC$.
    \item[VII.] With Consent 2 granted, the government initiates the creation of the citizen's $VP$ by performing the following operations:
    \begin{enumerate}[label = \arabic*)]
        \item Creates the citizen's $VP$.
        \item Computes the message digest of the $VP$ (i.e., $MD_{VP}$).
        \item Signs digitally on the $MD_{VP}$.
        \item Keeps the $MD_{VP}$ and its signature on the $BC$.
        \item Encrypts the $VP$ using $PK_C$.
        \item Uploads the encrypted $VP$ of the citizen to IPFS.
        \item Retrieves the $cID$ from IPFS.
        \item Records the $cID$ on the $BC$.
    \end{enumerate}
            \eqnspacebefore
            \begin{align*}
                MD_{VP} &\leftarrow H(VP) \\
                \sigma &\leftarrow SIG_{SK_{Govt}}(MD_{VP}) \\
                BC &\leftarrow\ <MD_{VP}, \sigma> \\ 
                Enc\_VP &\leftarrow Encrypt_{PK_C}(VP) \\
                cID &\leftarrow\ Uploads\ Enc\_VP\ to\ IPFS\\
                BC &\leftarrow cID
            \end{align*}
            \eqnspaceafter
    \item[VIII.] Once the citizen's VP is generated and $cID$ is recorded on the $BC$, the $SC$ unlocks the security deposits for both parties. It is important to note that, at this time, the citizen receives the money not only for this particular protocol but also for the preceding protocol conducted between the citizen and the vaccination center.\\
    \end{enumerate}
    
    \noindent Refer to Figure~\ref{Fig7} for an illustration of the process of creating and storing a vaccine passport for a citizen.

    \noindent\textbf{Module 6: Verification of Vaccine Passport}\\
    When a verifier proceeds to check a user's $VP$, s/he must pass through a protocol as shown in Figure~\ref{Fig9}.
    \begin{enumerate}[leftmargin=*]
        \item[I.] VF seeks permission from the citizen to check their $VP$. Once the citizen grants so, VF obtains the $cID$ through $SC$ interaction. Obtaining the $cID$, VF can fetch the citizen's encrypted $VP$ from $IPFS$.
        \item[II.] Since $VP$ was encrypted under the citizen's public key (i.e. $PK_C$), VF needs the citizen's secret key (i.e. $SK_C$) to decrypt it. However, sharing $SK$ compromises system security. So, instead of sharing a secret key, we are introducing the Proxy Re-encryption technique here. However, the proxy is absent here; the proxy's job is delegated to the end parties.
        \item[III.] The citizen generates a re-encryption key and then shares the key with the $VF$, while keeping the key's commitment on the $BC$.
        \eqnspacebefore
        \begin{align*}
            RK_{C \rightarrow VF} &\leftarrow GenReencKey(SK_C, PK_{VF}) \\
            commit_{RK} &\leftarrow H(RK_{C \rightarrow VF}) \\
            BC &\leftarrow commit_{RK}
        \end{align*}
        \eqnspaceafter
        \item[IV.] Receiving the $RK_{C \rightarrow VF}$, VF will re-encrypt the encrypted file. This re-encryption enables the VF to decrypt the file using its own secret key ($SK_{VF}$).
        \item[V.] After the decryption, VF fetches the message digest of the citizen's $VP$ (i.e., $MD_{VP}$) and checks if it complies with the decrypted file. VF also verifies the issuing VC's signature $\sigma$ from $BC$. And then, the VF decides if the citizen's $VP$ is valid. A record containing the VF's details, timestamp, and verification result is put on the $BC$.
    \end{enumerate}

\subsection{Implementation \& Technical Details}
As per the protocols described in Subsection \ref{Subsection: Protocol Design}, the entire system can be divided into six main modules. Each module facilitates the interaction between multiple parties and records the transactions in the blockchain. These six main modules serve the following purposes:
\begin{enumerate}
    \item \textbf{Registration of New Vaccination Centers}
    \item \textbf{Refilling Vaccine Stock at Vaccination Centers} 
    \item \textbf{Obtaining TokenID by Citizen}
    \item \textbf{Injecting Vaccine to Citizen by Vaccination Centers}
    \item \textbf{Generating \& Storing Vaccine Passport by Government}
    \item \textbf{Verifying Vaccine Passport by Verifier}
\end{enumerate}

Each module is written as an algorithm, converted into smart contract codes, and deployed on the Sepolia Test Network. Each smart contract consists of a set of \textit{structures}, \textit{mappings}, and \textit{methods}. We have documented \textit{timestamps} corresponding to various events or instances when an entity invokes a function, and these are detailed in Table~\ref{tab:Table_Timestamps}. Readers can refer to Table~\ref{tab:Table_Structs} and Table~\ref{tab:Table_Mappings}, respectively, for the necessary structures and mapping definitions used in our algorithms.

\begin{table}[!ht]
		\centering
		\caption{Timestamp Definitions Across Various Modules of our Vaccine Passport System}
		\label{tab:Table_Timestamps}
		\scalebox{0.7}{
			\begin{tabular}{r|l}
                \hline \hline
                \rowcolor{gray!120}
                    \multicolumn{2}{|l|}{\hspace{2.1cm}\textbf{\textcolor{white}{\normalsize Module 1:}} \textit{\textcolor{white}{\normalsize Registration of Vaccination Centers}}}\\
				\hline \hline
				\rowcolor{gray!30}
				\textbf{Abbreviation} & \textbf{Interpretation} \\ \hline \hline 
				$T_{regAppl}$ & Timestamp when \textit{VC} submits the registration application \\ \hline
				$T_{hashAppl}$ & Timestamp when \textit{Govt} submits hash of the received application \\ \hline
				$T_{decideOnHash}$ & Timestamp when \textit{VC} consents/descents on the hash value \\ \hline
                $T_{decideOnAppl}$ & Timestamp when \textit{Govt} accepts/rejects the registration application \\ \hline \hline

                \rowcolor{gray!120}
				\multicolumn{2}{|l|}{\hspace{2.1cm}{\textbf{\textcolor{white}{\normalsize Module 2:}} \textit{\textcolor{white}{\normalsize Refilling of Vaccine Stock}}}} \\
				\hline \hline
				\rowcolor{gray!30}
				\textbf{Abbreviation} & \textbf{Interpretation} \\ \hline \hline
                $T_{refillAppl}$ & Timestamp when \textit{VC} submits the refill application \\ \hline
                $T_{commitment}$ & Timestamp when \textit{Govt} commits $MR$ of the vaccine set \\ \hline
                \hline
    
                \rowcolor{gray!120}
				\multicolumn{2}{|l|}{\hspace{2.1cm}{\textbf{\textcolor{white}{\normalsize Module 3:}} \textit{\textcolor{white}{\normalsize Obtaining TokenID}}}} \\
				\hline \hline
                \rowcolor{gray!30}
				\textbf{Abbreviation} & \textbf{Interpretation} \\ \hline \hline 
				$T_{Appl}$ & Timestamp when \textit{C} submits token application \\ \hline
				$T_{Verification}$ & Timestamp when \textit{Govt} provides result verifying application \\ \hline \hline

                \rowcolor{gray!120}
				\multicolumn{2}{|l|}{\hspace{2.1cm}{\textbf{\textcolor{white}{\normalsize Module 4:}} \textit{\textcolor{white}{\normalsize Injecting Vaccine}}}} \\
				\hline \hline
                \rowcolor{gray!30}
				\textbf{Abbreviation} & \textbf{Interpretation} \\ \hline \hline 
				$T_{protocolBegins}$ & Timestamp when \textit{C} initiates the protocol \\ \hline 
                $T_{lockMoneyByVC}$ & Timestamp when \textit{VC} locks money \\ \hline
                $T_{lockMoneyByC}$ & Timestamp when \textit{C} locks money \\ \hline 
                $T_{commit_{MT\_Proof}}$ & Timestamp when \textit{VC} commits \textit{MT\_Proof} \\ \hline 
                $T_{consent1}$ & Timestamp when \textit{C} sends consent1/descent1 \\ \hline
                $T_{commit_{vID}}$ & Timestamp when \textit{VC} commits to \textit{vID} \\ \hline 
                $T_{consent2}$ & Timestamp when \textit{C} sends consent2/descent2 \\ \hline 
                $T_{consent3}$ & Timestamp when \textit{C} sends consent3/descent3 \\ \hline 
                $T_{moneyReceivedByC}$ & Timestamp when \textit{C} receives it's locked money \\ \hline 
                $T_{moneyReceivedByVC}$ & Timestamp when \textit{VC} receives it's locked money \\ \hline 
                $T_{protocolEnds}$ & Timestamp when \textit{VC} ends the protocol successfully \\ \hline
                \hline

                \rowcolor{gray!120}
				\multicolumn{2}{|l|}{\hspace{2.1cm}{\textbf{\textcolor{white}{\normalsize Module 5:}} \textit{\textcolor{white}{\normalsize Generating and Storing Vaccine Passport}}}} \\
				\hline \hline
                \rowcolor{gray!30}
				\textbf{Abbreviation} & \textbf{Interpretation} \\ \hline \hline 
                $T_{lockMoneyByC}$ & Timestamp when \textit{C} locks money \\ \hline
                $T_{lockMoneyByGovt}$ & Timestamp when \textit{Govt} locks money \\ \hline 
                $T_{provideVaccinationProof}$ & Timestamp when \textit{C} sends the vaccination proof \\ \hline 
                $T_{consent1}$ & Timestamp when \textit{Govt} sends consent1/descent1 \\ \hline
                $T_{consent2}$ & Timestamp when \textit{Govt} sends consent2/descent2 \\ \hline 
                $T_{issueVP}$ & Timestamp when \textit{Govt} issues \textit{VP} \\ \hline 
                $T_{moneyReceivedByC}$ & Timestamp when \textit{C} receives it's locked money \\ \hline 
                $T_{moneyReceivedByGovt}$ & Timestamp when \textit{Govt} receives it's locked money \\ \hline 
                \hline

                \rowcolor{gray!120}
				\multicolumn{2}{|l|}{\hspace{2.1cm}{\textbf{\textcolor{white}{\normalsize Module 6:}} \textit{\textcolor{white}{\normalsize Verifying Vaccine Passport}}}} \\
				\hline \hline
                \rowcolor{gray!30}
				\textbf{Abbreviation} & \textbf{Interpretation} \\ \hline \hline 
                $T_{lockMoneyByVF}$ & Timestamp when \textit{VF} locks money \\ \hline
                $T_{lockMoneyAndCommitRkByC}$ & Timestamp when \textit{C} locks money and commits Re-encryption Key \\ \hline 
                $T_{provideConsent}$ & Timestamp when \textit{VF} provides consent \\ \hline 
                $T_{grantAccessByC}$ & Timestamp when \textit{C} grants access to \textit{VF} \\ \hline
                $T_{fetchVPInfo}$ & Timestamp when \textit{VF} fetches \textit{VP} information \\ \hline 
                $T_{verificationResult}$ & Timestamp when \textit{VF} sends verification result \\ \hline 
                $T_{unlockMoney}$ & Timestamp when security money gets unlocked \\ \hline 
                \hline
                
			\end{tabular}
		}
	\end{table}
    
    \begin{table}[!ht]
		\centering
		\caption{Structs Used in our Implementation}
		\label{tab:Table_Structs}
		\scalebox{0.8}{
			\begin{tabular}{r|m{7cm}}
				\hline \hline
                \rowcolor{gray!120}
				\multicolumn{2}{|l|}{\hspace{1.2cm}{\textbf{\textcolor{white}{\normalsize Module 1:}} \textit{\textcolor{white}{\normalsize Registration of Vaccine Centers}}}} \\
				\hline \hline
				\rowcolor{gray!30}
				\textbf{Struct Name} & \textbf{Members} \\ \hline \hline 
				$VC$ & \textit{vcID}, \textit{currentStockID}, \textit{vialsInStock}, \textit{moneyEarned} \\ \hline
				$RegAppl$ & \textit{underReview}, \textit{$T_{regAppl}$}, \textit{$T_{hashAppl}$}, \textit{hash}, \textit{$T_{decideOnHash}$}, \textit{decision}, \textit{$T_{decideOnAppl}$}, \textit{regApplID} \\ \hline \hline

                \rowcolor{gray!120}
				\multicolumn{2}{|l|}{\hspace{1.2cm}{\textbf{\textcolor{white}{\normalsize Module 2:}} \textit{\textcolor{white}{\normalsize Refilling of Vaccine Stock}}}} \\
				\hline \hline
				\rowcolor{gray!30}
				\textbf{Struct Name} & \textbf{Members} \\ \hline \hline
                $ReStockAppl$ & \textit{refillApplID}, \textit{$T_{refillAppl}$}, \textit{underProcess}, \textit{vialsCount}, \textit{commitment}, \textit{$T_{commitment}$} \\ \hline 
                $VaccineStock$ & \textit{stockID}, \textit{owner}, \textit{vialsCount}, \textit{stockMR} \\ \hline \hline
    
                \rowcolor{gray!120}
				\multicolumn{2}{|l|}{\hspace{1.2cm}{\textbf{\textcolor{white}{\normalsize Module 3:}} \textit{\textcolor{white}{\normalsize Obtaining TokenID}}}} \\
				\hline \hline
                \rowcolor{gray!30}
				\textbf{Struct Name} & \textbf{Parameters} \\ \hline \hline 
				$Citizen$ & \textit{citizenInfoDigest}, \textit{tokenID}, \textit{vaccinationStatus}, \textit{vpStatus}, \textit{cID} \\ \hline
				$TokenAppl$ & \textit{tokenApplID}, \textit{citizenInfoDigest}, \textit{underReview}, \textit{$T_{tokenAppl}$}, \textit{result}, \textit{$T_{result}$} \\ \hline \hline

                \rowcolor{gray!120}
				\multicolumn{2}{|l|}{\hspace{1.2cm}{\textbf{\textcolor{white}{\normalsize Module 4:}} \textit{\textcolor{white}{\normalsize Injecting Vaccine}}}} \\
				\hline \hline
                \rowcolor{gray!30}
				\textbf{Struct Name} & \textbf{Parameters} \\ \hline \hline 
				$InjectingProtocol$ & \textit{protocolID}, \textit{underProcess}, \textit{tokenID}, \textit{vcID}, \textit{$T_{protocolBegins}$}, \textit{$T_{lockMoneyByVC}$}, \textit{$T_{lockMoneyByC}$}, \textit{$commit_{MT\_Proof}$}, \textit{$T_{commit_{MT\_Proof}}$}, \textit{consent1}, \textit{$T_{consent1}$}, \textit{$commit_{vID}$}, \textit{$T_{commit_{vID}}$}, \textit{consent2}, \textit{$T_{consent2}$}, \textit{consent3}, \textit{$T_{consent3}$}, \textit{$T_{vaccination}$} , \textit{$T_{moneyReceivedByC}$}, \textit{$T_{moneyReceivedByVC}$}, \textit{acknowledgement},  \textit{$T_{acknowledgement}$} 
				\\ \hline \hline

                \rowcolor{gray!120}
				\multicolumn{2}{|l|}{\hspace{1.2cm}{\textbf{\textcolor{white}{\normalsize Module 5:}} \textit{\textcolor{white}{\normalsize Generating \& Storing Vaccine Passport}}}} \\
				\hline \hline
                \rowcolor{gray!30}
				\textbf{Struct Name} & \textbf{Parameters} \\ \hline \hline
                    $VP$ & \textit{$MD_{VP}$}, \textit{$\sigma$}, \textit{cID} \\ \hline 
                    $VPAppl$ & \textit{vpApplID}, \textit{applicantTokenID}, \textit{$T_{lockMoneyByC}$}, \textit{$T_{lockMoneyByGovt}$}, \textit{$T_{provideVaccinationProof}$}, \textit{consent1}, \textit{$T_{consent1}$}, \textit{consent2}, \textit{$T_{consent2}$}, \textit{$T_{issueVP}$}, \textit{$T_{moneyReceivedByC}$}, \textit{$T_{moneyReceivedByGovt}$} \\ \hline \hline

                \rowcolor{gray!120}
				\multicolumn{2}{|l|}{\hspace{1.2cm}{\textbf{\textcolor{white}{\normalsize Module 6:}} \textit{\textcolor{white}{\normalsize Verifying Vaccine Passport}}}} \\
				\hline \hline
                \rowcolor{gray!30}
				\textbf{Struct Name} & \textbf{Parameters} \\ \hline \hline
                    $VerificationProtocol$ & \textit{vfProtocolID}, \textit{underExecution}, \textit{tokenID}, \textit{vfAddr}, \textit{$T_{lockMoneyByVF}$}, \textit{$T_{lockMoneyAndCommitRkByC}$}, \textit{consent}, \textit{$T_{provideConsent}$}, \textit{$T_{grantAccessByC}$}, \textit{$T_{fetchVPInfo}$}, \textit{verificationResult}, \textit{$T_{verificationResult}$}, \textit{$T_{unlockMoney}$} \\ \hline \hline
                    
			\end{tabular}
		}
	\end{table}
 
    \begin{table}[!ht]
		\centering
		\caption{Mappings Used in our Implementation}
		\label{tab:Table_Mappings}
		\scalebox{0.72}{
			\begin{tabular}{r|l}
				\hline \hline
                \rowcolor{gray!120}
				\multicolumn{2}{|l|}{\hspace{2.6cm}{\textbf{\textcolor{white}{\normalsize Module 1:}} \textit{\textcolor{white}{\normalsize Registration of Vaccine Centers}}}} \\
				\hline \hline
				\rowcolor{gray!30}
				\textbf{Mapping Name} & \textbf{Relations} \\ \hline \hline 
				$currentRegistrationAppl$ & Maps \textit{RegAppl} $\leftarrow$ \textit{VC} Address   \\ \hline
                $RegApplBelongsTo$ & Maps \textit{VC} Address $\leftarrow$ \textit{regApplID} \\ \hline
                $vcAddrTovcID$ & Maps \textit{vcID} $\leftarrow$ \textit{VC} Address \\ \hline 
                $vcIDToVCDetails$ & Maps \textit{VC} $\leftarrow$ \textit{vcID} \\ \hline \hline

                \rowcolor{gray!120}
				\multicolumn{2}{|l|}{\hspace{2.6cm}{\textbf{\textcolor{white}{\normalsize Module 2:}} \textit{\textcolor{white}{\normalsize Refilling Vaccine Stock}}}} \\
				\hline \hline
				\rowcolor{gray!30}
				\textbf{Mapping Name} & \textbf{Relations} \\ \hline \hline 
                $currentRefillAppl$ & Maps \textit{ReStockAppl} $\leftarrow$ \textit{VC} Address \\ \hline
                $lockedServiceCharge$ & Maps Locked Amount $\leftarrow$ \textit{VC} Address\\ \hline
                $vaccineStockDetails$ & Maps \textit{VaccineStock} $\leftarrow$ \textit{stockID} \\
                \hline \hline
                
                \rowcolor{gray!120}
				\multicolumn{2}{|l|}{\hspace{2.6cm}{\textbf{\textcolor{white}{\normalsize Module 3:}} \textit{\textcolor{white}{\normalsize Obtaining TokenID}}}} \\
				\hline \hline
                \rowcolor{gray!30}
				\textbf{Mapping Name} & \textbf{Relations} \\ \hline \hline 
                $currentTokenAppl$ & Maps \textit{TokenAppl} $\leftarrow$ \textit{citizenInfoDigest} \\ \hline
				$tokenAppl$ & Maps \textit{TokenAppl} $\leftarrow$ \textit{tokenApplID}\\ \hline
                $citizenAddrTocitizenInfoDigest$ & Maps \textit{citizenInfoDigest}  $\leftarrow$ \textit{Citizen} Address \\ \hline
                $citizenInfoDigestToTokenID$ & Maps \textit{tokenID} $\leftarrow$ \textit{citizenInfoDigest} \\ \hline
                $tokenIDToCitizenDetails$ & Maps \textit{Citizen} $\leftarrow$ \textit{tokenID} \\ \hline \hline

                \rowcolor{gray!120}
				\multicolumn{2}{|l|}{\hspace{2.6cm}{\textbf{\textcolor{white}{\normalsize Module 4:}} \textit{\textcolor{white}{\normalsize Injecting Vaccine}}}} \\
				\hline \hline
				\rowcolor{gray!30}
				\textbf{Mapping Name} & \textbf{Relations} \\ \hline \hline 
                $currentInjectingProtocol$ & Maps \textit{InjectingProtocol} $\leftarrow$ \textit{C} Address \\ \hline 
                $informationAboutVP$ & Maps \textit{VP} $\leftarrow$ \textit{tokenID} \\ \hline 
                $vialState$ & Maps \{\textit{"Used"}, \textit{"Reserved"}, \textit{"Unused"}\} $\leftarrow$ \textit{vial ID commitment} \\ \hline \hline

                \rowcolor{gray!120}
				\multicolumn{2}{|l|}{\hspace{2.6cm}{\textbf{\textcolor{white}{\normalsize Module 5:}} \textit{\textcolor{white}{\normalsize Generating and Storing Vaccine Passport}}}} \\
				\hline \hline
				\rowcolor{gray!30}
				\textbf{Mapping Name} & \textbf{Relations} \\ \hline \hline 
                $currentVPAppl$ & Maps \textit{VPAppl} $\leftarrow$ \textit{C} Address \\ \hline \hline

                \rowcolor{gray!120}
				\multicolumn{2}{|l|}{\hspace{2.6cm}{\textbf{\textcolor{white}{\normalsize Module 6:}} \textit{\textcolor{white}{\normalsize Verifying Vaccine Passport}}}} \\
				\hline \hline
				\rowcolor{gray!30}
				\textbf{Mapping Name} & \textbf{Relations} \\ \hline \hline 
                $verificationProtocolDetails$ & Maps \textit{VerificationProtocol} $\leftarrow$ \textit{vfProtocolID} \\ \hline 
                $accessControl$ & Maps \textit{Boolean (true/false)} $\leftarrow$ \textit{tokenID} $\times$ \textit{vfAddr} \\ \hline \hline
			\end{tabular}
		}
	\end{table}

\begin{enumerate}[leftmargin=*]
    \item \textbf{Algorithm for Registration of New Vaccination Centers:}
    The algorithm~\ref{algo:Algorithm for VC Registration} overseeing the \textit{VC} registration process (as discussed in \emph{Subsection~\ref{Subsection: Protocol Design} Module 1}) comprises specific methods that are invoked in sequence by the alternating parties (\textit{VC} and \textit{Govt}) at specific time intervals.

    This algorithm~\ref{algo:Algorithm for VC Registration} allows new vaccination centers \textit{(VC)} to register themselves onto the blockchain network. Once registered, they are provided with a unique ID \textit{(vcID)} and can start administering vaccines.\\
    \textbf{Sequence of methods in algorithm~\ref{algo:Algorithm for VC Registration}:}
    \begin{itemize}
        \item \textit{timestampRegAppl} $\rightarrow$ \textit{VC} creates \textit{RegAppl} and sets \textit{$T_{Appl}$}.
        \item \textit{regApplHash} $\rightarrow$ \textit{Govt} submits the hash of the application.
        \item \textit{decideOnAcceptanceHash} $\rightarrow$ \textit{VC} accepts/rejects the hash.
        \item \textit{decideOnAcceptanceRegAppl} $\rightarrow$ \textit{Govt}  accepts/rejects application.
    \end{itemize}
    
            \begin{algorithm}[!ht]
                \scriptsize
                \DontPrintSemicolon
                \SetKwProg{Fn}{Function}{}{}
                \SetKw{KwEnd}{end}
               
                \SetKwFunction{timestampRegAppl}{timestampRegAppl}
                \Fn{\timestampRegAppl{} \Comment{\textbf{Caller: $VC$}}}{
                \textbf{Fetch}: Current \textit{RegAppl} of \textit{VC}\\
                \textbf{Check}: If \textit{VC} not yet registered \\
                \textbf{Check}: If \textit{RegAppl.underReview} == false\\
                \textit{RegAppl} $\leftarrow$ new \textit{RegAppl} \\ 
                \textit{RegAppl.$T_{regAppl}$} $\leftarrow$ \textit{block.timestamp} \\
                \textit{RegAppl.underReview} $\leftarrow$ true \\
                \textbf{Update}: Mapping entries of \textit{currentRegistrationAppl} \\
                \textbf{Store}: \textit{RegAppl}
                }
                \KwEnd
                \; \\

                \SetKwFunction{regApplHash}{regApplHash}
                \Fn{\regApplHash{vcAddr, hashAppl} \Comment{\textbf{Caller: $Govt$}}}{
                \textbf{Fetch}: Current \textit{RegAppl} of \textit{VC} having address \textit{vcAddr}\\
                \textbf{Check}: If \textit{VC} not yet registered \\
                \textbf{Check}: If \textit{RegAppl.underReview} == true \\
                \textbf{Check}: If \textit{RegAppl.$T_{regAppl}$} $\neq$ 0 \\
                \textbf{Check}: If ($block.timestamp$$-$\textit{RegAppl.$T_{regAppl}$}) $\leq$ timeout \\
                \textit{RegAppl.$T_{hashAppl}$} $\leftarrow$ \textit{block.timestamp} \\
                \textit{RegAppl.hash}  $\leftarrow$  \textit{hashAppl}  \\
                \textbf{Update}: \textit{RegAppl}
                }
                \KwEnd 
                \; \\

                \SetKwFunction{decideOnAcceptanceHash}{decideOnAcceptanceHash}
                \Fn{\decideOnAcceptanceHash{decision} \Comment{\textbf{Caller: $VC$}} }{
                \textbf{Fetch}: Current \textit{RegAppl} of \textit{VC}\\
                \textbf{Check}: If \textit{VC} not yet registered \\
                \textbf{Check}: If \textit{RegAppl.underReview} == true\\
                \textbf{Check}: If \textit{RegAppl.$T_{hashAppl}$} $\neq$ 0 \\
                \textbf{Check}: If ($block.timestamp$$-$\textit{RegAppl.$T_{hashAppl}$}) $\leq$ timeout \\
                \If{decision == true}
                {
                    \textbf{Generate}: a unique \textit{regApplID} \\
                    \textit{RegAppl.regApplID} $\leftarrow$ \textit{regApplID} \\
                    \textbf{Update}: Mapping entries of \textit{regApplBelongsTo}
                } \Else 
                {
                    \textit{RegAppl.underReview} $\leftarrow$ false
                }   
                \textit{RegAppl.$T_{decideOnHash}$} $\leftarrow$ \textit{block.timestamp}\\
                \textbf{Update}: \textit{RegAppl}
                }
                \KwEnd 
                \; \\

                \SetKwFunction{decideOnAcceptanceRegAppl}{decideOnAcceptanceRegAppl}
                \Fn{\decideOnAcceptanceRegAppl{regApplID, decision} \Comment{\textbf{Caller: $Govt$}}}{
                \textbf{Check}: If \textit{regApplID} valid \\
                \textit{vcAddr} $\leftarrow$ \textit{regApplBelongsTo}[\textit{regApplID}]\\
                \textbf{Fetch}: Current \textit{RegAppl} of \textit{VC} having address \textit{vcAddr}\\
                \textbf{Check}: If \textit{VC} not yet registered \\
                \textbf{Check}: If \textit{RegAppl.underReview} == true\\
                \textbf{Check}: If \textit{RegAppl.$T_{decideOnHash}$} $\neq$ 0 \\
                \textbf{Check}: If ($block.timestamp$$-$\textit{RegAppl.$T_{decideOnHash}$}) $\leq$ timeout\\
                \If{decision == true}
                {
                    \textbf{Generate}: a unique \textit{vcID}\\
                    \textbf{Assign} the \textit{vcID} to the \textit{VC}\\
                    \textit{VC} $\leftarrow$ new \textit{VC}\\
                    \textit{VC.vcID} $\leftarrow$ \textit{vcID}; \textit{VC.vialsInStock} $\leftarrow$ 0 \\
                    \textbf{Update}: \textit{VC}\\
                    \textbf{Update}: Mapping entries of \textit{vcAddrTovcID} \& \textit{vcIDToVCDetails}
                }
                \textit{RegAppl.decision} $\leftarrow$ \textit{decision}\\
                \textit{RegAppl.$T_{decideOnAppl}$} $\leftarrow$ \textit{block.timestamp}\\
                \textit{RegAppl.underReview} $\leftarrow$ false\\
                \textbf{Update}: \textit{RegAppl}
                }
                \KwEnd 
                \; \\
                
                \caption{Algorithm for VC Registration}
                \label{algo:Algorithm for VC Registration}
            \end{algorithm}
        
    \item \textbf{Algorithm for Refilling Vaccine Stock at Vaccination Centers:} The algorithm~\ref{algo:Algorithm for Refilling Vaccine Stock} facilitates the replenishment of vaccine vials or doses at vaccination centers, with crucial information securely stored on the blockchain (as discussed in \emph{Subsection~\ref{Subsection: Protocol Design} Module 2}). This ensures transparency and security in the vaccine distribution process. The algorithm comprises specific functions that are alternately invoked by \textit{VC} and \textit{Govt} in a timely manner.\\
    \textbf{Sequence of methods in algorithm~\ref{algo:Algorithm for Refilling Vaccine Stock}:}
    \begin{itemize}
        \item \textit{refillStockAppl} $\rightarrow$ \textit{VC} submits \textit{ReStockAppl} to \textit{Govt}.
        \item \textit{commitVaccineSet} $\rightarrow$ \textit{Govt} commits the $MR$ of the vaccine set to be delivered and also locks the service charge for \textit{VC} on \textit{SC}.
        \item \textit{decideOnAcceptanceVaccineSet} $\rightarrow$ \textit{VC} provides its consent if the $MR$ matches with the received vaccine vials set; otherwise, it declines.
        \item \textit{takeAwayLockedMoney} $\rightarrow$ \textit{Govt} can withdraw the locked amount if \textit{VC} denies accepting the vaccine set or becomes unresponsive.
    \end{itemize}

            \begin{algorithm}[!ht]
                \scriptsize 
                \DontPrintSemicolon
                \SetKwProg{Fn}{Function}{}{}
                \SetKw{KwEnd}{end}
               
                \SetKwFunction{refillStockAppl}{refillStockAppl}
                \Fn{\refillStockAppl{} \Comment{\textbf{Caller: $VC$}} }{
                \textbf{Check}: If \textit{VC} is registered \\
                \textbf{Check}: If \textit{VC.vialsInStock} == 0 \\
                \textbf{Fetch}: Current \textit{ReStockAppl} of \textit{VC} \\
                \textbf{Check}: If \textit{ReStockAppl.underProcess} == false \\
                \textit{ReStockAppl} $\leftarrow$ new \textit{ReStockAppl} \\
                \textbf{Generate}: a unique \textit{refillApplID} \\
                \textit{ReStockAppl.refillApplID} $\leftarrow$ \textit{refillApplID} \\
                \textit{ReStockAppl.$T_{refillAppl}$} $\leftarrow$ \textit{block.timestamp} \\
                \textit{ReStockAppl.underProcess} $\leftarrow$ true \\
                \textbf{Update}: \textit{ReStockAppl} \\
                \textbf{Update}: Mapping entries of \textit{currentRefillAppl}
                }
                \KwEnd 
                \; \\

                \SetKwFunction{commitVaccineSet}{commitVaccineSet}
                \Fn{\commitVaccineSet{vialsCount, $MR$, vcAddr} \Comment{\textbf{Caller: $Govt$}} }{
                \textbf{Check}: If \textit{VC} having address \textit{vcAddr} is registered \\
                \textbf{Check}: If \textit{vialsCount} > 0 \\
                \textbf{Fetch}: Current \textit{ReStockAppl} of \textit{VC}. \\
                \textbf{Check}: If \textit {ReStockAppl.underProcess} == true \\
                \textbf{Check}: \textit{ReStockAppl.$T_{refillAppl}$} $\neq$ 0\\
                \textbf{Check}: If ($block.timestamp$$-$\textit{ReStockAppl.$T_{refillAppl}$}) $\leq$ timeout\\
                \textbf{Check}: If correct amount (as \textit{serviceCharge} of \textit{VC}) is locked \\
                \textbf{Update}: Mapping entries of \textit{lockedServiceCharge} \\
                \textit{ReStockAppl.vialsCount} $\leftarrow$ \textit{vialsCount} \\
                \textit{ReStockAppl.commitment} $\leftarrow$ \textit{$MR$} \\
                \textit{ReStockAppl.$T_{commitment}$} $\leftarrow$ \textit{block.timestamp} \\
                \textbf{Update}: \textit{ReStockAppl}
                }
                \KwEnd 
                \; \\

                \SetKwFunction{decideOnAcceptanceVaccineSet}{decideOnAcceptanceVaccineSet}
                \Fn{\decideOnAcceptanceVaccineSet{decision} \Comment{\textbf{Caller: $VC$}} }{
                \textbf{Check}: If \textit{VC} is registered. \\
                \textbf{Fetch}: Current \textit{ReStockAppl} of \textit{VC} \\
                \textbf{Check}: If \textit {ReStockAppl.underProcess} == true \\
                \textbf{Check}: If \textit{ReStockAppl.$T_{commitment}$} $\neq$ 0\\
                \textbf{Check}: If ($block.timestamp$$-$\textit{ReStockAppl.$T_{commitment}$}) $\leq$ timeout\\
                \If{\textit{decision} == true}{
                    \textbf{Generate}: a unique \textit{stockID} \\
                    Instantiate new \textit{VaccineStock} and populate the members \\
                    \textit{VC.currentStockID} $\leftarrow$ \textit{stockID} \\
                    \textbf{Update}: \textit{VC} \\
                    \textbf{Update}: Mapping entries of \textit{vaccineStockDetails}
                }
                \Else{
                    \textbf{Transfer}: locked money to \textit{Govt} \\
                    \textbf{Update}: Mapping entries of \textit{lockedServiceCharge}
                }
                \textit{ReStockAppl.$T_{acceptVaccineSet}$} $\leftarrow$ \textit{block.timestamp} \\
                \textit{ReStockAppl.underProcess} $\leftarrow$ false \\
                \textbf{Update}: \textit{ReStockAppl}
                }
                \KwEnd 
                \; \\

                \SetKwFunction{takeAwayLockedMoney}{takeAwayLockedMoney}
                \Fn{\takeAwayLockedMoney{vcAddr} \Comment{\textbf{Caller: $Govt$}} }{
                \textbf{Check}: If \textit{VC} having address \textit{vcAddr} is registered \\
                \textbf{Fetch}: Current \textit{ReStockAppl} of \textit{VC}. \\
                \textbf{Check}: If \textit {ReStockAppl.underProcess} == true \\
                \textbf{Check}: If \textit{ReStockAppl.$T_{commitment}$} $\neq$ 0 \\
                \textbf{Check}: If \textit{ReStockAppl.$T_{acceptVaccineSet}$} == 0 \\
                \textbf{Check}: If ($block.timestamp$$-$\textit{ReStockAppl.$T_{commitment}$}) $>$ timeout \\
                \textit{ReStockAppl.underProcess} $\leftarrow$ false \\
                \textbf{Transfer:} locked money to $Govt$\\
                \textbf{Update}: Mapping entries of \textit{lockedServiceCharge}
                }
                \KwEnd 
                \; \\
                
                \caption{Algorithm for Refilling Vaccine Stock}
                \label{algo:Algorithm for Refilling Vaccine Stock}
            \end{algorithm}
    
    \item \textbf{Algorithm for Obtaining TokenID by Citizen:} The algorithm~\ref{algo:Algorithm for Obtaining Citizen Token} enables citizens to obtain a unique TokenID, a prerequisite for accessing vaccination services. As mentioned in \emph{Subsection~\ref{Subsection: Protocol Design} Module 3}), the process begins with a citizen (\textit{C}) applying for the \textit{TokenID} from the \textit{Govt}. To protect privacy, the citizen's private information is initially transmitted to the \textit{Govt} through an off-chain mode. After this, all subsequent transactions take place on the blockchain.\\
    \textbf{Sequence of methods in algorithm~\ref{algo:Algorithm for Obtaining Citizen Token}:}
    \begin{itemize}
        \item \textit{applForTokenID} $\rightarrow$ \textit{C} applies for a \textit{TokenID}. The function parameter \textit{citizenInfoDigest} represents the commitment of personal data (mentioned as \textit{$commit_m$} in Figure~\ref{Fig5}).
        \item \textit{verifyAppl} $\rightarrow$ \textit{Govt} decides whether to accept or reject the application based on the provided information.
    \end{itemize}
 
            \begin{algorithm}[!ht]
                \scriptsize
                \DontPrintSemicolon
                \SetKwProg{Fn}{Function}{}{}
                \SetKw{KwEnd}{end}
               
                \SetKwFunction{applForTokenID}{applForTokenID}
                \Fn{\applForTokenID{citizenInfoDigest} \Comment{\textbf{Caller: $C$}} }{
                \textbf{Check}: $C$ with given \textit{citizenInfoDigest} not yet received \textit{tokenID} \\
                \textbf{Fetch}: Current \textit{TokenAppl} of $C$ \\
                \textbf{Check}: If \textit{TokenAppl.underReview} == false \\
                \textit{TokenAppl} $\leftarrow$ new \textit{TokenAppl} \\
                \textbf{Generate}: a unique \textit{tokenApplID} \\
                \textit{TokenAppl.tokenApplID} $\leftarrow$ \textit{tokenApplID} \\
                \textit{TokenAppl.citizenInfoDigest} $\leftarrow$ \textit{citizenInfoDigest} \\
                \textit{TokenAppl.underReview} $\leftarrow$ \textit{true}   \\
                \textit{TokenAppl.$T_{tokenAppl}$} $\leftarrow$ \textit{block.timestamp}   \\
                \textbf{Update}: \textit{TokenAppl} \\
                \textbf{Update}: Mapping entries of  \textit{currentTokenAppl}, \textit{tokenAppl} \& \textit{citizenAddrTocitizenInfoDigest}  
                }
                \KwEnd 
                \; \\

                \SetKwFunction{verifyAppl}{verifyAppl}
                \Fn{\verifyAppl{tokenApplID, decision} \Comment{\textbf{Caller: $Govt$}} }{
                \textbf{Fetch}: \textit{TokenAppl} corresponding to \textit{tokenApplID} \\
                \textbf{Check}: If \textit{TokenAppl.underReview} == true \\
                \textbf{Check}: If \textit{TokenAppl.$T_{tokenAppl}$} $\neq$ 0 \\
                \textbf{Check}: If \textit{TokenAppl.$T_{result}$} == 0 \\
                \textbf{Check}: If ($block.timestamp$$-$\textit{TokenAppl.$T_{tokenAppl}$}) $\leq$ timeout\\
                \textbf{Check}: If $C$ with \textit{TokenAppl.citizenInfoDigest} not yet received \textit{tokenID}\\
                \If{$decision$ == true}{
                    \textbf{Generate}: a unique \textit{tokenID} \\
                    \textit{Citizen} $\leftarrow$ \textit{new Citizen} \\
                    \textit{Citizen.citizenInfoDigest} $\leftarrow$ \textit{TokenAppl.citizenInfoDigest} \\
                    \textit{Citizen.tokenID} $\leftarrow$ \textit{tokenID} \\
                    \textit{Citizen.vaccinationStatus} $\leftarrow$ \textit{false} \\
                    \textit{Citizen.cID} $\leftarrow$ \textit{NULL} \\
                    \textbf{Update}: Citizen \\
                    \textbf{Update}: Mapping entries of \textit{citizenInfoDigestToTokenID} \& \textit{tokenIDToCitizenDetails}
                }
                \textit{TokenAppl.result} $\leftarrow$ \textit{decision}\\
                \textit{TokenAppl.$T_{result}$} $\leftarrow$ \textit{block.timestamp}\\
                \textit{TokenAppl.underReview} $\leftarrow$ false\\
                \textbf{Update}: \textit{TokenAppl}
                }
                
                \KwEnd 
                \; \\
                
                \caption{Algorithm for Obtaining Citizen Token}
                \label{algo:Algorithm for Obtaining Citizen Token}
            \end{algorithm}

    \item \textbf{Algorithm for Injecting Vaccine to Citizen: }The Algorithm~\ref{algo:Algorithm for Injecting Vaccine} outlines the secure and transparent process through which a \textit{VC} administers a vaccine dose to a \textit{C}. The details of the protocol have been discussed in \emph{Subsection~\ref{Subsection: Protocol Design} Module 4}. \\
    \textbf{Sequence of methods in algorithm~\ref{algo:Algorithm for Injecting Vaccine}:}
    \begin{itemize}
        \item \textit{beginProtocol} $\rightarrow$ $C$ begins the protocol mentioning its desired $vcID$.
        \item \textit{lockMoneyByVC} $\rightarrow$ $VC$ locks security money.
        \item \textit{lockMoneyByC} $\rightarrow$ $C$ locks security money.
        \item \textit{commitMTProof} $\rightarrow$ $VC$ commits $MT\_Proof$.
        \item \textit{provideConsent1} $\rightarrow$ $C$ provides its first consent.
        \item \textit{commitVialID} $\rightarrow$ $VC$ commits the vial ID (\textit{vID}) to ensure traceability.
        \item \textit{provideConsent2} $\rightarrow$ $C$ provides its second consent.
        \item \textit{provideConsent3} $\rightarrow$ $C$ provides its third consent, verifying if the given $vID$ satisfies the membership proof.
        \item \textit{registerVaxTimestamp} $\rightarrow$ $VC$ registers the timestamp of vaccination on the $BC$.
        \item \textit{acknowledgeVaccination} $\rightarrow$ $C$ acknowledges the vaccination, completing the process.
    \end{itemize}
    
    \begin{algorithm}[!ht]
                \scriptsize
                \DontPrintSemicolon
                \SetKwProg{Fn}{Function}{}{}
                \SetKw{KwEnd}{end}
               
                \SetKwFunction{beginProtocol}{beginProtocol}
                \Fn{\beginProtocol{vcID} \Comment{\textbf{Caller: $C$}} }{
                \textbf{Check}: If \textit{vcID} is valid \\
                \textbf{Check}: If \textit{C} has a valid \textit{tokenID} and not yet vaccinated \\
                \textbf{Fetch}: Current \textit{InjectingProtocol} of \textit{C} \\
                \textbf{Check}: If \textit{InjectingProtocol.underProcess} == false \\
                \textit{InjectingProtocol} $\leftarrow$ new \textit{InjectingProtocol} \\
                \textbf{Generate}: a unique \textit{protocolID} \\
                \textit{InjectingProtocol.protocolID} $\leftarrow$ \textit{protocolID} \\
                \textit{InjectingProtocol.underProcess} $\leftarrow$ true \\
                \textit{InjectingProtocol.tokenID} $\leftarrow$ \textit{tokenID} \\
                \textit{InjectingProtocol.vcID} $\leftarrow$ \textit{vcID} \\
                \textit{InjectingProtocol.$T_{protocolBegins}$} $\leftarrow$ \textit{block.timestamp} \\
                \textbf{Update}: \textit{InjectingProtocol} \\
                \textbf{Update}: Mapping entries of \textit{currentInjectingProtocol}
                }
                \KwEnd 
                \; \\

                \SetKwFunction{lockMoneyByVC}{lockMoneyByVC}
                \Fn{\lockMoneyByVC{cAddr} \Comment{\textbf{Caller: $VC$}} }{
                \textbf{Check}: If \textit{VC} has a valid \textit{vcID} \\
                \textbf{Check}: If \textit{C} with \textit{cAddr} has a valid \textit{tokenID} and is not vaccinated \\
                \textbf{Fetch}: Current \textit{InjectingProtocol} of \textit{C} \\
                \textbf{Check}: If \textit{InjectingProtocol.underProcess} == true \\
                \textbf{Check}: If \textit{InjectingProtocol.vcID} == \textit{vcID}\\
                \textbf{Check}: If \textit{InjectingProtocol.tokenID} == \textit{tokenID}\\
                \textbf{Check}: If \textit{InjectingProtocol.$T_{protocolBegins}$} $\neq$ 0 \\
                \textbf{Check}: If \textit{InjectingProtocol.$T_{lockMoneyByVC}$} == 0 \\
                \textbf{Check}: If ($block.timestamp$$-$\textit{InjectingProtocol.$T_{protocolBegins}$}) $\leq$ timeout\\
                \textbf{Check}: If correct amount is locked \\
                \textit{InjectingProtocol.$T_{lockMoneyByVC}$} $\leftarrow$ \textit{block.timestamp} \\
                \textbf{Update}: \textit{InjectingProtocol}
                }
                \KwEnd 
                \; \\

                \SetKwFunction{lockMoneyByC}{lockMoneyByC}
                \Fn{\lockMoneyByC{vcID} \Comment{\textbf{Caller: $C$}} }{
                \textbf{Check}: If \textit{C} has a valid \textit{tokenID} and is not vaccinated \\
                \textbf{Check}: If \textit{vcID} is valid \\
                \textbf{Fetch}: Current \textit{InjectingProtocol} of \textit{C} \\
                \textbf{Check}: If \textit{InjectingProtocol.underProcess} == true \\
                \textbf{Check}: If \textit{InjectingProtocol.vcID} == \textit{vcID}\\
                \textbf{Check}: If \textit{InjectingProtocol.tokenID} == \textit{tokenID}\\
                \textbf{Check}: If \textit{InjectingProtocol.$T_{lockMoneyByVC}$} $\neq$ 0 \\
                \textbf{Check}: If \textit{InjectingProtocol.$T_{lockMoneyByC}$} == 0 \\
                \textbf{Check}: If ($block.timestamp$$-$\textit{InjectingProtocol.$T_{lockMoneyByVC}$}) $\leq$ timeout\\
                \textbf{Check}: If correct amount is locked \\
                \textit{InjectingProtocol.$T_{lockMoneyByC}$} $\leftarrow$ \textit{block.timestamp} \\
                \textbf{Update}: \textit{InjectingProtocol}
                }
                \KwEnd 
                \; \\

                \SetKwFunction{commitMTProof}{commitMTProof}
                \Fn{\commitMTProof{cAddr, $commit_{MT\_Proof}$} \Comment{\textbf{Caller: $VC$}} }{
                \textbf{Check}: If \textit{VC} has a valid \textit{vcID} \\
                \textbf{Check}: If \textit{C} with \textit{cAddr} has a valid \textit{tokenID} and is not vaccinated \\
                \textbf{Fetch}: Current \textit{InjectingProtocol} of \textit{C} \\
                \textbf{Check}: If \textit{InjectingProtocol.underProcess} == true \\
                \textbf{Check}: If \textit{InjectingProtocol.vcID} == \textit{vcID} \\
                \textbf{Check}: If \textit{InjectingProtocol.tokenID} == \textit{tokenID}\\
                \textbf{Check}: If \textit{InjectingProtocol.$T_{lockMoneyByC}$} $\neq$ 0 \\
                \textbf{Check}: If \textit{InjectingProtocol.$T_{commit_{MT\_Proof}}$} == 0 \\
                \textbf{Check}: If ($block.timestamp$$-$\textit{InjectingProtocol.$T_{lockMoneyByC}$}) $\leq$ timeout\\
                \textit{InjectingProtocol.$commit_{MT\_Proof}$} $\leftarrow$ $commit_{MT\_Proof}$ \\
                \textit{InjectingProtocol.$T_{commit_{MT\_Proof}}$} $\leftarrow$ \textit{block.timestamp} \\
                \textbf{Update}: \textit{InjectingProtocol}
                }
                \KwEnd 
                \; \\

                \caption{Algorithm for Injecting Vaccine}
                \label{algo:Algorithm for Injecting Vaccine}
                
        \end{algorithm}

        \begin{algorithm}[!ht]

                \ContinuedFloat
                \scriptsize
                \DontPrintSemicolon
                \SetKwProg{Fn}{Function}{}{}
                \SetKw{KwEnd}{end}
        
                \SetKwFunction{provideConsentOne}{provideConsent1}
                \Fn{\provideConsentOne{vcID, consent1} \Comment{\textbf{Caller: $C$}} }{
                \textbf{Check}: If \textit{C} has a valid \textit{tokenID} and is not vaccinated \\
                \textbf{Check}: If \textit{vcID} is valid \\
                \textbf{Fetch}: Current \textit{InjectingProtocol} of \textit{C} \\
                \textbf{Check}: If \textit{InjectingProtocol.underProcess} == true \\
                \textbf{Check}: If \textit{InjectingProtocol.vcID} == \textit{vcID} \\
                \textbf{Check}: If \textit{InjectingProtocol.tokenID} == \textit{tokenID}\\
                \textbf{Check}: If \textit{InjectingProtocol.$T_{commit_{MT\_Proof}}$} $\neq$ 0 \\
                \textbf{Check}: If \textit{InjectingProtocol.$T_{consent1}$} == 0 \\
                \textbf{Check}: If ($block.timestamp$$-$\textit{InjectingProtocol.$T_{commit_{MT\_Proof}}$}) $\leq$ timeout \\
                \If{consent1 == false}{
                    \textbf{Transfer Money}: \textit{C} and \textit{VC} get back their locking amount \\
                    $T_{moneyReceivedByC}$ = $T_{moneyReceivedByVC}$ $\leftarrow$ \textit{block.timestamp} \\
                    \textit{InjectingProtocol.underProcess} $\leftarrow$ false
                }
                \textit{InjectingProtocol.consent1} $\leftarrow$ \textit{consent1} \\
                \textit{InjectingProtocol.$T_{consent1}$} $\leftarrow$ \textit{block.timestamp} \\
                \textbf{Update}: \textit{InjectingProtocol}
                }
                \KwEnd 
                \; \\

                \SetKwFunction{commitVialID}{commitVialID}
                \Fn{\commitVialID{cAddr, $commit_{vID}$} \Comment{\textbf{Caller: $VC$}} }{
                \textbf{Check}: If \textit{VC} has a valid \textit{vcID} \\
                \textbf{Check}: If \textit{C} with \textit{cAddr} has a valid \textit{tokenID} and is not vaccinated \\
                \textbf{Fetch}: Current \textit{InjectingProtocol} of \textit{C} \\
                \textbf{Check}: If \textit{InjectingProtocol.underProcess} == true \\
                \textbf{Check}: If \textit{InjectingProtocol.vcID} == \textit{vcID} \\
                \textbf{Check}: If \textit{InjectingProtocol.tokenID} == \textit{tokenID}\\
                \textbf{Check}: If \textit{InjectingProtocol.$T_{consent1}$} $\neq$ 0 \\
                \textbf{Check}: If \textit{InjectingProtocol.$T_{commit_{vID}}$} == 0 \\
                \textbf{Check}: If ($block.timestamp$$-$\textit{InjectingProtocol.$T_{consent1}$}) $\leq$ timeout \\
                \textbf{Check}: If \textit{vialState} of the vial with \textit{$commit_{vID}$} is \textit{"Unused"} \\
                \textit{InjectingProtocol.$commit_{vID}$} $\leftarrow$ \textit{$commit_{vID}$} \\
                \textit{InjectingProtocol.$T_{commit_{vID}}$} $\leftarrow$ \textit{block.timestamp} \\
                \textbf{Update}: \textit{InjectingProtocol} \\
                \textbf{Update}: Mapping entries of \textit{vialState} (change vial state to \textit{"Reserved"})
                }
                \KwEnd 
                \; \\

                \SetKwFunction{provideConsentTwo}{provideConsent2}
                \Fn{\provideConsentTwo{vcID, consent2} \Comment{\textbf{Caller: $C$}} }{
                \textbf{Check}: If \textit{C} has a valid \textit{tokenID} and is not vaccinated \\
                \textbf{Check}: If \textit{vcID} is valid \\
                \textbf{Fetch}: Current \textit{InjectingProtocol} of \textit{C} \\
                \textbf{Check}: If \textit{InjectingProtocol.underProcess} == true \\
                \textbf{Check}: If \textit{InjectingProtocol.vcID} == \textit{vcID} \\
                \textbf{Check}: If \textit{InjectingProtocol.tokenID} == \textit{tokenID}\\
                \textbf{Check}: If \textit{InjectingProtocol.$T_{commit_{vID}}$} $\neq$ 0 \\
                \textbf{Check}: If \textit{InjectingProtocol.$T_{consent2}$} == 0 \\
                \textbf{Check}: If ($block.timestamp$$-$\textit{InjectingProtocol.$T_{commit_{vID}}$}) $\leq$ timeout\\
                \If{consent2 == false}{
                    \textbf{Transfer Money}: \textit{C} and \textit{VC} get back their locking amount \\
                    $T_{moneyReceivedByC}$ = $T_{moneyReceivedByVC}$ $\leftarrow$ \textit{block.timestamp} \\
                    \textit{InjectingProtocol.underProcess} $\leftarrow$ false
                }
                \textit{InjectingProtocol.consent2} $\leftarrow$ \textit{consent2} \\
                \textit{InjectingProtocol.$T_{consent2}$} $\leftarrow$ \textit{block.timestamp} \\
                \textbf{Update}: \textit{InjectingProtocol}
                }
                \KwEnd 
                \; \\
                
                \caption{Algorithm for Injecting Vaccine (Contd.)}
                
            \end{algorithm}

            \begin{algorithm}[!ht]

                \ContinuedFloat
                \scriptsize
                \DontPrintSemicolon
                \SetKwProg{Fn}{Function}{}{}
                \SetKw{KwEnd}{end}

                \SetKwFunction{provideConsentThree}{provideConsent3}
                \Fn{\provideConsentThree{vcID, consent3} \Comment{\textbf{Caller: $C$}} }{
                \textbf{Check}: If \textit{C} has a valid \textit{tokenID} and is not vaccinated \\
                \textbf{Check}: If \textit{vcID} is valid \\
                \textbf{Fetch}: Current \textit{InjectingProtocol} of \textit{C} \\
                \textbf{Check}: If \textit{InjectingProtocol.underProcess} == true \\
                \textbf{Check}: If \textit{InjectingProtocol.vcID} == \textit{vcID} \\
                \textbf{Check}: If \textit{InjectingProtocol.tokenID} == \textit{tokenID}\\
                \textbf{Check}: If \textit{InjectingProtocol.$T_{consent2}$} $\neq$ 0 \\
                \textbf{Check}: If \textit{InjectingProtocol.$T_{consent3}$} == 0 \\
                \textbf{Check}: If ($block.timestamp$$-$\textit{InjectingProtocol.$T_{consent2}$}) $\leq$ timeout\\
                \textit{InjectingProtocol.consent3} $\leftarrow$ \textit{consent3} \\
                \textit{InjectingProtocol.$T_{consent3}$} $\leftarrow$ \textit{block.timestamp} \\
                \textbf{Update}: \textit{InjectingProtocol}
                }
                \KwEnd 
                \; \\
                
                \SetKwFunction{registerVaxTimestamp}{registerVaxTimestamp}
                \Fn{\registerVaxTimestamp{cAddr} \Comment{\textbf{Caller: $VC$}} }{
                \textbf{Check}: If \textit{VC} has a valid \textit{vcID} \\
                \textbf{Check}: If \textit{C} with \textit{cAddr} has a valid \textit{tokenID} and is not vaccinated \\
                \textbf{Fetch}: Current \textit{InjectingProtocol} of \textit{C} \\
                \textbf{Check}: If \textit{InjectingProtocol.underProcess} == true \\
                \textbf{Check}: If \textit{InjectingProtocol.vcID} == \textit{vcID} \\
                \textbf{Check}: If \textit{InjectingProtocol.tokenID} == \textit{tokenID}\\
                \textbf{Check}: If \textit{InjectingProtocol.$T_{consent3}$} $\neq$ 0 \\
                \textbf{Check}: If \textit{InjectingProtocol.consent3} == true \\
                \textbf{Check}: If \textit{InjectingProtocol.$T_{vaccination}$} == 0 \\
                \textbf{Check}: If ($block.timestamp$$-$\textit{InjectingProtocol.$T_{consent3}$}) $\leq$ timeout \\
                \textit{InjectingProtocol.$T_{vaccination}$} $\leftarrow$ \textit{block.timestamp} \\
                \textbf{Update}: \textit{InjectingProtocol}
                }
                \KwEnd 
                \; \\

                \SetKwFunction{acknowledgeVaccination}{acknowledgeVaccination}
                \Fn{\acknowledgeVaccination{vcID, ack} \Comment{\textbf{Caller: $C$}} }{
                \textbf{Check}: If \textit{C} has a valid \textit{tokenID} and is not vaccinated \\
                \textbf{Check}: If \textit{vcID} is valid \\
                \textbf{Fetch}: Current \textit{InjectingProtocol} of \textit{C} \\
                \textbf{Check}: If \textit{InjectingProtocol.underProcess} == true \\
                \textbf{Check}: If \textit{InjectingProtocol.vcID} == \textit{vcID} \\
                \textbf{Check}: If \textit{InjectingProtocol.tokenID} == \textit{tokenID}\\
                \textbf{Check}: If \textit{InjectingProtocol.$T_{vaccination}$} $\neq$ 0 \\
                \textbf{Check}: If \textit{InjectingProtocol.$T_{acknowledgement}$} == 0 \\
                \textbf{Check}: If ($block.timestamp$$-$\textit{InjectingProtocol.$T_{vaccination}$}) $\leq$ timeout \\
                \If{ack == true}{
                    \textbf{Transfer Money}: \textit{VC} gets its locking amount and also the \textit{serviceCharge} \\
                    $T_{moneyReceivedByVC}$ $\leftarrow$ \textit{block.timestamp} \\
                    \textit{C.vaccinationStatus} $\leftarrow$ true \\
                    \textit{VC.vialsInStock} $\leftarrow$ \textit{VC.vialsInStock} - 1 \\
                    \textit{VC.moneyEarned} $\leftarrow$ $VC.moneyEarned$ + $serviceCharge$ \\
                    \textit{InjectingProtocol.underProcess} $\leftarrow$ false \\
                    \textbf{Update}: \textit{VC}, \textit{C} \\
                    \textbf{Update}: Mapping entries of \textit{vialState} (change vial state to \textit{"Used"})
                }
                \textit{InjectingProtocol.acknowledgment} $\leftarrow$ \textit{ack} \\
                \textit{InjectingProtocol.$T_{acknowledgement}$} $\leftarrow$ \textit{block.timestamp} \\
                \textbf{Update}: \textit{InjectingProtocol}
                }
                \KwEnd 
                \; \\
                
                \caption{Algorithm for Injecting Vaccine (Contd.)}
                
            \end{algorithm}

    \item \textbf{Algorithm for Generating and Storing Vaccine Passport of Citizen: }
    After successfully receiving the vaccine, the \textit{C} must apply for the \textit{VP} to the \textit{Govt}. The vaccine passport includes information about the vaccine name, target disease, timestamp of vaccination, and other relevant details. Due to the large size of the file, it is not stored directly on the \textit{BC}. Instead, \textit{Govt} uploads the encrypted vaccine passport (\textit{VP}) to IPFS and then stores the essential security information on the \textit{BC} invoking \textit{SC} functions. The Algorithm~\ref{algo:Algorithm for Generating and Storing Vaccine Passport} corresponding to the protocol \emph{Subsection~\ref{Subsection: Protocol Design} Module 5} depicts the entire process of how the \textit{Govt} issues and stores \textit{VP} of a citizen who received the vaccine.\\
    \textbf{Sequence of methods in algorithm~\ref{algo:Algorithm for Generating and Storing Vaccine Passport}:}
    \begin{itemize}
        \item \textit{initiateVPApplAndLockMoney}: \textit{C} initiates the \textit{VP} application by locking a certain amount on \textit{SC}.       
        \item \textit{lockMoneyByGovt}: \textit{Govt} also locks the same amount on \textit{SC}.
        \item \textit{sendVaccinationProof}: \textit{C} submits vaccination proof, specifies vial ID - \textit{vID} and commits \textit{MT\_proof} on-chain. 
        \item \textit{sendConsent1}: Upon offline verification of the Merkle tree proof against the on-chain commitment, the \textit{Govt} issues its initial consent.
        \item \textit{sendConsent2}: \textit{Govt} provides its second consent if the $vID$ satisfies the given membership proof.
        \item \textit{uploadVPInfo}: Finally, \textit{Govt} uploads the encrypted \textit{VP} on IPFS and uploads essential security parameters onchain by calling this function. At the same time, \textit{SC} releases the locked amount to both \textit{Govt} and \textit{C}.
    \end{itemize}
            
            \begin{algorithm}[!ht]
                \scriptsize
                \DontPrintSemicolon
                \SetKwProg{Fn}{Function}{}{}
                \SetKw{KwEnd}{end}

                \SetKwFunction{initiateVPApplAndLockMoney}{initiateVPApplAndLockMoney}
                \Fn{\initiateVPApplAndLockMoney{} \Comment{\textbf{Caller: $C$}} }{
                \textbf{Check}: If \textit{C} has a valid \textit{tokenID} and is vaccinated \\
                \textbf{Check}: If \textit{C} has not obtained \textit{VP} \\
                \textbf{Check}: If correct amount is locked \\
                \textbf{Fetch}: Current \textit{VPAppl} of \textit{C} \\
                \textbf{Check}: If \textit{VPAppl.underprocess} == false \\
                \textit{VPAppl} $\leftarrow$ new \textit{VPAppl} \\
                \textbf{Generate}: a unique \textit{vpApplID} \\
                \textit{VPAppl.vpApplID} $\leftarrow$ \textit{vpApplID} \\
                \textit{VPAppl.applicantTokenID} $\leftarrow$ \textit{tokenID} \\
                \textit{VPAppl.$T_{lockMoneyByC}$} $\leftarrow$ \textit{block.timestamp} \\
                \textbf{Update}: \textit{VPAppl} \\
                \textbf{Update}: Mapping entries of \textit{currentVPAppl}
                }
                \KwEnd 
                \; \\

                \SetKwFunction{lockMoneyByGovt}{lockMoneyByGovt}
                \Fn{\lockMoneyByGovt{cAddr} \Comment{\textbf{Caller: $Govt$}} }{
                \textbf{Check}: If \textit{C} with \textit{cAddr} has a valid \textit{tokenID} and is vaccinated \\
                \textbf{Check}: If \textit{C} has not obtained \textit{VP} \\
                \textbf{Check}: If correct amount is locked \\
                \textbf{Fetch}: Current \textit{VPAppl} of \textit{C} \\
                \textbf{Check}: If \textit{VPAppl.underprocess} == true \\
                \textbf{Check}: If \textit{VPAppl.applicantTokenID} == \textit{tokenID} \\
                \textbf{Check}: If \textit{VPAppl.$T_{lockMoneyByC}$} $\neq$ 0 \\
                \textbf{Check}: If \textit{VPAppl.$T_{lockMoneyByGovt}$} == 0 \\
                \textbf{Check}: If ($block.timestamp$$-$\textit{VPAppl.$T_{lockMoneyByC}$}) $\leq$ timeout\\
                \textit{VPAppl.$T_{lockMoneyByGovt}$} $\leftarrow$ \textit{block.timestamp} \\
                \textbf{Update}: \textit{VPAppl}
                }
                \KwEnd 
                \; \\

                \SetKwFunction{sendVaccinationProof}{sendVaccinationProof}
                \Fn{\sendVaccinationProof{vID, $commit_{MT\_Proof}$} \Comment{\textbf{Caller: $C$}} }{
                \textbf{Check}: If \textit{C} has a valid \textit{tokenID} and is vaccinated \\
                \textbf{Check}: If \textit{C} has not obtained \textit{VP} \\
                \textbf{Fetch}: Current \textit{VPAppl} of \textit{C} \\
                \textbf{Check}: If \textit{VPAppl.underprocess} == true \\
                \textbf{Check}: If \textit{VPAppl.applicantTokenID} == \textit{tokenID} \\
                \textbf{Check}: If \textit{VPAppl.$T_{lockMoneyByGovt}$} $\neq$ 0 \\
                \textbf{Check}: If \textit{VPAppl.$T_{provideVaccinationProof}$} == 0 \\
                \textbf{Check}: If ($block.timestamp$$-$\textit{VPAppl.$T_{lockMoneyByGovt}$})$\leq$timeout\\
                \textbf{Fetch}: Latest \textit{InjectingProtocol} of \textit{C} \\
                \textbf{Check}: If \textit{InjectingProtocol.$commit_{MT\_Proof}$}==$commit_{MT\_Proof}$\\
                \textbf{Check}: If \textit{InjectingProtocol.$commit_{vID}$} == $h(vID)$ \\
                \textbf{Check}: If vial with ID - \textit{vID} is "Used," employing Map \textit{vialState} \\
                \textit{VPAppl.$T_{provideVaccinationProof}$} $\leftarrow$ \textit{block.timestamp} \\
                \textbf{Update}: \textit{VPAppl}
                }
                \KwEnd 
                \; \\

                \SetKwFunction{sendConsentOne}{sendConsent1}
                \Fn{\sendConsentOne{cAddr, consent1} \Comment{\textbf{Caller: $Govt$}} }{
                \textbf{Check}: If \textit{C} with \textit{cAddr} has a valid \textit{tokenID} and is vaccinated \\
                \textbf{Check}: If \textit{C} has not obtained \textit{VP} \\
                \textbf{Fetch}: Current \textit{VPAppl} of \textit{C} \\
                \textbf{Check}: If \textit{VPAppl.underprocess} == true \\
                \textbf{Check}: If \textit{VPAppl.applicantTokenID} == \textit{tokenID} \\
                \textbf{Check}: If \textit{VPAppl.$T_{provideVaccinationProof}$} $\neq$ 0 \\
                \textbf{Check}: If \textit{VPAppl.$T_{consent1}$} == 0 \\
                \textbf{Check}: If ($block.timestamp$$-$\textit{VPAppl.$T_{provideVaccinationProof}$}) $\leq$ timeout\\
                \If{consent1 == false}{
                    \textbf{Transfer Money}: \textit{Govt} and \textit{C} get back their locking amount \\
                    \textit{VPAppl.$T_{moneyReceivedByGovt}$} $\leftarrow$ \textit{block.timestamp} \\
                    \textit{VPAppl.$T_{moneyReceivedByC}$} $\leftarrow$ \textit{block.timestamp} \\
                    \textit{VPAppl.underProcess} $\leftarrow$ false
                }
                \textit{VPAppl.consent1} $\leftarrow$ \textit{consent1} \\
                \textit{VPAppl.$T_{consent1}$} $\leftarrow$ \textit{block.timestamp} \\
                \textbf{Update}: \textit{VPAppl}
                }
                \KwEnd 
                \; \\
                
                \caption{Algorithm for Generating and Storing Vaccine Passport}
                \label{algo:Algorithm for Generating and Storing Vaccine Passport}
                
        \end{algorithm}

        \begin{algorithm}
                \ContinuedFloat
                \scriptsize
                \DontPrintSemicolon
                \SetKwProg{Fn}{Function}{}{}
                \SetKw{KwEnd}{end}

                \SetKwFunction{sendConsentTwo}{sendConsent2}
                \Fn{\sendConsentTwo{cAddr, consent2} \Comment{\textbf{Caller: $Govt$}} }{
                \textbf{Check}: If \textit{C} with \textit{cAddr} has a valid \textit{tokenID} and is vaccinated \\
                \textbf{Check}: If \textit{C} has not obtained \textit{VP} \\
                \textbf{Fetch}: Current \textit{VPAppl} of \textit{C} \\
                \textbf{Check}: If \textit{VPAppl.underprocess} == true \\
                \textbf{Check}: If \textit{VPAppl.applicantTokenID} == \textit{tokenID} \\
                \textbf{Check}: If \textit{VPAppl.$T_{consent1}$} $\neq$ 0 \\
                \textbf{Check}: If \textit{VPAppl.$T_{consent2}$} == 0 \\
                \textbf{Check}: If ($block.timestamp$$-$\textit{VPAppl.$T_{consent1}$}) $\leq$ timeout\\
                \textit{VPAppl.consent2} $\leftarrow$ \textit{consent2} \\
                \textit{VPAppl.$T_{consent2}$} $\leftarrow$ \textit{block.timestamp} \\
                \textbf{Update}: \textit{VPAppl}
                }
                \KwEnd 
                \; \\

                \SetKwFunction{uploadVPInfoAndGetPayment}{uploadVPInfoAndGetPayment}
                \Fn{\uploadVPInfoAndGetPayment{cAddr, $MD_{VP}$, sign, cID} \Comment{\textbf{Caller: $Govt$}} }{
                \textbf{Check}: If \textit{C} with \textit{cAddr} has a valid \textit{tokenID} and is vaccinated \\
                \textbf{Check}: If \textit{C} has not obtained \textit{VP} \\
                \textbf{Fetch}: Current \textit{VPAppl} of \textit{C} \\
                \textbf{Check}: If \textit{VPAppl.underprocess} == true \\
                \textbf{Check}: If \textit{VPAppl.applicantTokenID} == \textit{tokenID} \\
                \textbf{Check}: If \textit{VPAppl.$T_{consent2}$} $\neq$ 0 \\
                \textbf{Check}: If \textit{VPAppl.$T_{issueVP}$} == 0 \\
                \textbf{Check}: If ($block.timestamp$$-$\textit{VPAppl.$T_{consent2}$}) $\leq$ timeout \\
                \textbf{Check}: If \textit{VPAppl.consent2} == true \\
                \textbf{Transfer Money}: \textit{Govt} and \textit{C} get back their locking amount \\
                \textbf{Transfer Money}: \textit{C} also gets back it's locking amount for \textit{InjectingProtocol} \\
                \textit{VPAppl.$T_{moneyReceivedByGovt}$} $\leftarrow$ \textit{block.timestamp} \\
                \textit{VPAppl.$T_{moneyReceivedByC}$} $\leftarrow$ \textit{block.timestamp} \\
                \textit{VPAppl.underProcess} $\leftarrow$ false \\
                \textbf{Update}: \textit{VPAppl} \\
                \textit{VP} $\leftarrow$ new \textit{VP} \\
                \textit{VP.$MD_{VP}$} $\leftarrow$ $MD_{VP}$ \\
                \textit{VP.$\sigma$} $\leftarrow$ \textit{sign} \\
                \textit{VP.cID} $\leftarrow$ \textit{cID} \\
                \textbf{Update}: \textit{VP} \\
                \textbf{Assign}: \textit{VP} to \textit{C} \\
                \textbf{Update}: Mapping entries of \textit{informationAboutVP}
                }
                \KwEnd 
                \; \\
                
                \caption{Algorithm for Generating and Storing Vaccine Passport (Contd.)}
                
        \end{algorithm}
    \item \textbf{Algorithm for Verifying Vaccine Passport:} This algorithm (corresponding to the protocol described in \textit{Subsection~\ref{Subsection: Protocol Design} Module 6}) allows healthcare providers and other authorized parties to verify the authenticity of a citizen's vaccine passport. This helps prevent fraud and ensures that only vaccinated individuals are granted access to certain services.\\
    \textbf{Sequence of methods in algorithm~\ref{algo:Algorithm for Verification of Vaccine Passport}:}
    \begin{itemize}
        \item \textit{lockMoneyByVF} $\rightarrow$ \textit{VF} locks money on \textit{SC} specifying the \textit{C}'s address for which passport verification is sought, and it marks the start of the verification protocol.
        \item \textit{lockMoneyAndCommitRK} $\rightarrow$ \textit{C} also locks the same amount on \textit{SC} and commits re-encryption key - \textit{RK}.
        \item \textit{provideConsent} $\rightarrow$ \textit{VF} provides its consent if the offline received \textit{RK} matches with its commitment.
        \item \textit{grantAccessPermission} $\rightarrow$ \textit{C} provides access permission to \textit{VF} to fetch \textit{cID} for encrypted \textit{VP}.
        \item \textit{fetchVPInfo} $\rightarrow$ \textit{VF} retrieves \textit{C}'s encrypted \textit{VP} details.
        \item \textit{verificationResult} $\rightarrow$ \textit{VF} sends the verification result.
    \end{itemize}
        \begin{algorithm}
                \scriptsize
                \DontPrintSemicolon
                \SetKwProg{Fn}{Function}{}{}
                \SetKw{KwEnd}{end}

                \SetKwFunction{lockMoneyByVF}{lockMoneyByVF}
                \Fn{\lockMoneyByVF{cAddr} \Comment{\textbf{Caller: $VF$}} }{
                \textbf{Check}: If \textit{C} with \textit{cAddr} has a valid \textit{tokenID} \\
                \textbf{Check}: If \textit{C} is vaccinated and holds a \textit{VP} \\
                \textbf{Check}: If correct amount is locked \\
                \textbf{Fetch}: Latest \textit{VerificationProtocol} between \textit{C} and \textit{VF} \\
                \textbf{Check}: If \textit{VerificationProtocol.underExecution} == false \\
                \textit{VerificationProtocol} $\leftarrow$ new \textit{VerificationProtocol} \\
                \textbf{Generate}: a unique \textit{vfProtocolID} \\
                \textit{VerificationProtocol.vfProtocolID} $\leftarrow$ \textit{vfProtocolID} \\
	        \textit{VerificationProtocol.underExecution} $\leftarrow$ true \\
	        \textit{VerificationProtocol.tokenID} $\leftarrow$ \textit{tokenID} \\
	        \textit{VerificationProtocol.vfAddr} $\leftarrow$ \textit{msg.sender} \\
	        \textit{VerificationProtocol.$T_{lockMoneyByVF}$} $\leftarrow$ \textit{block.timestamp} \\
                \textbf{Update}: \textit{VerificationProtocol} \\
                \textbf{Update}: Mapping entries of \textit{verificationProtocolDetails}
                }
                \KwEnd 
                \; \\

                \SetKwFunction{lockMoneyAndCommitRK}{lockMoneyAndCommitRK}
                \Fn{\lockMoneyAndCommitRK{vfProtocolID, commitRK} \Comment{\textbf{Caller: $C$}} }{
                \textbf{Check}: If correct amount is locked \\
                \textbf{Fetch}: Latest \textit{VerificationProtocol} corresponding to \textit{vfProtocolID} \\
                \textbf{Check}: If \textit{VerificationProtocol.underExecution} == true \\
                \textbf{Check}: If \textit{VerificationProtocol.tokenID} == \textit{C}'s \textit{tokenID} \\
	        \textbf{Check}: If \textit{VerificationProtocol.$T_{lockMoneyByVF}$} != 0 \\
                \textbf{Check}: If \textit{VerificationProtocol.$T_{lockMoneyAndCommitRkByC}$} == 0 \\
                \textbf{Check}: If ($block.timestamp$$-$\textit{VPAppl.$T_{lockMoneyByVF}$})$\leq$timeout \\
                \textit{VerificationProtocol.commitRK} $\leftarrow$ \textit{commitRK} \\
	        \textit{VerificationProtocol.$T_{lockMoneyAndCommitRkByC}$}$\leftarrow$\textit{block.timestamp} \\
                \textbf{Update}: \textit{VerificationProtocol} 
                }
                \KwEnd 
                \; \\

                \SetKwFunction{provideConsent}{provideConsent}
                \Fn{\provideConsent{vfProtocolID, decision} \Comment{\textbf{Caller: $VF$}} }{
                \textbf{Fetch}: Latest \textit{VerificationProtocol} corresponding to \textit{vfProtocolID} \\
                \textbf{Check}: If \textit{VerificationProtocol.underExecution} == true \\
                \textbf{Check}: If \textit{VerificationProtocol.vfAddr} == \textit{msg.sender} \\
	        \textbf{Check}: If \textit{VerificationProtocol.$T_{lockMoneyAndCommitRkByC}$} != 0 \\
                \textbf{Check}: If \textit{VerificationProtocol.$T_{provideConsent}$} == 0 \\
                \textbf{Check}: If ($block.timestamp$$-$\textit{VPAppl.$T_{lockMoneyAndCommitRkByC}$}) $\leq$ timeout \\
                \textit{VerificationProtocol.consent} $\leftarrow$ \textit{decision} \\
	        \textit{VerificationProtocol.$T_{provideConsent}$} $\leftarrow$ \textit{block.timestamp} \\
                \If{decision == false}{
                    \textbf{Transfer Money}: \textit{VF} and \textit{C} get back their locking amount \\
                    \textit{VerificationProtocol.$T_{unlockMoney}$} $\leftarrow$ \textit{block.timestamp} \\
                    \textit{VerificationProtocol.underExecution} $\leftarrow$ false
                }
                \textbf{Update}: \textit{VerificationProtocol}
                }
                \KwEnd 
                \; \\

                \SetKwFunction{grantAccessPermission}{grantAccessPermission}
                \Fn{\grantAccessPermission{vfProtocolID} \Comment{\textbf{Caller: $C$}} }{
                \textbf{Fetch}: Latest \textit{VerificationProtocol} corresponding to \textit{vfProtocolID} \\
                \textbf{Check}: If \textit{VerificationProtocol.underExecution} == true \\
                \textbf{Check}: If \textit{VerificationProtocol.tokenID} == \textit{C}'s \textit{tokenID} \\
	        \textbf{Check}: If \textit{VerificationProtocol.$T_{provideConsent}$} != 0 \\
                \textbf{Check}: If \textit{VerificationProtocol.$T_{grantPermission}$} == 0 \\
                \textbf{Check}: If ($block.timestamp$$-$\textit{VPAppl.$T_{provideConsent}$}) $\leq$ timeout \\
                \textbf{Check}: If \textit{VerificationProtocol.consent} == true \\
	        \textit{VerificationProtocol.$T_{grantAccessByC}$} $\leftarrow$ \textit{block.timestamp} \\
                \textbf{Update}: \textit{VerificationProtocol} \\
                \textbf{Update}: Mapping entries of \textit{accessControl}
                }
                \KwEnd 
                \; \\
                
                \caption{Algorithm for Verification of Vaccine Passport}
                \label{algo:Algorithm for Verification of Vaccine Passport}
                
        \end{algorithm}

        \begin{algorithm}
                \ContinuedFloat
                \scriptsize
                \DontPrintSemicolon
                \SetKwProg{Fn}{Function}{}{}
                \SetKw{KwEnd}{end}

                \SetKwFunction{fetchVPInfo}{fetchVPInfo}
                \Fn{\fetchVPInfo{vfProtocolID} \Comment{\textbf{Caller: $VF$}} }{
                \textbf{Fetch}: Latest \textit{VerificationProtocol} corresponding to \textit{vfProtocolID} \\
                \textbf{Check}: If \textit{VerificationProtocol.underExecution} == true \\
                \textbf{Check}: If \textit{VerificationProtocol.vfAddr} == \textit{msg.sender} \\
	        \textbf{Check}: If \textit{VerificationProtocol.$T_{grantAccessByC}$} != 0 \\
                \textbf{Check}: If \textit{VerificationProtocol.$T_{fetchVPInfo}$} == 0 \\
                \textbf{Check}: If ($block.timestamp$$-$\textit{VPAppl.$T_{grantAccessByC}$}) $\leq$ timeout \\
	        \textit{VerificationProtocol.$T_{fetchVPInfo}$} $\leftarrow$ \textit{block.timestamp} \\
                \textbf{Update}: \textit{VerificationProtocol} \\
                \textbf{Return}: \textit{VP} details of \textit{C} corresponding to \textit{tokenID} equals \textit{VerificationProtocol.tokenID}
                }
                \KwEnd 
                \; \\

                \SetKwFunction{verificationResult}{verificationResult}
                \Fn{\verificationResult{vfProtocolID} \Comment{\textbf{Caller: $VF$}} }{
                \textbf{Fetch}: Latest \textit{VerificationProtocol} corresponding to \textit{vfProtocolID} \\
                \textbf{Check}: If \textit{VerificationProtocol.underExecution} == true \\
                \textbf{Check}: If \textit{VerificationProtocol.vfAddr} == \textit{msg.sender} \\
	        \textbf{Check}: If \textit{VerificationProtocol.$T_{fetchVPInfo}$} != 0 \\
                \textbf{Check}: If \textit{VerificationProtocol.$T_{verificationResult}$} == 0 \\
                \textbf{Check}: If ($block.timestamp$$-$\textit{VPAppl.$T_{fetchVPInfo}$}) $\leq$ timeout \\
	        \textit{VerificationProtocol.verificationResult} $\leftarrow$ \textit{result} \\
                \textit{VerificationProtocol.$T_{verificationResult}$} $\leftarrow$ \textit{block.timestamp} \\
                \textbf{Transfer Money}: \textit{VF} and \textit{C} get back their locking amount \\
                \textit{VerificationProtocol.$T_{unlockMoney}$} $\leftarrow$ \textit{block.timestamp} \\
                \textit{VerificationProtocol.underExecution} $\leftarrow$ false \\
                \textbf{Update}: \textit{VerificationProtocol} 
                }
                \KwEnd 
                \; \\
                
                \caption{Algorithm for Verification of Vaccine Passport (Contd.)}
                
        \end{algorithm}
        
\end{enumerate}

\section{Security Analysis}
\label{Security Analysis}
Subsection~\ref{Security Goal} covers the system's security goals. Now, we will elaborate on how each goal is achieved using the blockchain platform and cryptographic techniques.

Blockchain technology incorporates various fundamental cryptographic primitives, such as hash functions and digital signatures. The security of the blockchain relies on the protection of these basic cryptographic elements. We assume that the fundamental cryptographic primitives are secure, which implies that our underlying blockchain platform is also protected. Consequently, the money held on the blockchain is safe, and therefore, the payments performed via the system are likewise secure.

With these considerations in mind, let us delve into the details of the security analysis of our system.

\begin{enumerate}[leftmargin = *]
    \item \textit{\textbf{Fairness}}
            \begin{itemize}[leftmargin = *]
                \item \textit{Unforgeability}. The $VP$ containing the vaccination status of an individual is stored in the IPFS by the vaccination centre. In order to successfully forge a vaccine passport, an adversary has to use a token-ID of a vaccinated citizen (say citizen $C$) to pass the verification protocol (say for some verifier $VF$). However, the usage of a proxy re-encryption mechanism would require the adversary to successfully generate the re-encryption key $RK_{C \xrightarrow{} VF}$. But the adversary has no knowledge of the secret key ${SK}_{C}$ of citizen $C$.\\
                Therefore, the protection against unforgeability is ensured by the following security properties of the proxy re-encryption scheme $PRE$, which is assumed to be secure 
                \begin{enumerate}[leftmargin = *]
                    \item The encryption scheme underlying $PRE$ is CPA- secure.
                     Any adversary $\mathcal{A}$ that is able to compute $RK_{C \xrightarrow{} VF}$ for any $VF$ and without the knowledge of ${SK}_{C}$ with high probability can be used to build an adversary $\mathcal{B}$ that breaks the $CPA$ security of the encryption scheme underlying the $PRE$ scheme.This can be illustrated using the following security game as depicted in figure~\ref{fig: Security Game}:
                     
                        \begin{figure}[!ht]
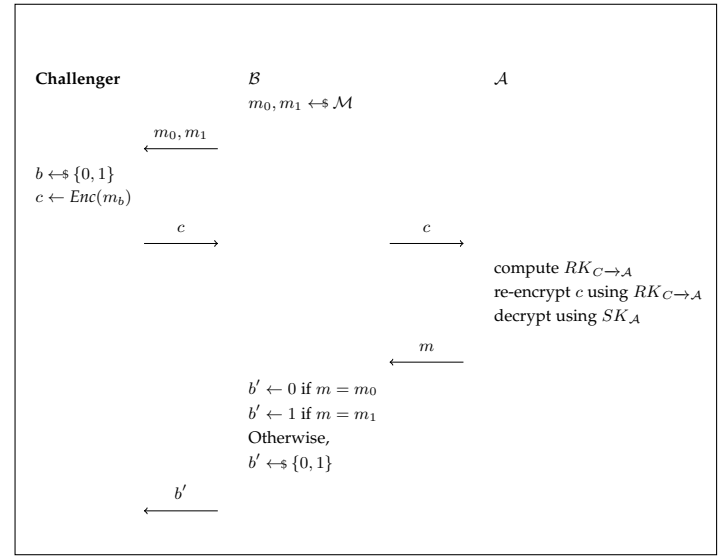

                        \centering
                        \fbox{
                        \begin{minipage}{0.95\columnwidth}
                        \scalebox{0.7}{
                        \begin{minipage}{\linewidth}
                        \pseudocode{%
                         \textbf{Challenger} \<\< \textbf{$\mathcal{B}$} \<\< \textbf{$\mathcal{A}$} \\
                          \<\< m_0, m_1\sample \mathcal{M}\\
                          \< \sendmessageleft*[1.4cm]{m_0, m_1} \< \\
                         b \sample \left\{0,1\right\} \<\< \\
                         c \gets \emph{Enc}(m_b) \<\< \\
                         \< \sendmessageright*[1.4cm]{c} \<\<\sendmessageright*[1.4cm]{c} \< \\
                         \<\<\<\< \text{compute } {RK}_{C \xrightarrow{} \mathcal{A}}\\
                         \<\<\<\< \text{re-encrypt $c$ using } {RK}_{C \xrightarrow{} \mathcal{A}}\\
                         \<\<\<\< \text{decrypt using } {SK}_{\mathcal{A}}\\
                         \<\<\< \sendmessageleft*[1.4cm]{m} \< \\
                         \<\< b' \gets 0 \text{ if } m = m_0\< \\
                         \<\< b' \gets 1 \text{ if } m = m_1\< \\
                         \<\< \text{Otherwise,}\\
                         \<\< b' \sample \left\{0,1\right\}\< \\
                         \< \sendmessageleft*[1.4cm]{b'} \< \\
                        }
                        \end{minipage}
                        }
                        \end{minipage}
                        }
                        \caption{Vaccine Passport Forgery Security Game}
                        \label{fig: Security Game}
                        \end{figure}

                     $\mathcal{B}$ wins if $b=b'$. So the winning probability of $\mathcal{B}$ is at least as much as that of $\mathcal{A}$.\\
                     
                    \item $PRE$ is secure against collusions. Several colluding proxies cannot gain any information about the secret key of $C$ by using the re-encryption keys issued to them by $C$ to be able to re-delegate decryption rights on behalf of $C$. That is, a set of re-encryption keys $RK_{C \xrightarrow{} {VF}_1}, \ldots, RK_{C \xrightarrow{} {VF}_k}$ leak no information about ${SK}_{C}$.
                    But the adversary with access to token-ID of $C$ and having seen $RK_{C \xrightarrow{} VF}$ can certainly re-use it to pass verification for $VF$ as no other information about the citizen is demanded at the verification stage. To prevent this, the re-encryption key can be shared over a secure channel instead of being sent over the public blockchain network.\\
                     
                \end{enumerate}
                
            \item \textit{Prevention of misuse}. Illegal distribution and misuse of vaccine doses are prevented by mandatory registration and application procedures, details of which are registered in the blockchain and thus publicly verifiable.
            Primitives like \emph{merkle tree} and \emph{collision-resistant hash functions} are used to facilitate validity checking of each dose. \\ \ \\
            Additionally, locking amounts in the smart contracts at the start of a protocol as and when required further enforces fairness and penalises malicious behaviour.
            \end{itemize}

    \item \textit{\textbf{Privacy}}
            \begin{itemize}
                \item While registering on the blockchain, a citizen does not disclose personally identifiable information (PII) such as name, age, address, phone number, etc. Instead, the citizen sends the hash of their PII to the smart contract during the token generation process. The pre-image resistance property of a cryptographic hash function ensures that personal details are computationally infeasible to retrieve from the hash.
                
                Once the token is generated for a citizen, it is bound to their public key. In subsequent protocols, only this token ID is used, and no other personal information is required, therefore preserving privacy.
                
                Citizens' vaccine passports contain only vaccination administration details and are stored on IPFS in an encrypted format. IPFS provides decentralized storage, while encryption ensures that only the citizen possesses the decryption keys, maintaining data confidentiality. For others to verify a citizen's vaccine passport, they must undergo a proxy re-encryption scheme.
                
                In the verification process, verifiers must obtain explicit permission from the citizen. Without consent, verifiers cannot proceed further. Citizens retain control over access permissions through an access control matrix stored on the blockchain. This matrix enables citizens to manage permissions for third parties, granting or revoking access as needed through interactions with smart contracts.
                
                In summary, these measures ensure robust privacy protection for citizens across all aspects of our blockchain-based vaccine passport system.
            \end{itemize}
    \item \textit{\textbf{Data Security}}
            \begin{itemize}
                \item Altering a citizen's vaccine passport records could have disastrous effects, potentially accelerating the spread of contagious diseases and endangering lives. In our context, citizens might attempt to falsify their vaccine passports to gain benefits reserved for vaccinated individuals without actually receiving the vaccine. Through meticulous system implementation, we ensure that no one, not even the citizen themselves, can forge or tamper with their stored vaccine passport record in our envisioned system and exploit it.
                
                Our protocol utilizes blockchain as the underlying framework, leveraging its inherent immutability to uphold the integrity of citizens' vaccine passport records. We store various cryptographic computation results, such as hash values, commitment values, and digital signatures, on the blockchain whenever necessary to protect sensitive data.
            \end{itemize}
    \item \textit{\textbf{Liveness}}
            \begin{itemize}
                \item Liveness is critical in distributed systems because it guarantees that the system remains active and responsive even in the presence of failures, delays, or malicious behavior. A blockchain-based system inherently supports liveness through decentralized consensus, ensuring continuous progress and transaction finality. The fault-tolerant nature of the network enables uninterrupted operation even during node failures or malicious behavior, contributing to implicit liveness.

                We ensure that our proposed blockchain-based system keeps progressing and continues to process and finalize transactions.

                While blockchain-based systems inherently possess liveness as an implicit property, ensuring the smooth and continuous functioning of the underlying network, application-level liveness becomes essential for scenarios where specific actions must be taken within defined time frames. In certain instances, the non-execution of one function can potentially block the execution of subsequent functions, making the system stagnant for an indefinite period of time. This creates a need for timely and autonomous actions to maintain system responsiveness and application-level liveness. 

                In our system, we have implemented an additional security measure. Each function must be called within a specific, predetermined time frame. If a function is not called within this time frame, the responsible party will be penalized by having their locked stake in the smart contract deducted and transferred to the other party as compensation. This action will result in the termination of the protocol. We have outlined and documented all potential exit functions to provide these features.

                By integrating these features into the smart contract design, we enhance the application-level liveness of the system, offering a reliable, autonomous, and dynamic environment for all stakeholders involved and ensuring the timely execution of various operations.
            \end{itemize}
\end{enumerate}

\section{Results and Discussions}
\label{Results and Discussions}
\textbf{Implementation Setup:} We successfully implemented the proposed \emph{Vaccine Passport System} on a system running Linux Ubuntu 22.04.2 LTS with a 12 Gen Intel(R) Core(TM) i7-1255U and 16.0 GiB of RAM. We deployed the smart contracts, written in Solidity, on the Ethereum Sepolia test network, utilizing the MetaMask crypto wallet for account creation and transaction initiation. The source code of our smart contracts is available on the GitHub repository\footnote{\tiny{\url{https://github.com/Debendranath-Das/Blockchain-Enabled-Secure-Vaccine-Passport-System/tree/}}}. The deployed contract addresses and deployment gas costs are detailed in Table~\ref{tab1}.

\begin{table}[htbp]
    \caption{Deployment Addresses and Cost of Smart Contracts}
    \begin{center}
        \scalebox{0.7}{
            \begin{tabular}{|c|c|c|}
            \hline
            \rowcolor{gray!30}
            \textbf{Smart Contract} & \textbf{Deployment Address} & \textbf{Deployment Gas Cost}\\ \hline \hline
            SC\_VC\_Govt & 0x98e8e9c40d7feab2d0b5a373c694e71de4310c6c & 1971190 \\ \hline
            SC\_C\_Govt\_1 & 0x2af1b8f42985cf9156cd15bb6ba539fe62e3aa25 & 1435723 \\ \hline
            SC\_C\_Govt\_2 & 0x33638204d0712448ca5490c66708091dceb87f4e & 3249238\\ \hline
            SC\_C\_VC\_1 & 0x7c63c824f2d50cb492949b6f293e5c2eda2031ac & 5104937\\ \hline
            SC\_C\_VC\_2 & 0x9b5bf13df8b35ada1eb073d268cd57b9400f2066 & 2388173\\ \hline
            SC\_C\_VF & 0x98aeb13345fc1250024543d9171dabf7bcd77c25 & 1847275\\ \hline
            SC\_Requirements\_Check & 0x605cfc9c7f146810cf5ed9c91ac51eaf3c051889 & 666361\\ \hline
            \end{tabular}
        }
        \label{tab1}
    \end{center}
\end{table}

\begin{figure*}[!ht]
    \centering
    \begin{minipage}{0.48\textwidth}
        \centering
        \includegraphics[width=\textwidth, height = 5 cm]{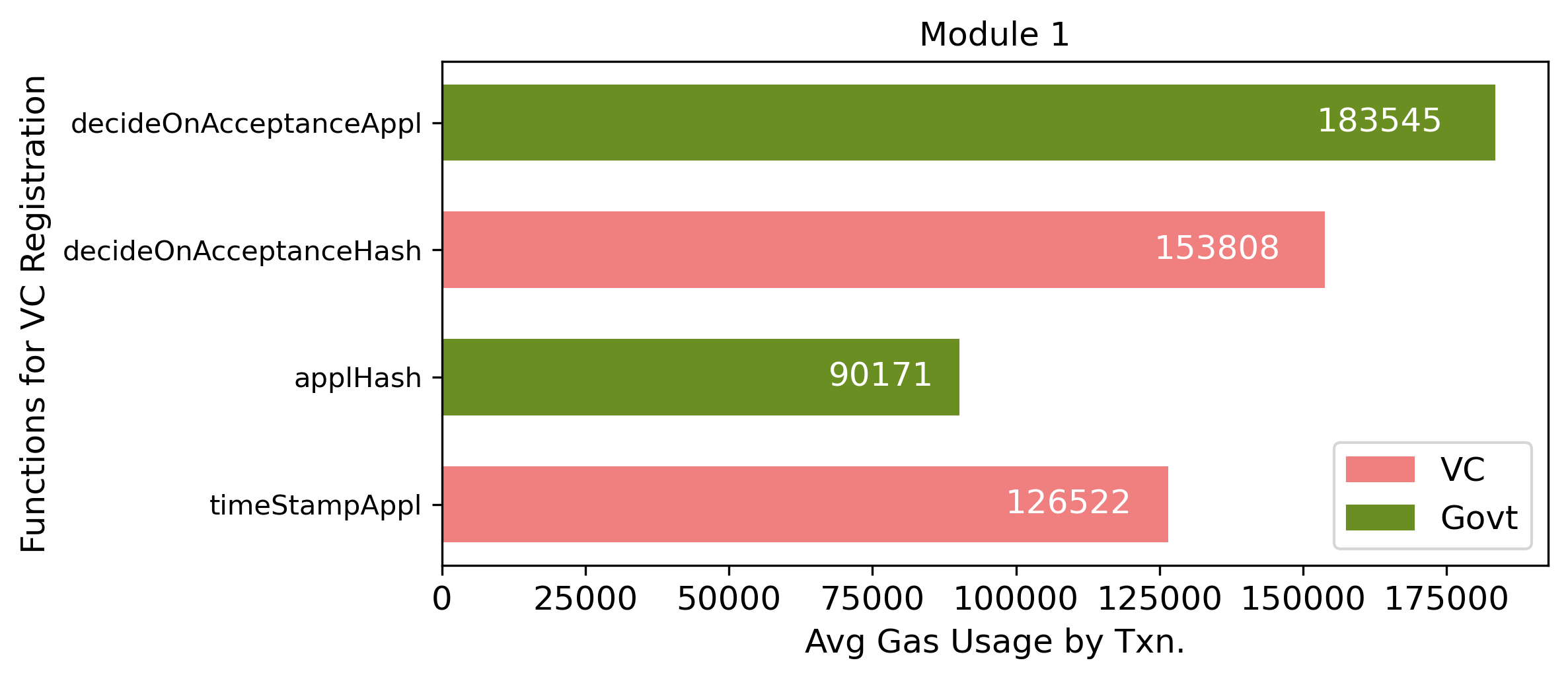}
        \caption{Gas Consumption of Transactions for VC Registration (Module 1)}
        \label{Fig11}
    \end{minipage}
    \hfill
    \begin{minipage}{0.48\textwidth}
        \centering
        \includegraphics[width=\textwidth, height = 5 cm]{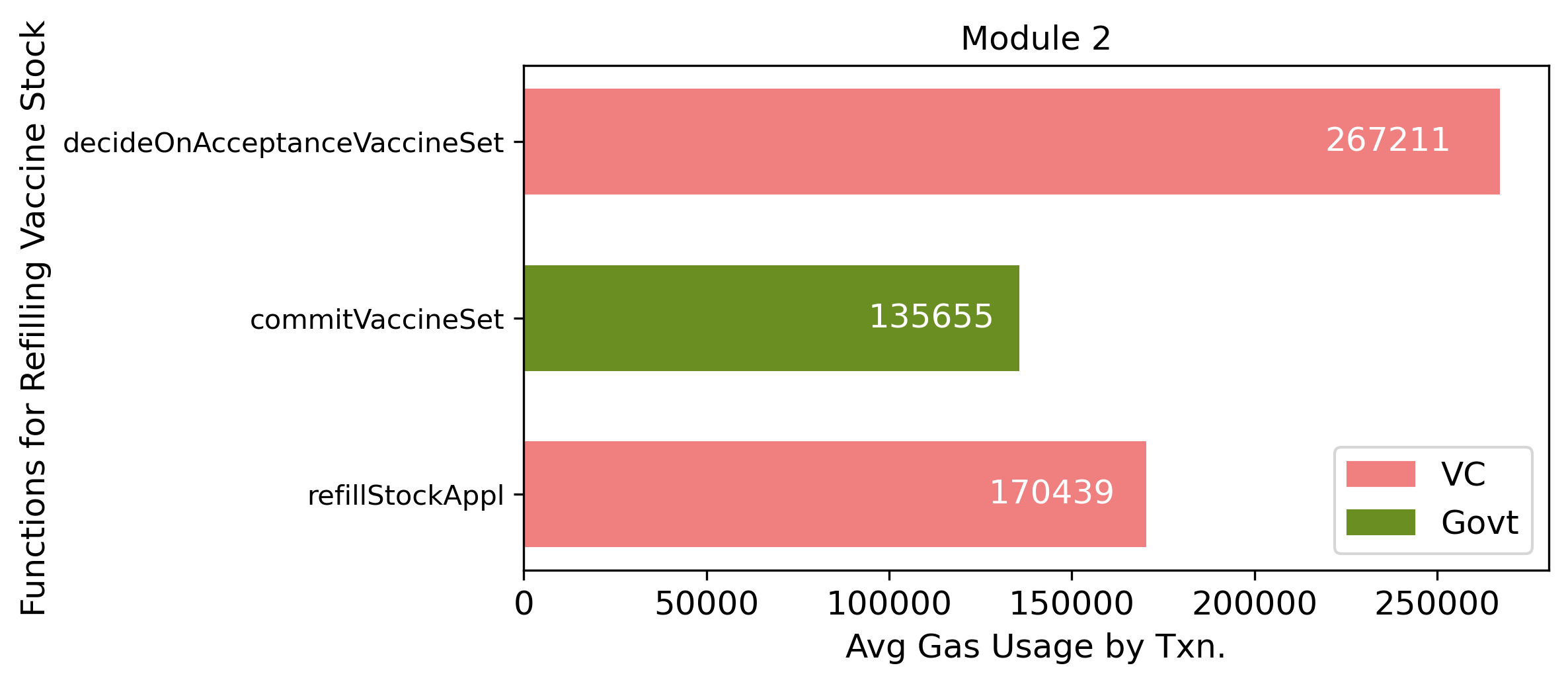}
        \caption{Gas Consumption of Transactions for Refilling Vaccine Stock (Module 2)}
        \label{Fig12}
    \end{minipage}

    \vspace{0.3cm} 

    \begin{minipage}{0.48\textwidth}
        \centering
        \includegraphics[width=\textwidth, height = 5 cm]{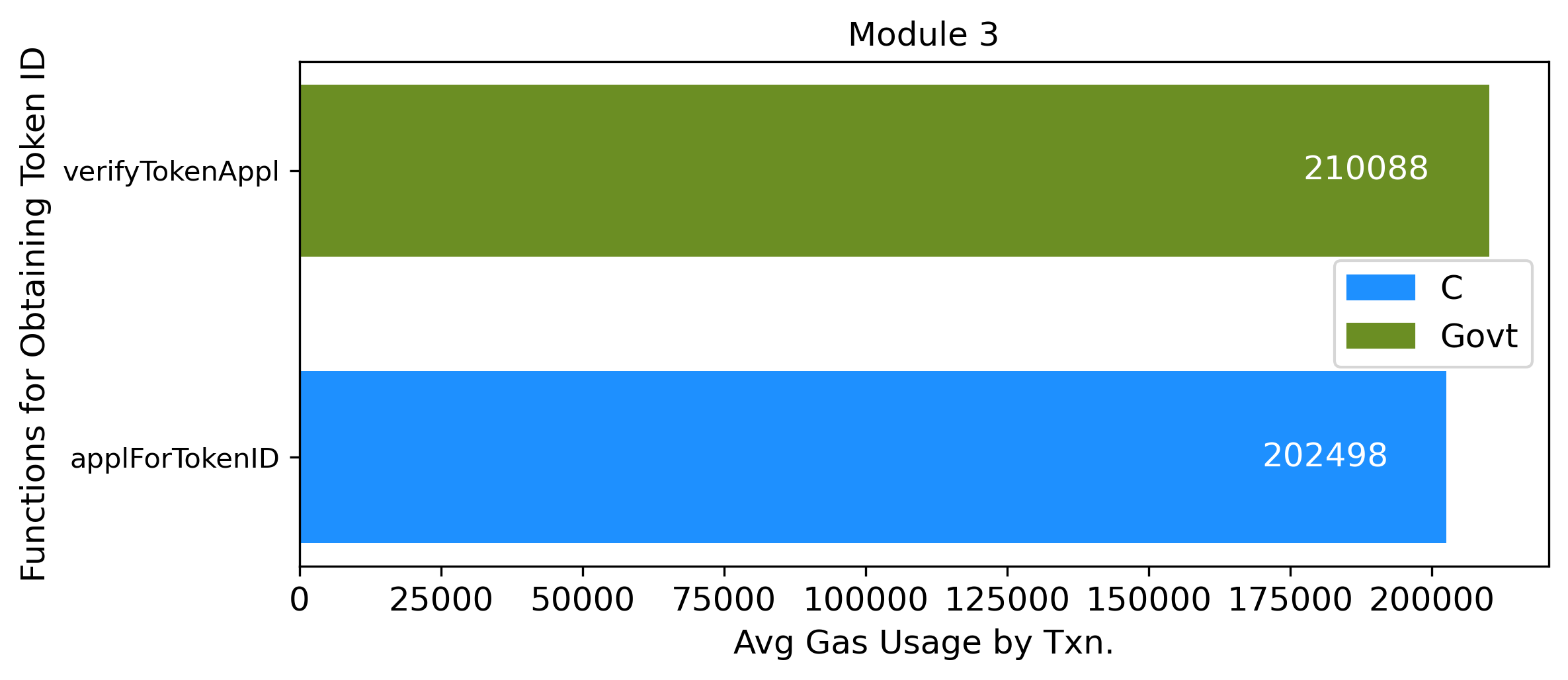}
        \caption{Gas Consumption of Transactions for Obtaining Token ID (Module 3)}
        \label{Fig13}
    \end{minipage}
    \hfill
    \begin{minipage}{0.48\textwidth}
        \centering
        \includegraphics[width=\textwidth, height = 5 cm]{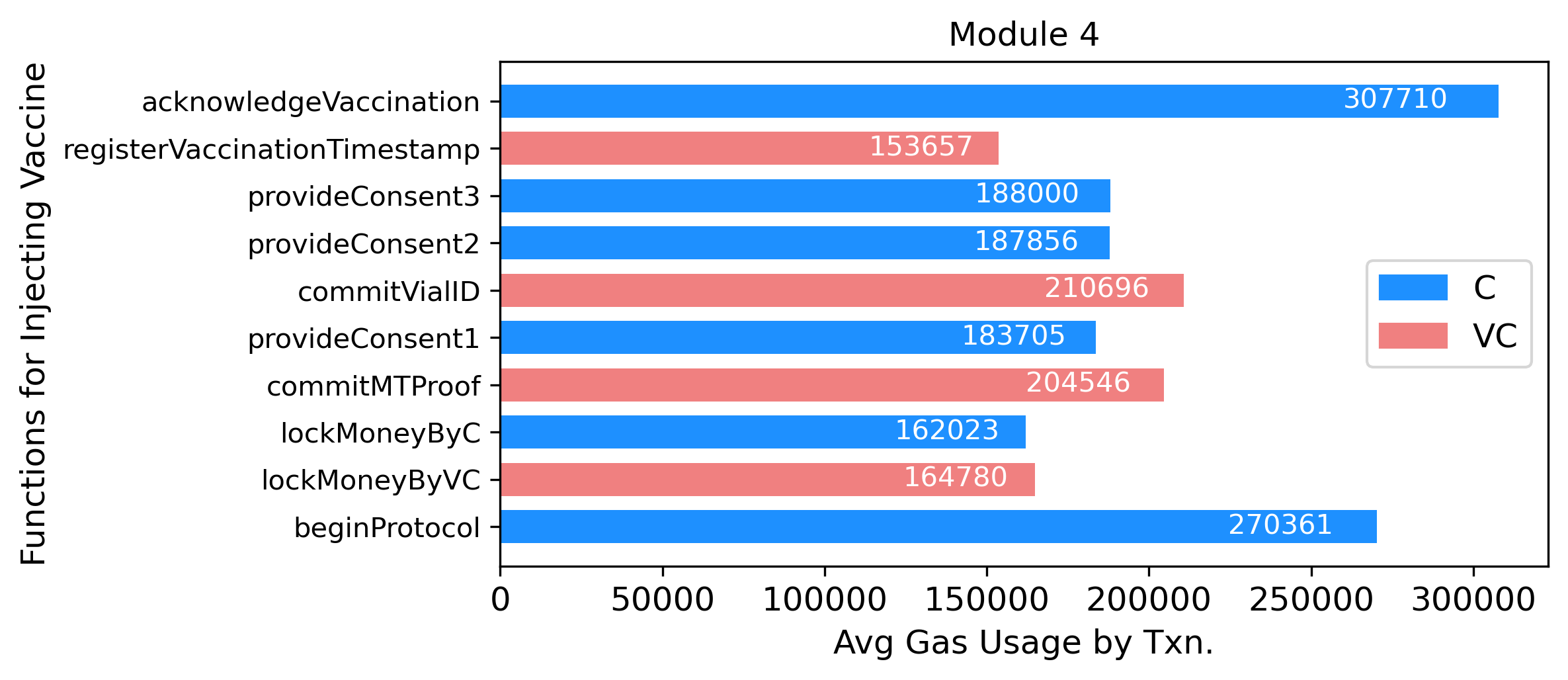}
        \caption{Gas Consumption of Transactions for Injecting Vaccine (Module 4)}
        \label{Fig14}
    \end{minipage}

    \vspace{0.3cm} 

    \begin{minipage}{0.48\textwidth}
        \centering
        \includegraphics[width=\textwidth, height = 5 cm]{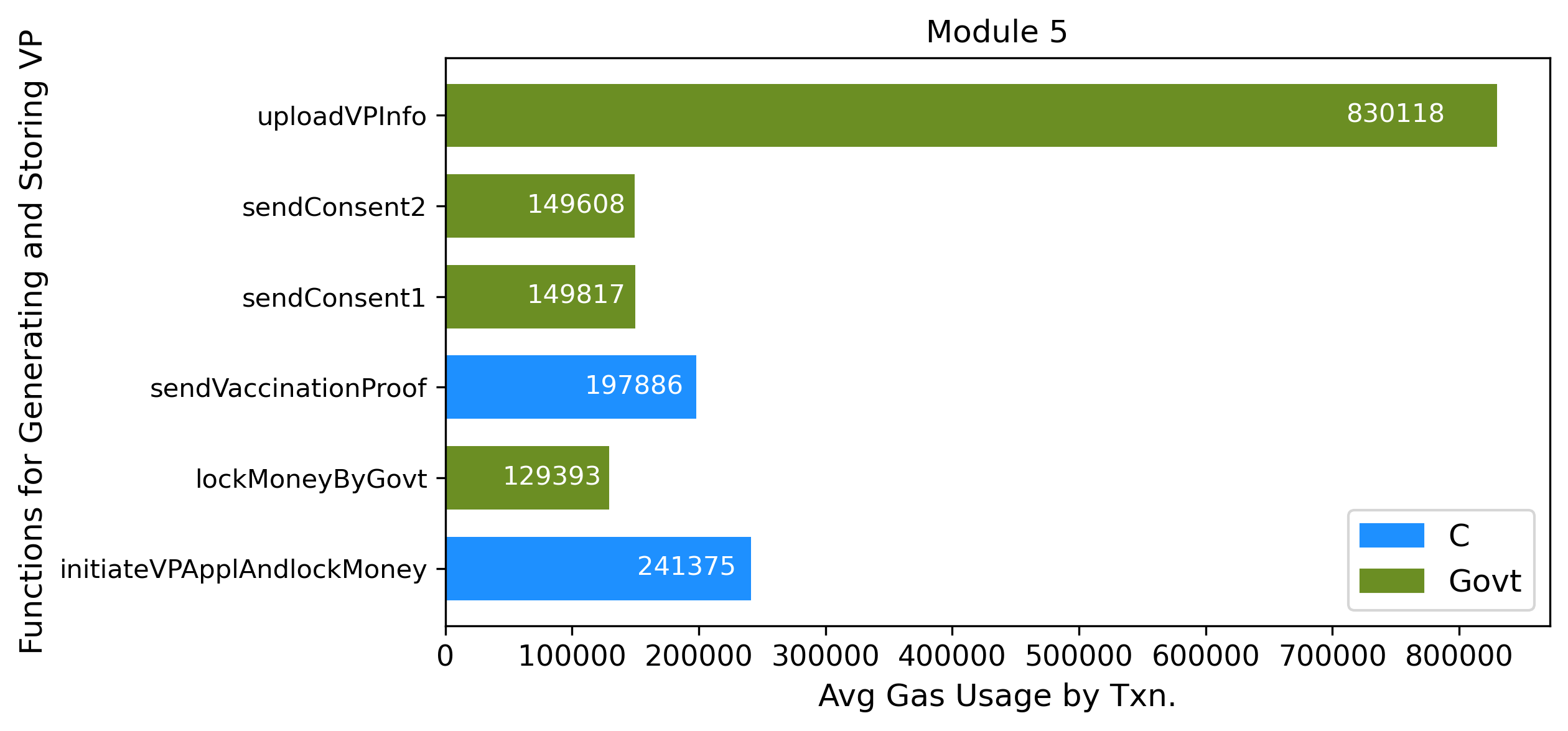}
        \caption{Gas Consumption of Transactions for Generating and Storing Vaccine Passport (Module 5)}
        \label{Fig15}
    \end{minipage}
    \hfill
    \begin{minipage}{0.48\textwidth}
        \centering
        \includegraphics[width=\textwidth, height = 5 cm]{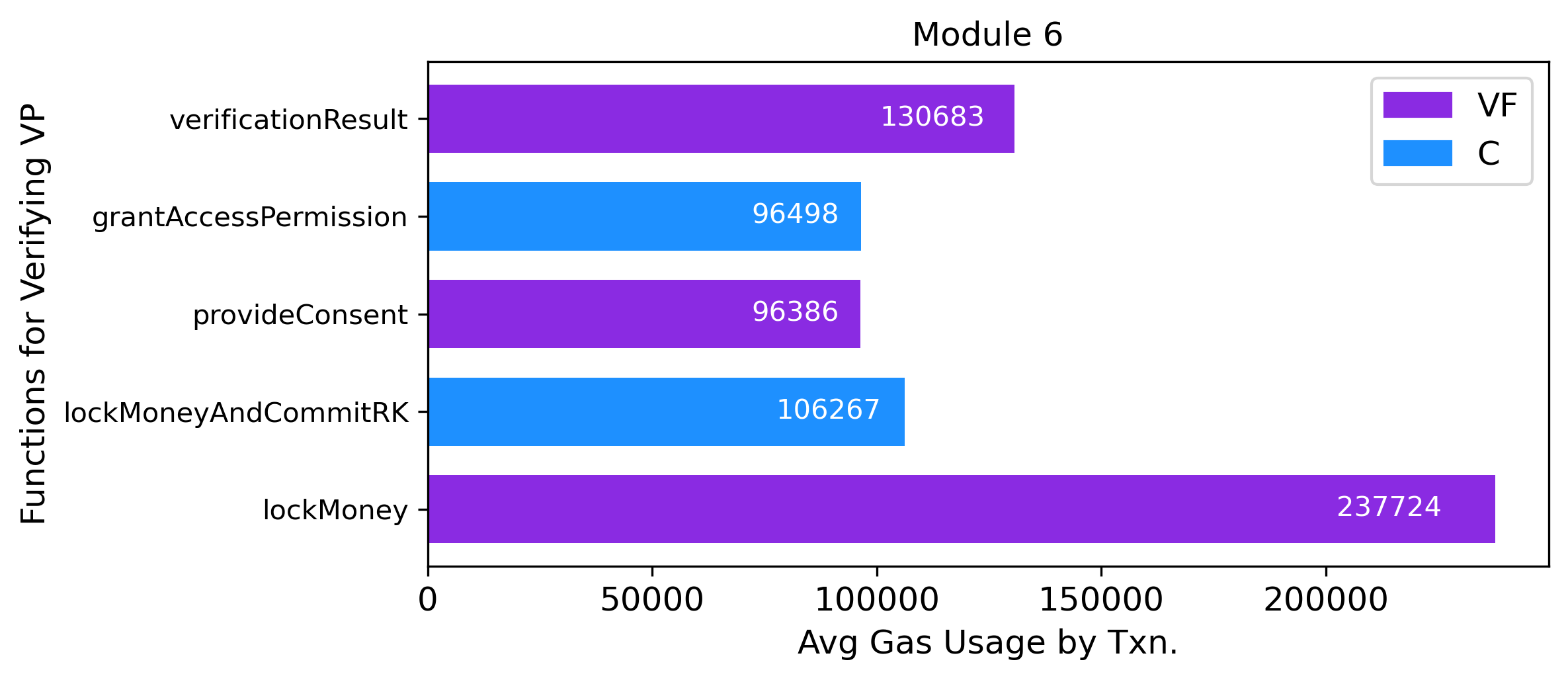}
        \caption{Gas Consumption of Transactions for Verifying Vaccine Passport (Module 6)}
        \label{Fig16}
    \end{minipage}
    
\end{figure*}

\begin{figure*}[!ht]
    \centering
    \begin{minipage}{0.48\textwidth}
        \centering
        \includegraphics[width=\textwidth, height = 5 cm]{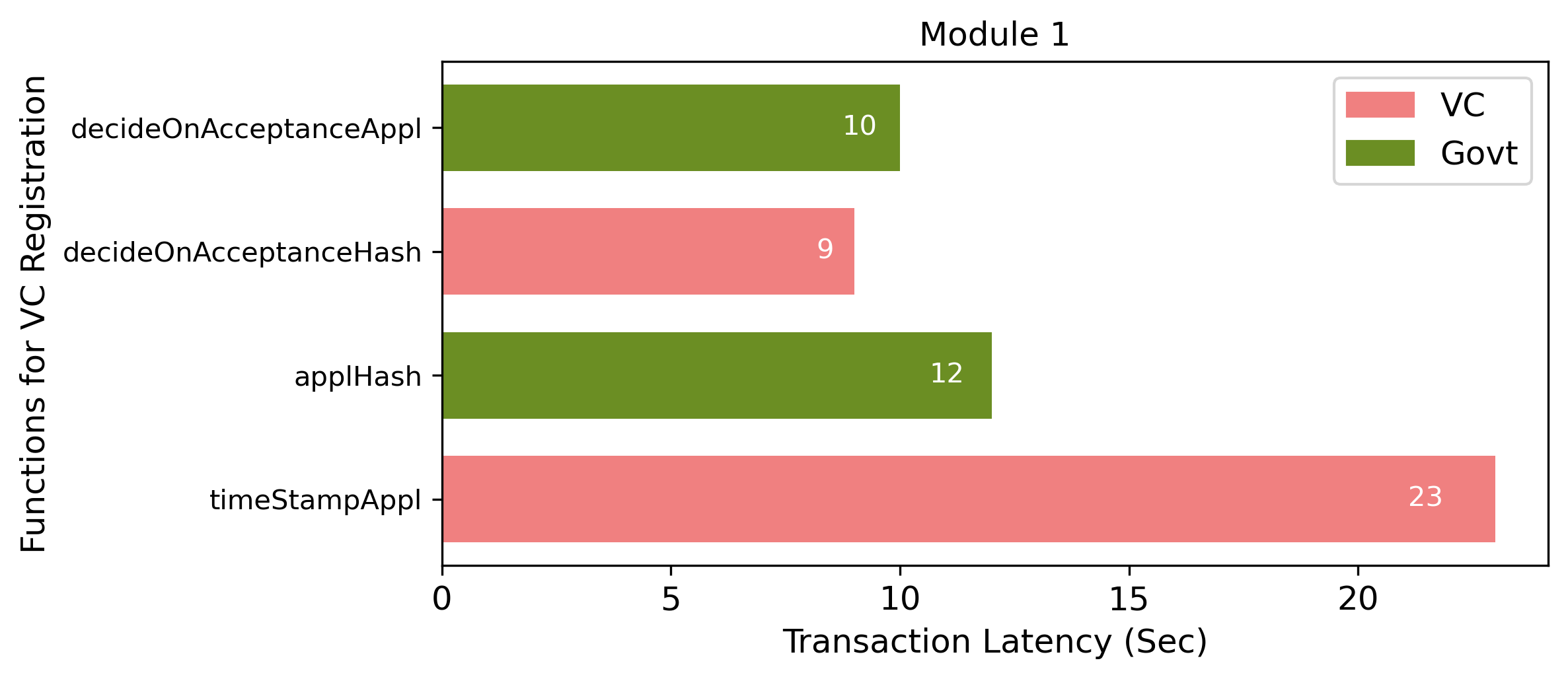}
        \caption{Transactions Latency for VC Registration (Module 1)}
        \label{Fig17}
    \end{minipage}
    \hfill
    \begin{minipage}{0.48\textwidth}
        \centering
        \includegraphics[width=\textwidth, height = 5 cm]{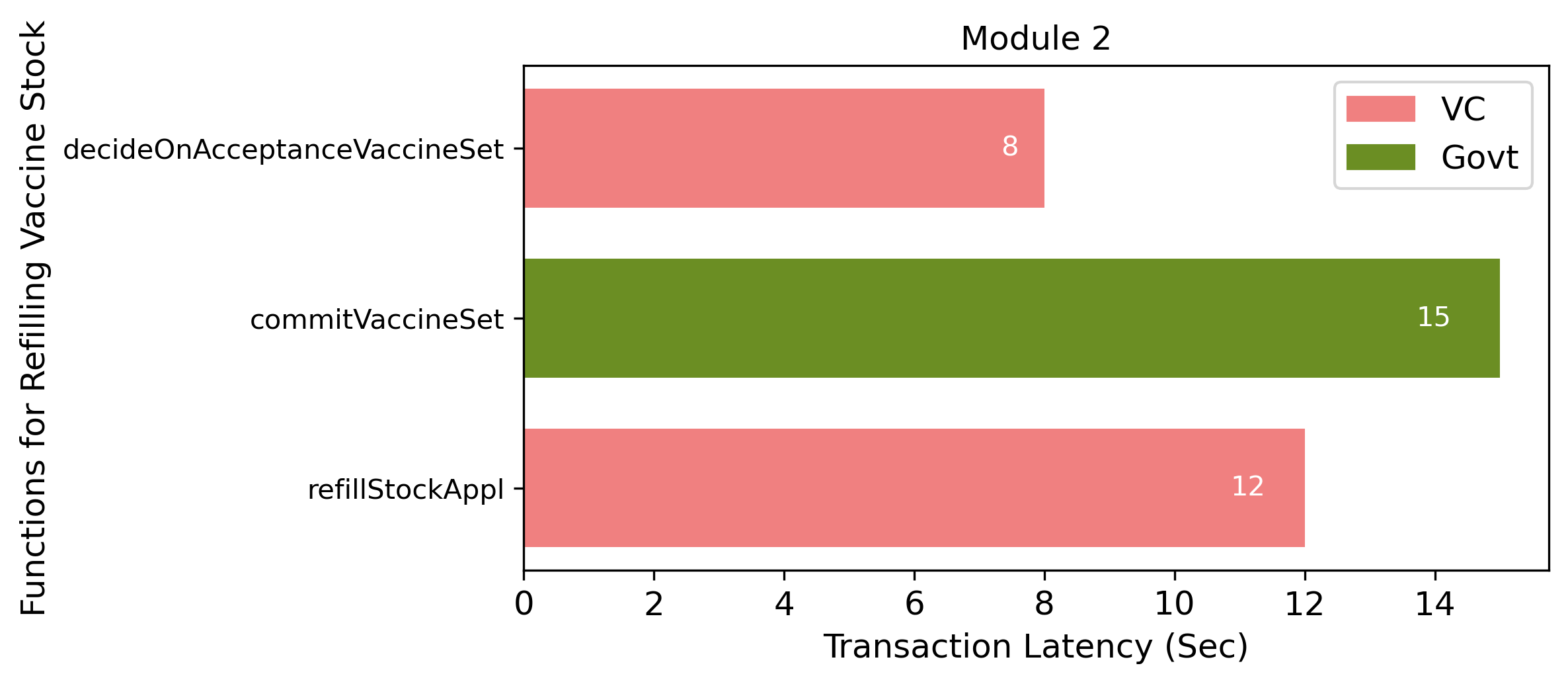}
        \caption{Transactions Latency for Refilling Vaccine Stock (Module 2)}
        \label{Fig18}
    \end{minipage}

    \vspace{0.3cm} 

    \begin{minipage}{0.48\textwidth}
        \centering
        \includegraphics[width=\textwidth, height = 5 cm]{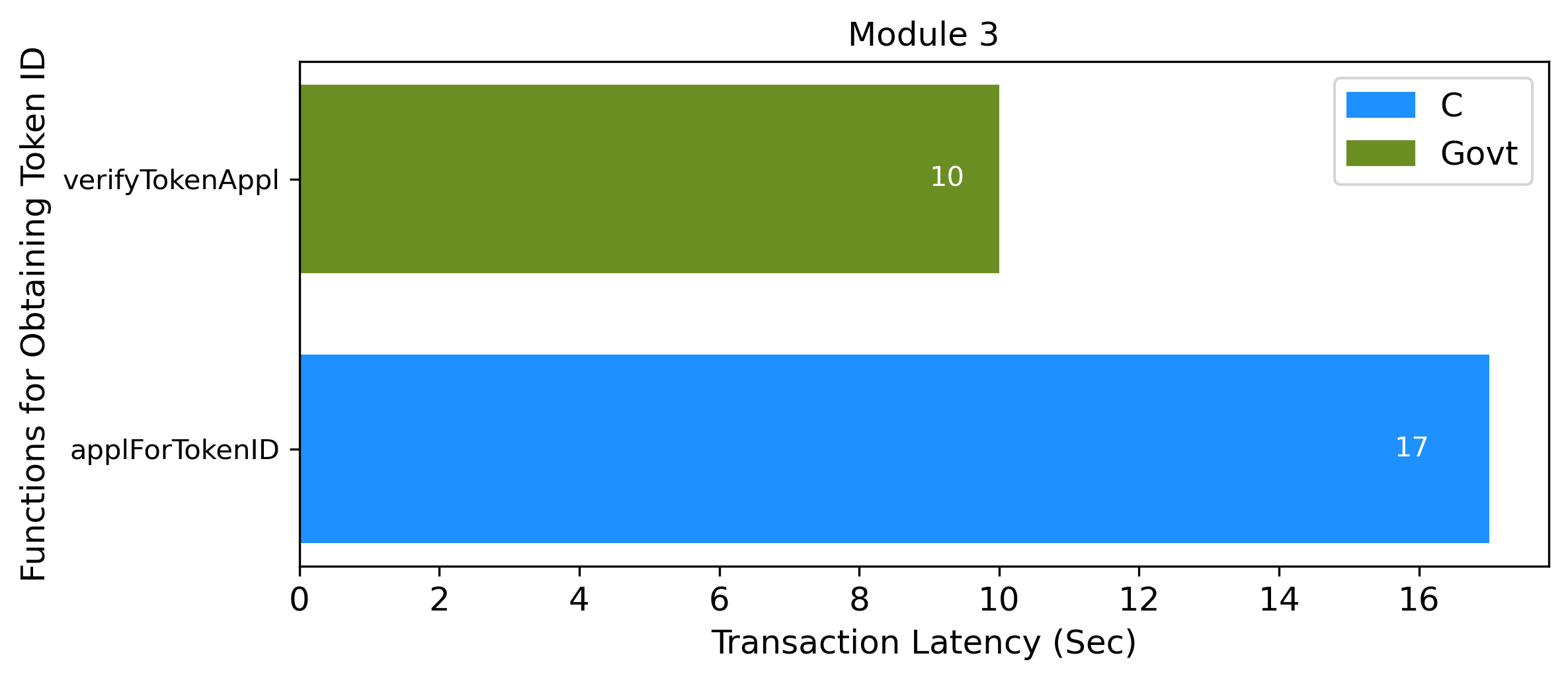}
        \caption{Transactions Latency for Obtaining Token ID (Module 3)}
        \label{Fig19}
    \end{minipage}
    \hfill
    \begin{minipage}{0.48\textwidth}
        \centering
        \includegraphics[width=\textwidth, height = 5 cm]{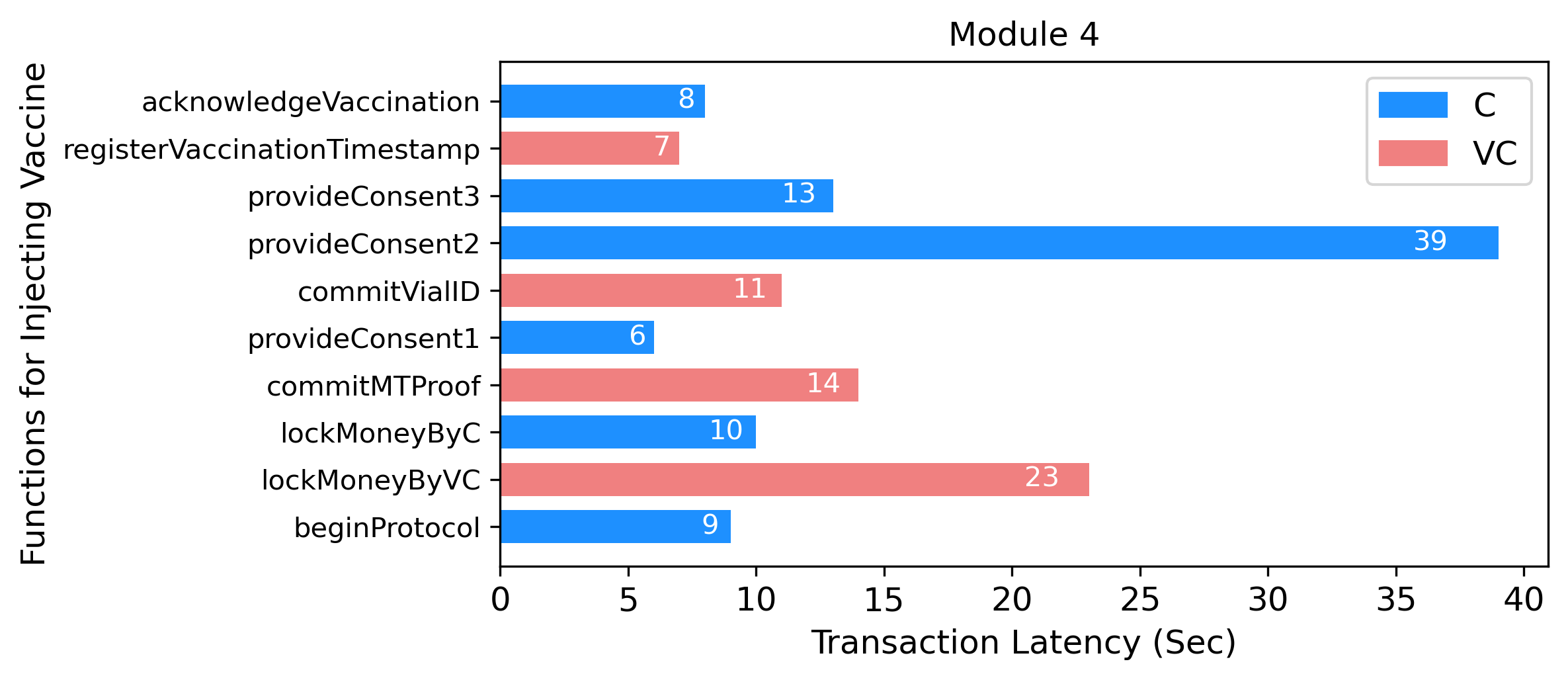}
        \caption{Transactions Latency for Injecting Vaccine (Module 4)}
        \label{Fig20}
    \end{minipage}

    \vspace{0.3cm} 

    \begin{minipage}{0.48\textwidth}
        \centering
        \includegraphics[width=\textwidth, height = 5 cm]{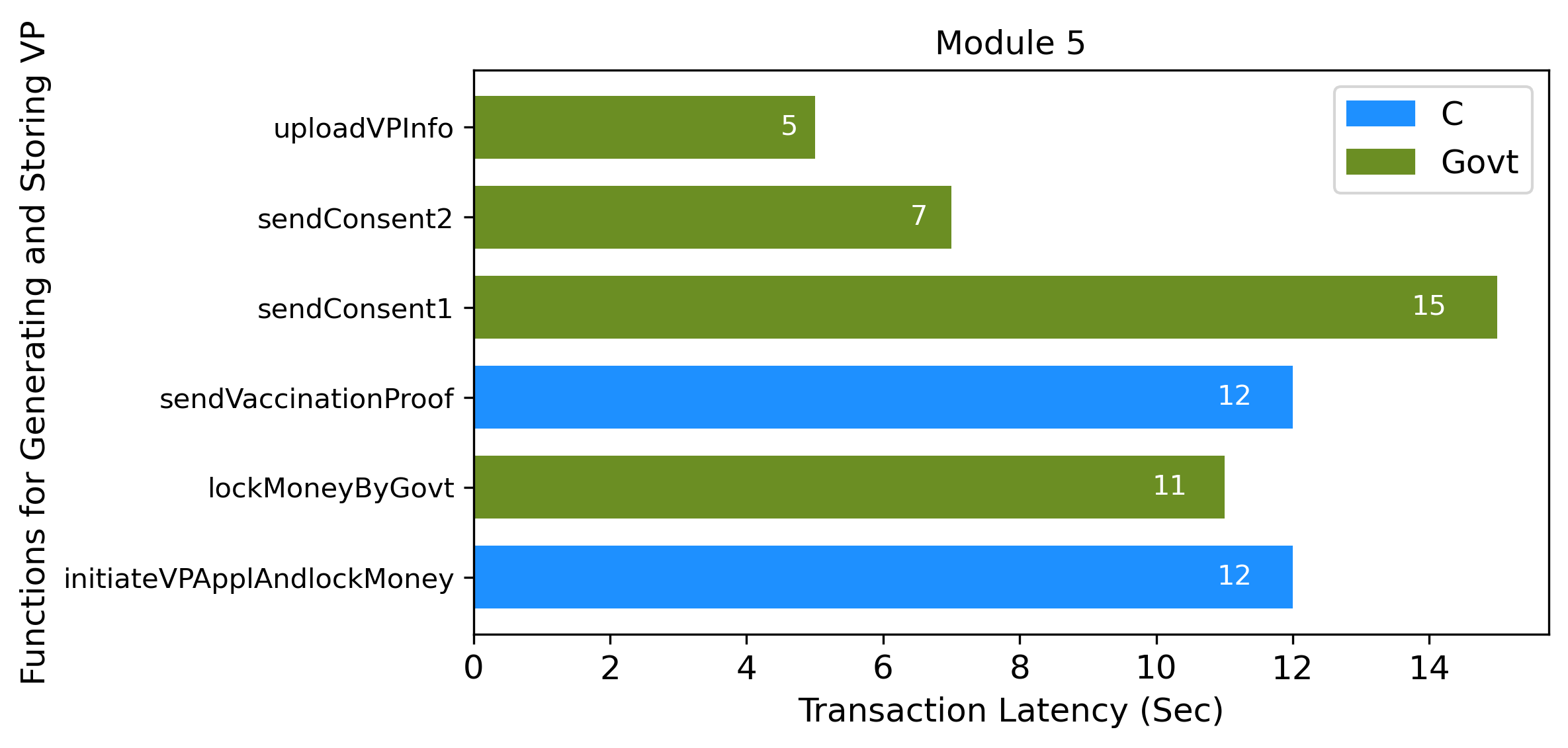}
        \caption{Transactions Latency for Generating and Storing Vaccine Passport (Module 5)}
        \label{Fig21}
    \end{minipage}
    \hfill
    \begin{minipage}{0.48\textwidth}
        \centering
        \includegraphics[width=\textwidth, height = 5 cm]{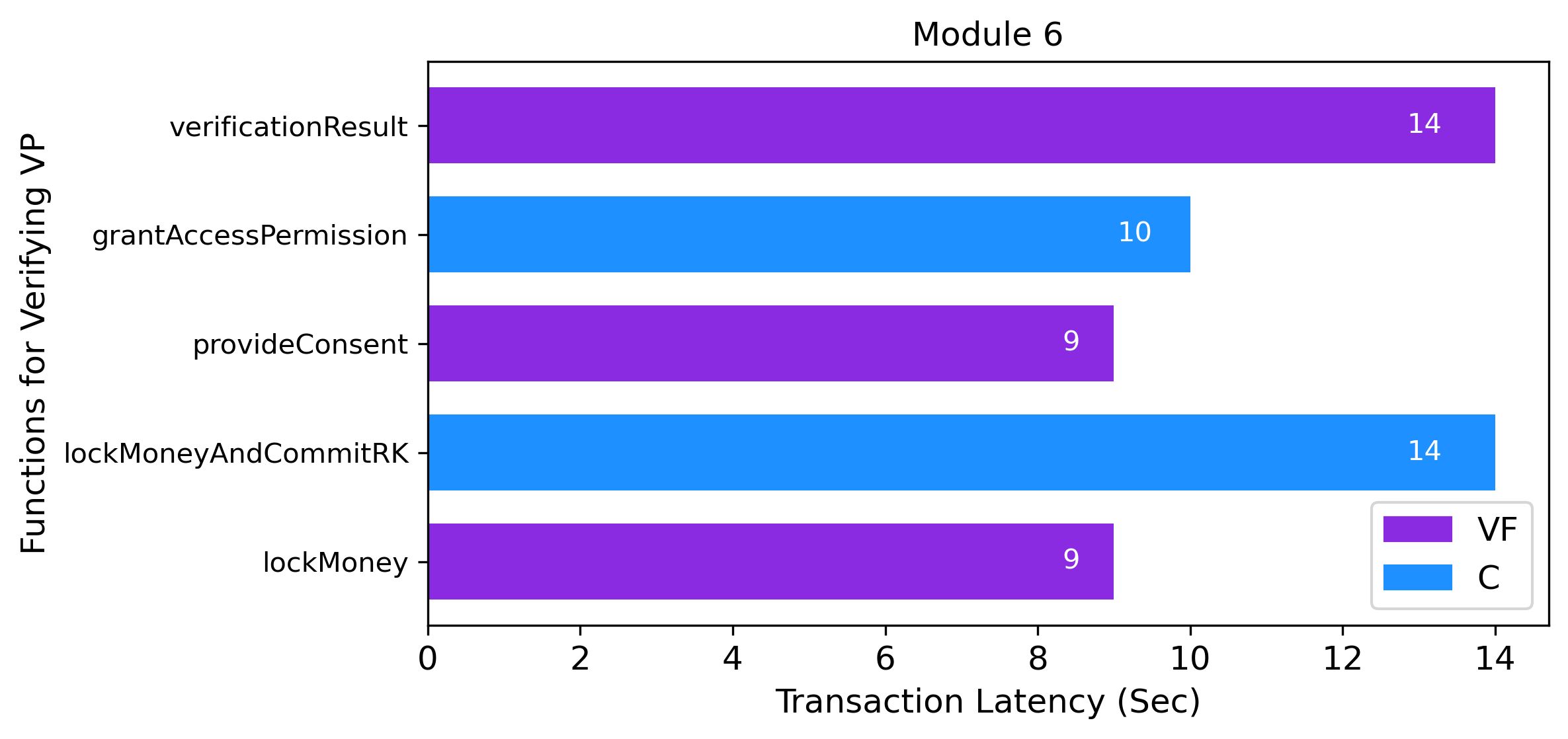}
        \caption{Transactions Latency for Verifying Vaccine Passport (Module 6)}
        \label{Fig22}
    \end{minipage}
    
\end{figure*}

As outlined in Table~\ref{tab1}, two contracts, namely \textit{SC\_C\_VC} and \textit{SC\_C\_Govt}, are split into two parts due to their length. Modules 1 and 2 (i.e., registration of VC and refilling of vaccine stock, respectively) belong to the smart contract \textit{SC\_VC\_Govt}. Module 3 (i.e., obtaining token ID) is part of \textit{SC\_C\_Govt\_1}, while Module 4 (i.e., injecting vaccine) is managed by \textit{SC\_C\_VC}. Module 5 (generating and storing vaccine passports) belongs to \textit{SC\_C\_Govt\_2}, and Module 6 (i.e., verifying a vaccine passport) is handled by \textit{SC\_C\_VF}. Additionally, the required check statements are managed using a separate contract named \textit{SC\_Requirements\_Check}.

The two main factors determining any blockchain model's feasibility are the cost of implementation and the time taken. Likewise, we considered these two main factors during the evaluation of our model.

\begin{figure*}[!ht]
    \centering
    \includegraphics[width=\textwidth]{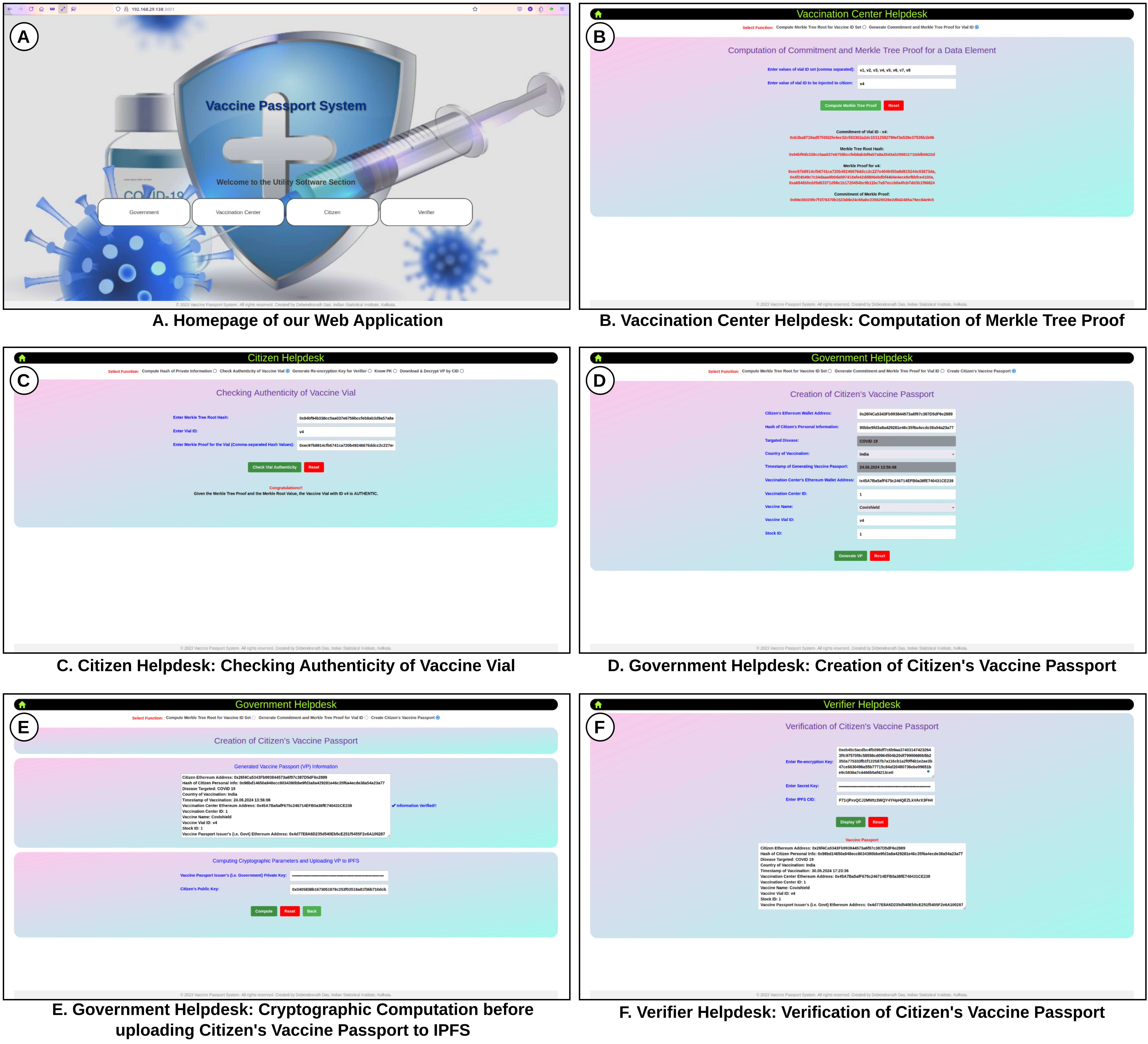}
    \caption{A few screenshots demonstrating the functionalities of our Web Application}
    \label{Fig23}
\end{figure*}

\textbf{Transaction Cost}: When conducting transactions on Ethereum, a gas fee is incurred. Every transaction executed on the Ethereum platform consumes a certain amount of gas, known as the gas cost, which depends on the complexity and computational resource requirements of the transaction. The transaction initiator needs to pay a transaction fee to the validator or miner for including the transaction on the blockchain. This fee is calculated by multiplying the gas cost by the gas price, which represents the unit price of gas in Gwei (1 Gwei = $10^{-9}$ ETH). While the gas cost is fixed, the gas price is a variable parameter chosen by the transaction initiator. Opting for higher gas prices provides a lucrative incentive for validators/miners to include transactions more quickly in a block. To calculate the correct gas price, one must monitor network congestion and current gas price trends. The average gas price in our case was 3.042282375575E-08 ETH, as determined by the MetaMask wallet. At the time of evaluating the results, the exchange rate was 1 ETH = 3831.140006 USD.
The average gas consumption for various transactions related to different modules, including transactions for vaccine registration, refilling vaccine stock, obtaining token IDs, administering vaccines, generating and storing vaccine passports, and verifying vaccine passports, is illustrated in Figures~\ref{Fig11}, \ref{Fig12}, \ref{Fig13}, \ref{Fig14}, \ref{Fig15}, \ref{Fig16}, respectively.

The bar charts depicting gas consumption for the transactions related to various modules of the blockchain-enabled vaccine passport system highlight significant differences based on the complexity and security requirements of each operation. Certain functions incur comparatively higher gas costs due to their more complex functionalities. The \textit{VC Registration} module exhibits moderate gas consumption, indicating a balanced approach to ensuring secure registration without excessive costs. \textit{Refilling Vaccine Stock}, on the other hand, shows relatively high gas consumption, reflecting the extensive validation processes necessary to maintain accurate stock records and prevent discrepancies. \textit{Obtaining a Token ID} is efficient, with lower gas costs, making it a cost-effective process for users to secure their unique identification for vaccine-related activities. \textit{Injecting Vaccines} demonstrates increased gas usage due to multiple verification steps that ensure the integrity and accuracy of vaccination records. The process of \textit{Generating and Storing Vaccine Passports} consumes significant gas, attributed to the intensive cryptographic operations and data storage requirements. Conversely, \textit{Verifying Vaccine Passports} is optimized for efficiency, with relatively low gas consumption, ensuring quick and cost-effective validation of vaccination status.

\textbf{Transaction Latency}: Transaction latency is defined as the time gap between initiating a transaction (i.e., when the initiator confirms the transaction and broadcasts it to the network) and its inclusion in a block of the blockchain after a distributed consensus validation. Public blockchains, such as Ethereum, experience latency due to this consensus process (e.g., Proof of Work (PoW), Proof of Stake (PoS), etc.), which can fluctuate based on network congestion and the computational effort required for validation. High transaction latency can impact the user experience by causing delays in transaction confirmation, which can be critical in time-sensitive applications. We have measured the average latency across a substantial number of successful transactions for our system. The transaction latency for different modules is illustrated in Figures~\ref{Fig17}, \ref{Fig18}, \ref{Fig19}, \ref{Fig20}, \ref{Fig21}, \ref{Fig22}. These measurements provide insights into the performance and reliability of our system under varying network conditions.

We have developed a web-based application to facilitate various offline cryptographic computations required for individual entities (Government, Citizen, Vaccination Center, and Verifier) in our system. These computations include the generation of Merkle tree root hashes using Keccak-256 (SHA-3), the creation of Merkle tree proofs, the commitment of personal information and vaccine vials, and the implementation of a re-encryption protocol based on proxy re-encryption.
The application has been developed using industry-standard web technologies: HTML5, CSS3, JavaScript (ES6+), and Node.js (v20.11.0) for the backend. We've implemented robust input validation and error handling to ensure the security and integrity of cryptographic operations. The user interface is designed to be intuitive, guiding users through complex processes with step-by-step instructions and real-time feedback.
For data storage, we've integrated an IPFS daemon hosted locally to store the encrypted vaccine passports. The web server is hosted locally. The entire source code for the application is available on GitHub repository\footnote{\scriptsize{\url{https://doi.org/10.5281/zenodo.12533359}}} Figure~\ref{Fig23} depicts several screens of our web application, showcasing its clean interface.

\section{Conclusion and Future Scope}
\label{Conclusion and Future Scope}
Our proposed \emph{Blockchain-Enabled Secure Vaccine Passport System} has successfully achieved the initial goals of ensuring secure, transparent, and efficient vaccine administration, creating vaccine passports, and verifying the same. Our system leverages blockchain's immutable ledger and integrates smart contracts to automate various processes, thereby minimizing fraud. While implementing the system, we also ensured that user privacy would not be compromised. Through comprehensive experimental evaluation, we have demonstrated that the system is robust and functions correctly, providing reliable and tamper-proof vaccination records. The use of IPFS for off-chain data storage further enhances security while maintaining accessibility. These results validate our approach and indicate the potential for widespread adoption across various jurisdictions, ultimately contributing to the modernization and security of public health infrastructure.

Moving forward, it is essential to focus on optimizing the scalability and performance of the system to handle increasing adoption and transaction volumes. Ensuring interoperability with existing health information systems and other blockchain networks for seamless integration and data exchange will be crucial. Additionally, the system should be adaptable to comply with diverse regulatory requirements across different regions, enhancing its global applicability. Improving the user experience, particularly for non-technical users, will be essential for widespread acceptance. By addressing these areas, the blockchain-enabled secure vaccine passport system can develop into a comprehensive solution, significantly advancing global public health infrastructure and vaccine management. With such a system in place, we can effectively manage global pandemics like COVID-19 without resorting to nationwide lockdowns and severe economic disruptions.

\bibliographystyle{ieeetr}
\bibliography{VaccinePassport}

\begin{thebibliography}{10}

\bibitem{ali2020covid}
I.~Ali and O.~M. Alharbi, ``Covid-19: Disease, management, treatment, and social impact,'' {\em Science of the total Environment}, vol.~728, p.~138861, 2020.

\bibitem{park2021fighting}
Y.~J. Park, J.~Farooq, J.~Cho, N.~Sadanandan, B.~Cozene, B.~Gonzales-Portillo, M.~Saft, M.~C. Borlongan, M.~C. Borlongan, R.~D. Shytle, {\em et~al.}, ``Fighting the war against covid-19 via cell-based regenerative medicine: lessons learned from 1918 spanish flu and other previous pandemics,'' {\em Stem cell reviews and reports}, vol.~17, pp.~9--32, 2021.

\bibitem{patel2010pandemic}
M.~Patel, A.~Dennis, C.~Flutter, and Z.~Khan, ``Pandemic (h1n1) 2009 influenza,'' {\em British journal of anaesthesia}, vol.~104, no.~2, pp.~128--142, 2010.

\bibitem{gatherer20142014}
D.~Gatherer, ``The 2014 ebola virus disease outbreak in west africa,'' {\em Journal of general virology}, vol.~95, no.~8, pp.~1619--1624, 2014.

\bibitem{covirestriction}
H.~Correspondent, ``{Covishield may not be eligible for ‘vaccine passport’ by the EU}.'' \url{http://tinyurl.com/y8sja52s}, June 28, 2021.

\bibitem{hasan2020blockchain}
H.~R. Hasan, K.~Salah, R.~Jayaraman, J.~Arshad, I.~Yaqoob, M.~Omar, and S.~Ellahham, ``Blockchain-based solution for covid-19 digital medical passports and immunity certificates,'' {\em Ieee Access}, vol.~8, pp.~222093--222108, 2020.

\bibitem{barati2021privacy}
M.~Barati, W.~J. Buchanan, O.~Lo, and O.~Rana, ``A privacy-preserving distributed platform for covid-19 vaccine passports,'' in {\em Proceedings of the 14th IEEE/ACM international conference on utility and cloud computing companion}, pp.~1--6, 2021.

\bibitem{haque2021towards}
A.~B. Haque, B.~Naqvi, A.~N. Islam, and S.~Hyrynsalmi, ``Towards a gdpr-compliant blockchain-based covid vaccination passport,'' {\em Applied Sciences}, vol.~11, no.~13, p.~6132, 2021.

\bibitem{shaikh2022block}
R.~N. Shaikh, C.~G. Jadhav, V.~R. Bhogawade, G.~Narang, and A.~M. Gangadhar, ``Block chain based electronic vaccination record storing system,'' in {\em 2022 8th International Conference on Advanced Computing and Communication Systems (ICACCS)}, vol.~1, pp.~272--276, IEEE, 2022.

\bibitem{agbedanu2022blocovid}
P.~Agbedanu, F.~U. Bawah, V.~Akoto-Adjepong, N.~Awarayi, I.~Nti, S.~Boateng, P.~Nimbe, and O.~Nyarko-Boateng, ``Blocovid: A blockchain-based covid-19 digital vaccination certificate verification system,'' in {\em 2022 International Conference on Engineering and Emerging Technologies (ICEET)}, pp.~1--6, IEEE, 2022.

\bibitem{bradish2023covichain}
P.~Bradish, S.~Chaudhari, M.~Clear, and H.~Tewari, ``Covichain: A blockchain based covid-19 vaccination passport,'' in {\em Future of Information and Communication Conference}, pp.~195--206, Springer, 2023.

\bibitem{wang2022blockchain}
R.~Wang, B.~Wu, and T.~Xia, ``A blockchain-based multiple-parties-involved vaccination passport system,'' in {\em 2022 3rd International Conference on E-commerce and Internet Technology (ECIT 2022)}, pp.~772--785, Atlantis Press, 2022.

\bibitem{nabil2022blockchain}
S.~S. Nabil, M.~S.~A. Pran, A.~A. Al~Haque, N.~R. Chakraborty, M.~J.~M. Chowdhury, and M.~S. Ferdous, ``Blockchain-based covid vaccination registration and monitoring,'' {\em Blockchain: Research and Applications}, vol.~3, no.~4, p.~100092, 2022.

\bibitem{shih2022international}
D.-H. Shih, P.-L. Shih, T.-W. Wu, S.-H. Liang, and M.-H. Shih, ``An international federal hyperledger fabric verification framework for digital covid-19 vaccine passport,'' in {\em Healthcare}, vol.~10, p.~1950, MDPI, 2022.

\bibitem{rashid2022block}
M.~M. Rashid, P.~Choi, S.-H. Lee, and K.-R. Kwon, ``Block-hpct: blockchain enabled digital health passports and contact tracing of infectious diseases like covid-19,'' {\em Sensors}, vol.~22, no.~11, p.~4256, 2022.

\bibitem{pericas2022highly}
R.~Peric{\`a}s-Gornals, M.~Mut-Puigserver, and M.~M. Payeras-Capell{\`a}, ``Highly private blockchain-based management system for digital covid-19 certificates,'' {\em International Journal of Information Security}, vol.~21, no.~5, pp.~1069--1090, 2022.

\bibitem{cao2022blockchain}
Y.~Cao, J.~Chen, and Y.~Cao, ``Blockchain-based privacy-preserving vaccine passport system,'' {\em Security and Communication Networks}, vol.~2022, no.~1, p.~4769187, 2022.

\bibitem{fugkeaw2023efficient}
S.~Fugkeaw, ``An efficient and scalable vaccine passport verification system based on ciphertext policy attribute-based encryption and blockchain,'' {\em Journal of Cloud Computing}, vol.~12, no.~1, p.~111, 2023.

\bibitem{koyama2023decentralized}
A.~Koyama, V.~C. Tran, M.~Fujimoto, V.~N.~Q. Bao, and T.~H. Tran, ``A decentralized covid-19 vaccine tracking system using blockchain technology,'' {\em Cryptography}, vol.~7, no.~1, p.~13, 2023.

\bibitem{masood2024developing}
F.~Masood and A.~R. Faridi, ``Developing a novel blockchain-based vaccine tracking and certificate system: An end-to-end approach,'' {\em Peer-to-Peer Networking and Applications}, pp.~1--19, 2024.

\bibitem{chakraborty2020contact}
P.~Chakraborty, S.~Maitra, M.~Nandi, and S.~Talnikar, ``Contact tracing in post-covid world,'' {\em Indian Statistical Institute Series}, 2020.

\bibitem{vci}
{Vaccination Credential Initiative}, ``{VCI}.'' \url{https://vci.org/}.

\bibitem{commonpass}
{The Commons Project Foundation}, ``{CommonPass}.'' \url{https://www.thecommonsproject.org/}.

\bibitem{ibmhealthpass}
{IBM}, ``{IBM Digital Health Pass}.'' \url{https://www.ibm.com/products/digital-health-pass}.

\bibitem{iata}
{International Air Transport Association}, ``{IATA Travel Pass Initiative}.'' \url{https://www.iata.org/en/programs/passenger/travel-pass/}.

\bibitem{dgc}
{European Commission}, ``{EU Digital COVID Certificate}.'' \url{https://ec.europa.eu/info/live-work-travel-eu/coronavirus-response/safe-covid-19-vaccines-europeans/eu-digital-covid-certificate_en}.

\bibitem{coronapass}
{Governement of Denmark}, ``{Corona Passport}.'' \url{https://www.sst.dk/en/English}.

\bibitem{greenpass}
{Government of Israel}, ``{Green Pass}.'' \url{https://www.gov.il/en/departments/guides/green-pass-info}.

\bibitem{das2024bisection}
D.~Das, ``Bisection: Blockchain-enabled secure health insurance processing,'' {\em International Journal of Ad Hoc and Ubiquitous Computing}, 2024.

\bibitem{merkle2001certified}
R.~C. Merkle, ``A certified digital signature,'' in {\em Advances in cryptology—CRYPTO’89 proceedings}, pp.~218--238, Springer, 2001.

\bibitem{kerry2013digital}
C.~F. Kerry and P.~D. Gallagher, ``Digital signature standard (dss),'' {\em FIPS PUB}, pp.~186--4, 2013.

\bibitem{Preneel2011}
B.~Preneel, {\em Hash Functions}, pp.~543--553.
\newblock Boston, MA: Springer US, 2011.

\bibitem{Liu2021MT}
H.~Liu, X.~Luo, H.~Liu, and X.~Xia, ``Merkle tree: A fundamental component of blockchains,'' in {\em 2021 International Conference on Electronic Information Engineering and Computer Science (EIECS)}, pp.~556--561, 2021.

\bibitem{merkle1988digital}
R.~C. Merkle, ``A digital signature based on a conventional encryption function,'' in {\em Advances in Cryptology—CRYPTO’87: Proceedings 7}, pp.~369--378, Springer, 1988.

\bibitem{jing2021merkle}
S.~Jing, X.~Zheng, and Z.~Chen, ``Review and investigation of merkle tree’s technical principles and related application fields,'' in {\em 2021 International Conference on Artificial Intelligence, Big Data and Algorithms (CAIBDA)}, pp.~86--90, 2021.

\bibitem{nabeels2011proxy}
N.~Blog, ``{Proxy Re-encryption}.'' \url{https://mohamednabeel.blogspot.com/2011/03/proxy-re-encryption.html}, March 2, 2011.

\bibitem{narayanan2016bitcoin}
A.~Narayanan, J.~Bonneau, E.~Felten, A.~Miller, and S.~Goldfeder, {\em Bitcoin and cryptocurrency technologies: a comprehensive introduction}.
\newblock Princeton University Press, 2016.

\bibitem{das2022blockchain}
D.~Das, A.~Muthaiah, and S.~Ruj, ``Blockchain-enabled secure and smart healthcare system,'' in {\em International Conference on Design Science Research in Information Systems and Technology}, pp.~97--109, Springer, 2022.

\bibitem{benet2014ipfs}
J.~Benet, ``Ipfs-content addressed, versioned, p2p file system,'' {\em arXiv preprint arXiv:1407.3561}, 2014.

\end{thebibliography}

\end{document}